\documentclass[twocolumn]{aastex631}

\usepackage{upgreek}
\usepackage{textcomp}
\usepackage{wasysym}
\usepackage{blindtext}
\usepackage{lineno}
\usepackage{rotating}
\usepackage{hyperref}
\usepackage{xcolor}
\usepackage{longtable}
\usepackage{tablefootnote}

\shorttitle{ch2/W2 Variability and Binary Brown Dwarfs}
\shortauthors{Brooks et al.}

\begin{document}

\graphicspath{{./}{figures/}}

\title{Long-term 4.6$\mu$m Variability in Brown Dwarfs and a New Technique for Identifying Brown Dwarf Binary Candidates}

\author[0000-0002-5253-0383]{Hunter Brooks}
\affiliation{Department of Astronomy and Planetary Science, Northern Arizona University, Flagstaff, AZ 86011, USA}
\affiliation{IPAC, Mail Code 100-22, Caltech, 1200 E. California Blvd., Pasadena, CA 91125, USA}

\author[0000-0003-4269-260X]{J.\ Davy Kirkpatrick}
\affiliation{IPAC, Mail Code 100-22, Caltech, 1200 E. California Blvd., Pasadena, CA 91125, USA}

\author[0000-0002-1125-7384]{Aaron M. Meisner}
\affiliation{NSF's National Optical-Infrared Astronomy Research Laboratory, 950 N. Cherry Ave., Tucson, AZ 85719, USA}

\author[0000-0001-5072-4574]{Christopher R. Gelino}
\affiliation{IPAC, Mail Code 100-22, Caltech, 1200 E. California Blvd., Pasadena, CA 91125, USA}

\author[0000-0001-8170-7072]{Daniella C. Bardalez Gagliuffi}
\affiliation{Physics and Astronomy Department, Amherst College, 21 Merrill Science Drive, Amherst, MA 01002, USA}
\affiliation{Department of Astrophysics, American Museum of Natural History, Central Park West at 79th Street, New York, NY 10024, USA}

\author[0000-0001-7519-1700]{Federico Marocco}
\affiliation{IPAC, Mail Code 100-22, Caltech, 1200 E. California Blvd., Pasadena, CA 91125, USA}

\author[0000-0002-6294-5937]{Adam C. Schneider}
\affiliation{United States Naval Observatory, Flagstaff Station, 10391 West Naval Observatory Rd., Flagstaff, AZ 86005, USA} 
\affiliation{Department of Physics and Astronomy, George Mason University, MS3F3, 4400 University Drive, Fairfax, VA 22030, USA}

\author[0000-0001-6251-0573]{Jacqueline K. Faherty}
\affiliation{Department of Astrophysics, American Museum of Natural History, Central Park West at 79th Street, New York, NY 10024, USA}

\author[0000-0003-2478-0120]{S.L.Casewell}
\affiliation{School of Physics and Astronomy, University of Leicester, University Road, Leicester, LE1 7RH, UK}

\author[0000-0001-9778-7054]{Yadukrishna Raghu}
\affiliation{Backyard Worlds: Planet 9}

\author[0000-0002-2387-5489]{Marc J. Kuchner}
\affiliation{NASA Goddard Space Flight Center, Exoplanets and Stellar Astrophysics Laboratory, Code 667, Greenbelt, MD 20771, USA}

\author{The Backyard Worlds:  Planet 9 Collaboration}

\begin{abstract}
Using a sample of 361 nearby brown dwarfs, we have searched for 4.6$\mu$m variability indicative of large-scale rotational modulations or large-scale long-term changes on timescales of over 10 years. Our findings show no statistically significant variability in \textit{Spitzer} ch2 or \textit{WISE} W2 photometry. For \textit{Spitzer} the ch2 1$\sigma$ limits are $\sim$8 mmag for objects at 11.5 mag and $\sim$22 mmag for objects at 16 mag. This corresponds to no variability above 4.5$\%$ at 11.5 mag and 12.5$\%$ at 16 mag. We conclude that highly variable brown dwarfs, at least two previously published examples of which have been shown to have 4.6$\mu$m variability above 80 mmag, are very rare. While analyzing the data, we also developed a new technique for identifying brown dwarfs binary candidates in \textit{Spitzer} data. We find that known binaries have IRAC ch2 PRF (point response function) flux measurements that are consistently dimmer than aperture flux measurements. We have identified 59 objects that exhibit such PRF versus apertures flux differences and are thus excellent binary brown dwarf candidates. 
\end{abstract}


\section{Introduction}
Brown dwarfs have similar chemical composition and size to large gaseous planets, indicating that brown dwarfs have the potential for cloud bands and large surface storms, like that of the Great Red Spot on Jupiter. It has been found that some brown dwarfs exhibit variability over short periods of a couple of hours to a few days, first reported in \cite{Bailer 1999} and \cite{Kolb 1999}. \cite{Artigau et al.(2009)} and \cite{Yang et al.(2016)} found variability as high as 14-420 mmag, mostly in the $J$, $H$, and $K$ bands over several rotational periods. This is a result of inhomogeneities in the brown dwarf's cloud bands or local storms, which induces observed variability as the object rotates. However, none of these papers explore long-term variability over timescales of years. Additionally, we might expect stellar flares in the warmest brown dwarfs as a result of their similarities to low-mass stars (\citealt{Gizis et al.(2017)}, \citealt{Ducrot et al.(2020)}). 

The goal of this paper is to examine long-term variability, on years to decade timescales, at $\sim$4.6$\mu$m for a large brown dwarf sample. This would not only be sensitive to rotationally modulated variability due to inhomogeneity over the surface, but would also be sensitive to long-term global changes in the atmosphere. 

While examining variability, we developed a new technique for identifying brown dwarf binary candidates in \textit{Spitzer} data. Our \textit{Spitzer} reductions measure both the aperture and point response function-fit (hereafter, PRF-fit) fluxes. If the PRF-fit fluxes are consistently dimmer than the aperture measurements, then it may be indicative of a binary brown dwarf system. We demonstrate this with known binaries and background-contaminated objects.

In \S\ref{expect} we discuss expectations of variability based on previous research. We then discuss the data acquisition and reduction in \S\ref{photometry} and its analysis in \S\ref{trust}. Long-term variability is explored in \S\ref{variability}, while the new technique for discovering brown dwarf binary candidates is discussed in \S\ref{binaries}. The results from these two sections are then compared to what was previously known about brown dwarf variability and binarity in \S\ref{dis}. 

\section{Variability Expectations}\label{expect}
Most previous papers on brown dwarf variability have focused on hemisphere-to-hemisphere inhomogeneity. As a result of brown dwarfs having a range of rotational periods from about an hour to a few weeks (\citealt{Tannock et al.(2021)}, \citealt{Luhman(2012)}), these studies have only concentrated on a few-day to few-week timescale. Some of these studies have repeated this analysis on return trips to look for anomalies on the surface of brown dwarfs that might change many rotational periods in the future (\citealt{Apai et al.(2017)}). 

\begin{figure*}
        \centering
        \includegraphics[scale = 0.72]{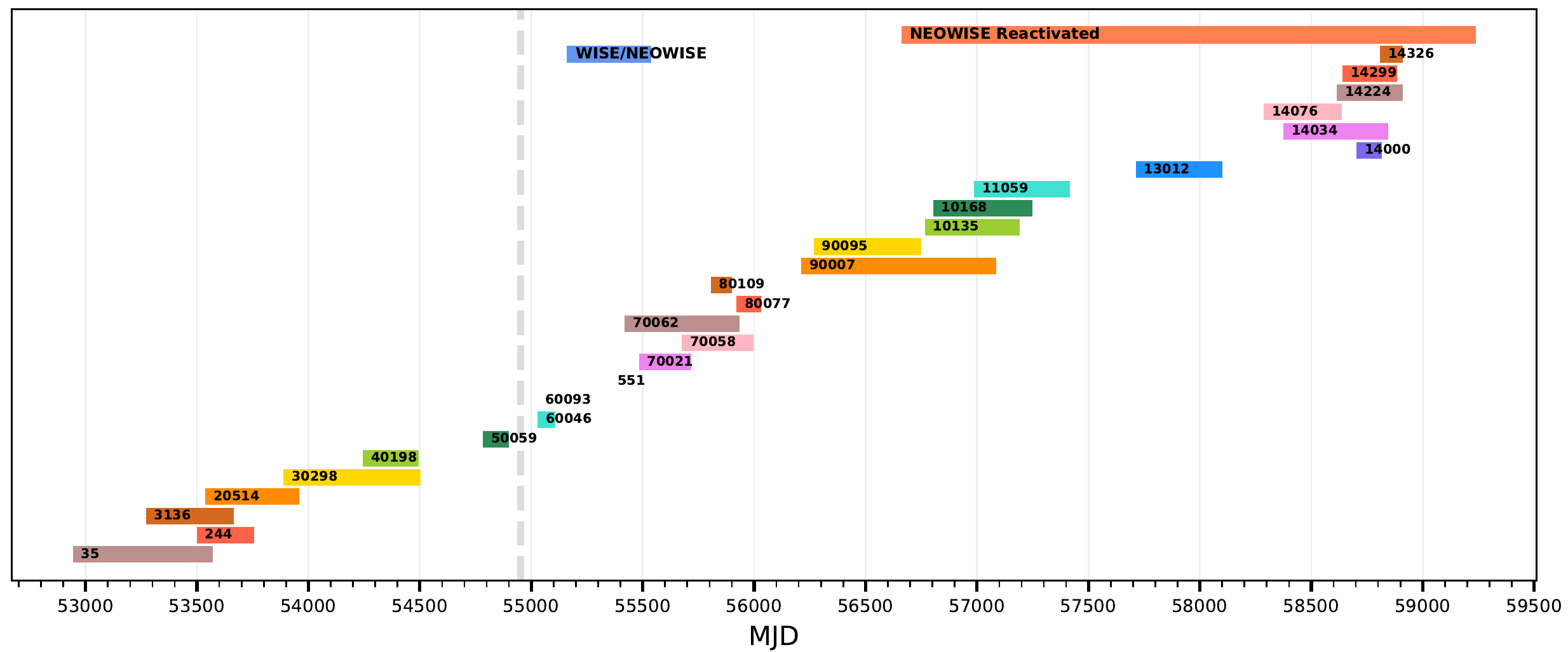}
        \caption{Observation windows of both the \textit{Spitzer} and \textit{WISE} data sets used in this paper. The \textit{Spitzer} program number (black numbers) and the time span that it surveyed are shown by the rainbow lines if the time period is only a handful of nights (Table 3 in \citealt{Kirkpatrick et al.(2021)}). The \textit{WISE} timescale is shown by the top two lines that include ``WISE/NEOWISE" and ``NEOWISE Reactivation" (\citealt{Mainzer et al.(2014)}). Finally, the light-gray dashed vertical line marks the transition from the \textit{Spitzer} cryogenic mission to the warm mission. For exmaple, 53000 MJD is December 27th, 2003 and 59500 MJD is October 13th, 2021. }
        \label{Figure 2}
\end{figure*}

Studies such as \cite{Kellogg et al.(2017)}, \cite{Radigan et al.(2012)}, \cite{Radigan et al.(2014)a}, and \cite{Radigan(2014)b} between 0.9-2.5$\mu$m and \cite{Vos et al.(2022)} and \cite{Yang et al.(2016)} between 3.6-4.5$\mu$m found variability in the L/T dwarf transition but did not sample the entire brown dwarf spectral range. Additional studies (\citealt{Metchev et al.(2015)} and \citealt{Cushing et al.(2016)}) found that large variations of \textgreater2$\%$ in \textit{Spitzer}/IRAC ch1 ($\sim$3.6$\mu$m) and ch2 ($\sim$4.5$\mu$m) exist across L/T/Y spectral types. It has been shown that large variations occur in the $J$ ($\sim$1.25$\mu$m) and $K$/$K_{s}$ ($\sim$2.2$\mu$m) bands, sometimes as large as 420mmag in $K_{s}$ (\citealt{Artigau et al.(2009)}). Studies at r-band ($\sim$0.658$\mu$m) for M and L dwarfs found that $\sim$50$\%$ vary at a level of \textgreater 2$\%$ (\citealt{Artigau(2018)}). 


Brown dwarf variability is believed to be caused by a variety of different physical phenomena. The most explored and discussed explanation is that brown dwarfs have a mixture of dusty clouds and hazes (e.g. \citealt{Marley et al.(2013)} and \citealt{Tsuji(2005)}). SIMP J013656.57+093347.3 is believed to have 75$\%$ of its surface covered by dusty clouds based on its $K_s$ to $J$ variability ratio, $\frac{\Delta K_s}{\Delta J}$ (\citealt{Artigau et al.(2009)}). An alternative idea is that variability is caused by differentially rotating cloud bands. Some cloud bands may appear more optically thin, depending on the filter used. This idea is explored in \cite{Yang et al.(2016)}, where six brown dwarfs were observed to have variability on the scale of a few percent. This variability was linked to differential cloud rotation and atmospheric optical depth, causing inhomogeneities in longitude.

Another scenario is local storms that can be seen on the top of an object's atmosphere, similar to the Great Red Spot on Jupiter. Jupiter and brown dwarfs have been juxtaposed multiple times throughout the history of brown dwarfs studies. \cite{Gelino Marley(2000)} explore the similarities that the gas giants of our solar system could have with brown dwarfs. As a storm rotates or increases/decreases in size, it could lead to variability seen in an unresolved disk. \cite{Gelino Marley(2000)} predicted that an unresolved storm similar to Jupiter’s Great Red Spot would create variability on the order of 0.2 mag at 4.78$\mu$m but only 0.04 mag at 0.410$\mu$m. \cite{Ge et al.(2019)} found that Jupiter exhibits peak-to-peak variability upwards of 20$\%$ at 5$\mu$m, which is where this variability is the greatest, indicating that brown dwarfs may exhibit such behavior and that using the \textit{WISE} W2 band would be an excellent band in which to search for such variability in brown dwarfs.

Our study uses a different approach than previous studies of brown dwarf variability, such as \cite{Metchev et al.(2015)} who used \textit{Spitzer}/IRAC's sweet spot for which corrections for the pixel phase effect (\citealt{Deming et al.(2015)}) enables extremely high levels of photometric precision. Such studies are thus sensitive to small amplitude variability (\textgreater0.2$\%$ for 3 to 5$\mu$m) at short timescales (typically \textless20 hours). While studies of variability over longer time baselines of several months have been performed (e.g., \citealt{Yang et al.(2016)}), an investigation of substellar object variability over several year time baselines has yet to be attempted. Because our observations are randomly sampled over several years, we expect significant coverage of all hemispheres, allowing for an investigation of large-amplitude variability.

\section{Photometry}\label{photometry}
\subsection{Spitzer/IRAC Channel 2}
The photometry used in this work is derived from the \textit{Spitzer} observations (\citealt{IRSA 2022}) used to compute the trigonometric parallaxes of 361 L, T, and Y dwarfs in \cite{Kirkpatrick et al.(2021)}, covering a long time baseline (Figure \ref{Figure 2}). This long time baseline, spanning upwards of seventeen years, provides an excellent dataset with which to study brown dwarfs at their peak wavelengths over a timescale that has never been explored before. The photometry comes from \textit{Spitzer's} Infrared Array Camera (hereafter, \textit{Spitzer}/IRAC; \citealt{Fazio et al.(2004)}), which, during the cryogenic mission,  was capable of observing at ch1 ($\sim$3.6$\mu$m), ch2 ($\sim$4.6$\mu$m), ch3 ($\sim$5.8$\mu$m), and ch4 ($\sim$8.0$\mu$m) bands (\citealt{Werner et al.(2004)}). Of these four, the one that was used is \textit{Spitzer} ch2, as this is the band in which brown dwarfs are brightest (\citealt{Mainzer et al.(2011)}). 

\textit{Spitzer's} cryogenic mission observed only 14 of the 361 objects in our sample. However, every object includes \textit{Spitzer} warm mission ch2 data, and most have a time baseline of \textgreater10 years.

\begin{figure}
        \centering
        \includegraphics[scale = 0.28]{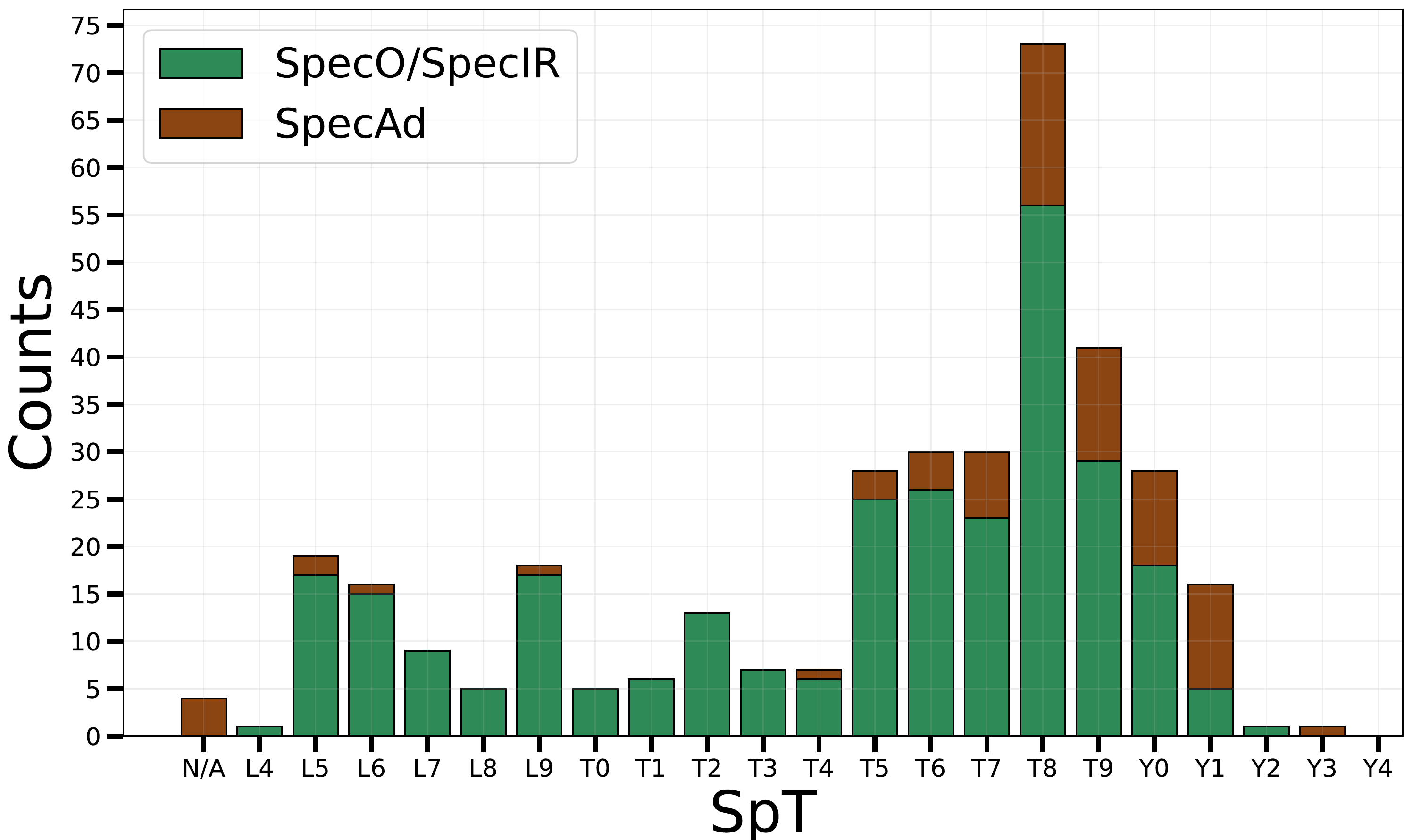}
        \caption{The spectral type distribution of the 361 objects from \cite{Kirkpatrick et al.(2021)}. Optical spectral types (SpecO) are used for L dwarfs and infrared types (SpecIR) are used for T and Y dwarfs. For objects without confirmed spectral types, we take the types estimated using available photometry (SpecAd).}
        \label{Figure 3}
\end{figure}

The spectral type distribution for these 361 brown dwarfs is seen in Figure \ref{Figure 3}. These spectral types come from \cite{Kirkpatrick et al.(2021)} where objects that have optical and infrared spectra are marked as ``SpecO/SpecIR" and objects that have spectral classifications based on colors are marked with ``SpecAd".

Most objects are at distances \textless20pc, which makes them the brightest members of their class. Hence, the photometric uncertainties will be as low as possible, allowing us the best chance to detect variability for these classes. Moreover, the majority of these objects were observed for several years and multiple times a year, meaning that each object will have an abundance of data points. 

\begin{figure}
        \centering
        \includegraphics[scale = 0.27]{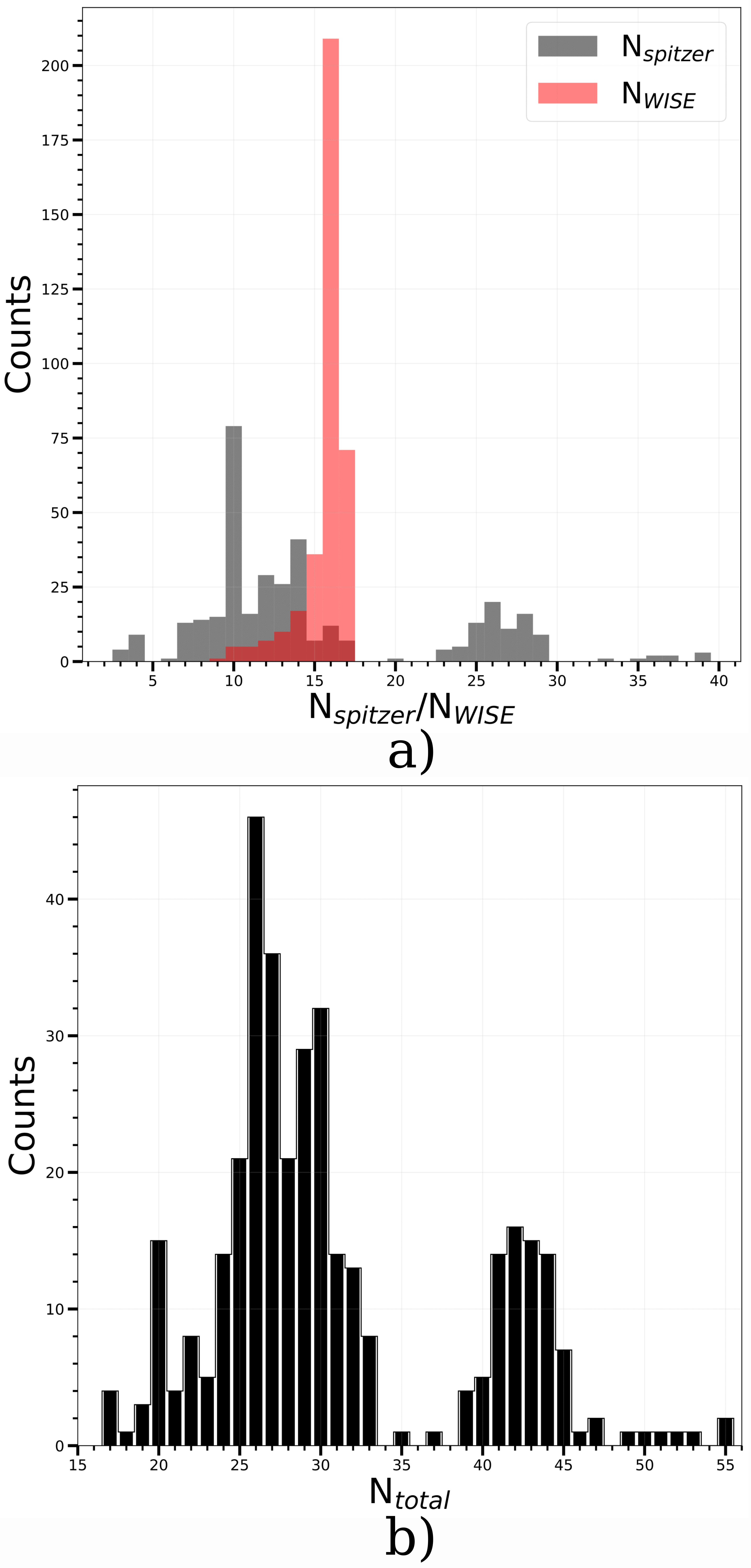}
        \caption{Subplot (a) displays a histogram of the number of AORs (N$_{spitzer}$) for each object in our sample of 361 brown dwarfs, displayed in gray. The distribution is multimodal, though most objects have between 6 and 16 AORs. The red distribution of subplot (a) is a histogram of the number of unTimely detections found for each object (N$_{WISE}$). Subplot (b) is a histogram of the total number of points for each object where N$_{total}$ = N$_{spitzer}$ + N$_{WISE}$.}
        \label{Figure 4}
\end{figure}

Figure \ref{Figure 4}a shows the number of Astronomical Observation Requests (hereafter, AOR), N$_{spitzer}$, for all 361 objects. Most objects have 6-16 AORs. However, many objects have \textgreater16 AORs and these will have the most data-rich light curves. 

\subsection{WISE W2}
To provide additional 4.6$\mu$m data, NASA's {\it Wide-field Infrared Survey Explorer} (hereafter, {\it WISE}; \citealt{Wright et al. 2010}) and {\it Near-Earth Object WISE} (hereafter, {\it NEOWISE}; \citealt{Mainzer et al.(2011)}) W2 photometry was considered. As Figure 15 of \cite{Kirkpatrick et al.(2021)} shows, \textit{Spitzer} ch2 and \textit{WISE} W2 filter bandpasses for brown dwarfs can be interchanged. \textit{WISE} data were obtained by cross-matching our objects with the unTimely Catalog (\citealt{Meisner et al.(2022)}), which contains detections from \textit{WISE}/\textit{NEOWISE} W2 coadds. The unTimely Catalog contains time-series W1 and W2 photometry from the beginning of the \textit{WISE} mission to the most current publicly released \textit{NEOWISE} data. The \textit{WISE} and \textit{NEOWISE} timescales can be seen at the top of Figure \ref{Figure 2}. The unTimely Catalog also contains W1 photometry over multiple epochs; this photometry could be explored for variability in future research. 

\begin{figure}
        \centering
        \includegraphics[scale = 0.3]{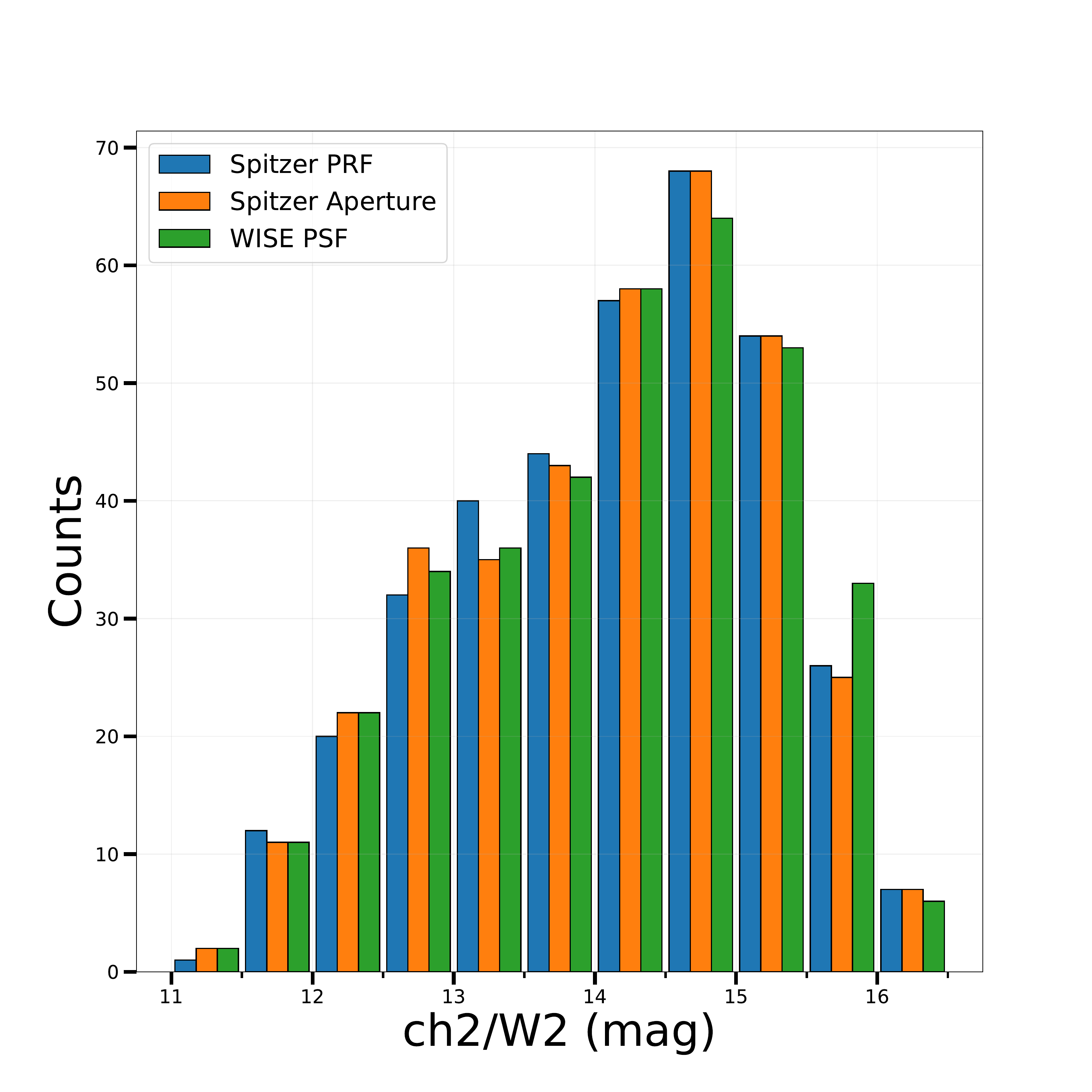}
        \caption{A histogram of the total number of Spitzer PRF (blue), Spitzer aperture (orange), and WISE Point Spread Function (green) average magnitudes for each 0.5mag wide bin.}
        \label{Figure ch2}
\end{figure}

\begin{figure}
        \centering
        \includegraphics[scale = 0.3]{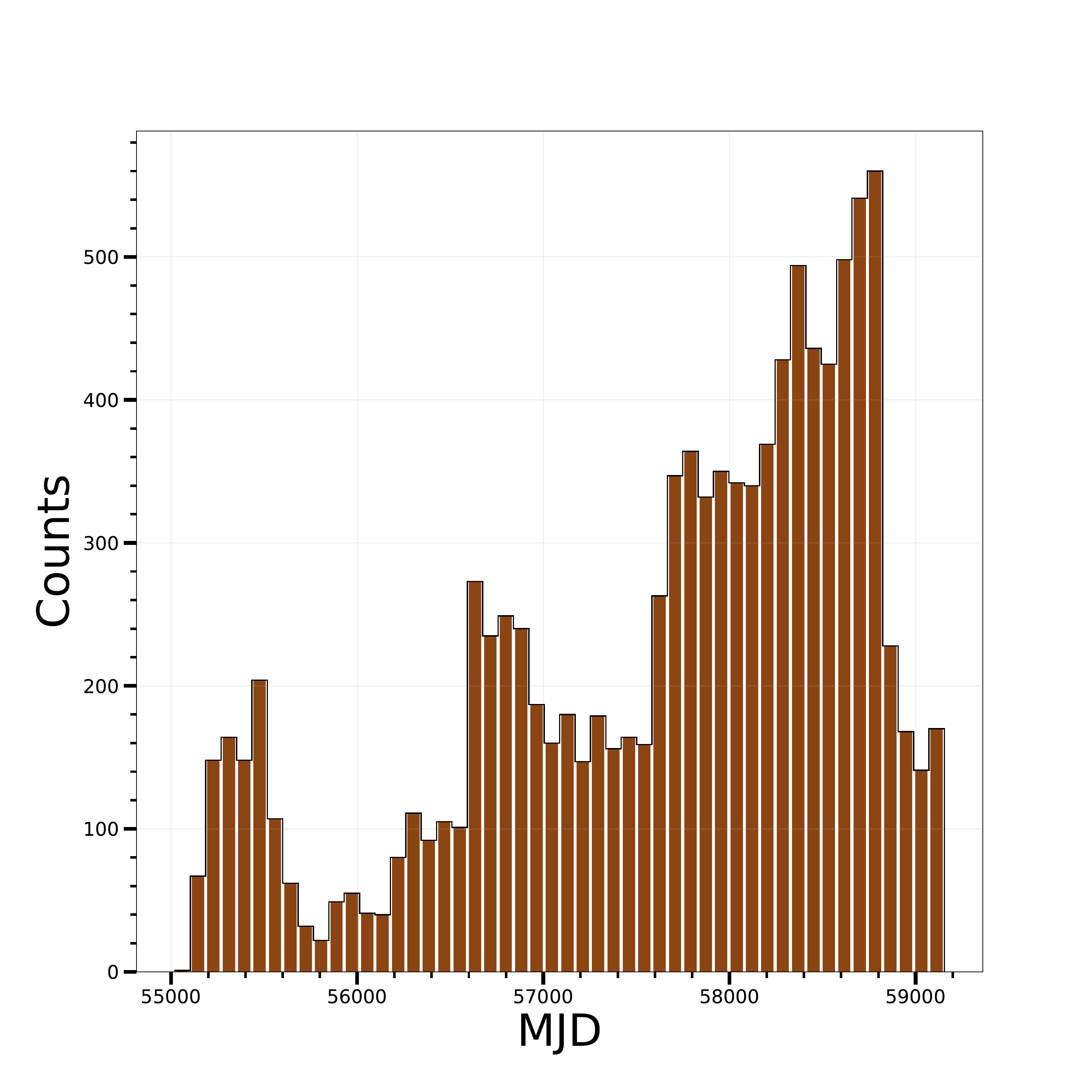}
        \caption{A histogram of the number of points for each MJD interval for both \textit{Spitzer} and \textit{WISE}. Each bin is 82 days each. }
        \label{Figure 12}
\end{figure}

Figure \ref{Figure 4}a shows the distribution of the number of unTimely detections, N$_{WISE}$, per object. There is much less variation compared to  N$_{spitzer}$ as a result of \textit{WISE}/\textit{NEOWISE} covering the whole sky every six months. Theoretically, every object should have 16 unTimely detections for each of the 16 sky passes made from classic \textit{WISE} up through \textit{NEOWISE} year 7. However, some objects have N$_{WISE}$ values of \textless16 because of background contamination problems. Due to the fact that the \textit{WISE} spacecraft precesses, some objects will have N$_{WISE}$ values of 17 because in \textit{NEOWISE} year 7 a portion of the sky already has its 17th coverage. The distribution of the number of points per object, combining N$_{spitzer}$ and N$_{WISE}$, can be seen in Figure \ref{Figure 4}b. Figure \ref{Figure ch2} displays the distribution for each photometry measurement per magnitude interval. Lastly, Figure \ref{Figure 12} shows the distribution of \textit{Spitzer} and unTimely points per MJD interval. We can see that the number of points per unit time increases with time. 

\section{Understanding the Photometry}\label{trust}

An object's median \textit{Spitzer} ch2 flux and its standard deviation were calculated for a single AOR by taking the median of the measured flux in all the dithers for that particular AOR. The magnitudes were calculated using these formulas: 

\begin{equation}
ch2_{aperture} = 2.5 \times \log_{10}(\frac{F_{0}}{F_{ap} \times C_{ap}})
\end{equation}

\begin{equation}
ch2_{PRF} = 2.5 \times \log_{10}(\frac{F_{0}}{\frac{F_{PRF}}{C_{PRF}}})
\end{equation}
where ch2$_{aperture}$ is the aperture magnitude, ch2$_{PRF}$ is the PRF-fit magnitude, F$_{0}$ is the flux at zero magnitude (179700000 $\mu$Jy, Table 4.2 of the \textit{Spitzer}/IRAC Handbook\footnote{\url{https://irsa.ipac.caltech.edu/data/SPITZER/docs/irac/iracinstrumenthandbook/IRAC_Instrument_Handbook.pdf}}), F$_{ap}$ is the measured aperture flux in $\mu$Jy from the MOPEX/APEX output (described below), F$_{PRF}$ is the measured PRF-fit flux in $\mu$Jy from the same output, C$_{ap}$ is the aperture correction (Table 4.8 of the \textit{Spitzer}/IRAC Handbook), and C$_{PRF}$ is the PRF-fit correction (Table C.1 of the \textit{Spitzer}/IRAC Handbook). This median technique was done to help us mitigate dithers that were affected by cosmic ray hits.

For the unTimely data points, the only measurement that is offered in the catalog is a Point Spread Function-fit (hereafter, PSF-fit; \citealt{Schlafly et al.(2019)}), which is analogous to the \textit{Spitzer} PRF-fit. Flux measurements were obtained directly from the catalog because \cite{Meisner Schlafly(2019)} calculated the flux and its uncertainties within the code. The magnitudes and their uncertainties were calculated using the formula: mag$_{WISE}$ = $22.5- 2.5 \times \log_{10}(F_{WISE})$ (\citealt{Meisner Schlafly(2019)}).  

Our \textit{Spitzer} ch2 data were reduced in \cite{Kirkpatrick et al.(2021)} using the MOPEX/APEX code (\citealt{Makovoz et al.(2006)}). Aperture photometry counts all the flux within a circle of fixed radius. For this study, the aperture radius used is $4{\farcs}8$. For background subtraction, the flux in a concentric annulus is medianed, and that value is subtracted off the aperture measurement to reveal the true target flux. The annulus has an inner radius of $28{\farcs}8$ and an outer radius of $48{\farcs}0$ (\textit{Spitzer}/IRAC Handbook\footnote{\url{https://irsa.ipac.caltech.edu/data/SPITZER/docs/irac/iracinstrumenthandbook/IRAC_Instrument_Handbook.pdf}}). The second photometry method is to perform a PRF fit (\citealt{Hora et al.(2012)}), which approximates the 3 parameter (x, y, and flux) distribution of our target flux to a known 3 parameter shape of a point source. The aperture and PRF-fit measurements were used to verify any variability seen, because if a variable point is seen in both measurements, it suggests that it is real and not an artifact of processing. For example, the aperture measurement is susceptible to cosmic ray hits, while the PRF-fit photometry ideally can remove this contamination. 

\begin{figure}
        \centering
        \includegraphics[scale = 0.29]{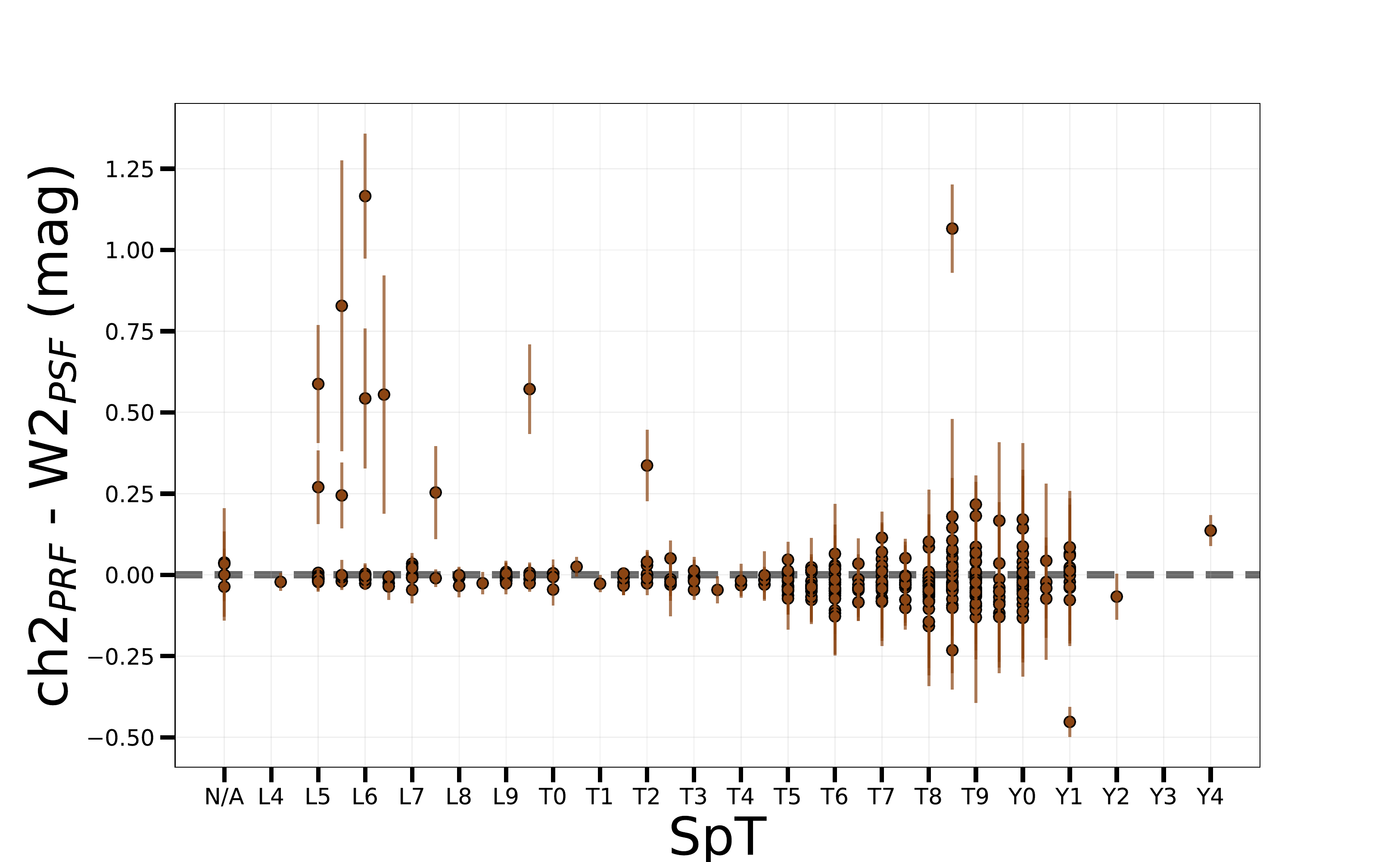}
        \caption{Comparison of spectral type versus the difference between the \textit{Spitzer} ch2$_{PRF}$ median and the unTimely W2$_{PSF}$ median. The large outliers are a result of the \textit{Spitzer} code trying to passively deblend the point source (further discussed in \S\ref{binary}). Otherwise, we see a general trend in which the \textit{Spitzer} ch2 and unTimely W2 photometry match extremely well, consistent with Figure 15 from \cite{Kirkpatrick et al.(2021)}.}
        \label{Figure 6}
\end{figure}

To further test the consistency of the \textit{Spitzer} PRF-fit ch2 measurements with those of the unTimely PSF-fit, we plotted the difference between the average ch2$_{PRF}$ and the average W2$_{PSF}$ for each object. This is plotted against spectral type in Figure \ref{Figure 6}. The trend hovers around 0 for objects before T8, which supports the compatibility of the \textit{Spitzer}/IRAC ch2 magnitudes and the \textit{WISE}/unTimely W2 magnitudes. Objects later than T8 have a considerable amount of scatter as a result of these objects having poor S/N in W2. Outliers earlier than spectral type T8 on this plot are a result of blending in \textit{Spitzer} or \textit{WISE} caused by background contamination or possibly binarity, as further discussed in \S\ref{binaries}.

\begin{figure}
        \centering
        \includegraphics[scale = 0.29]{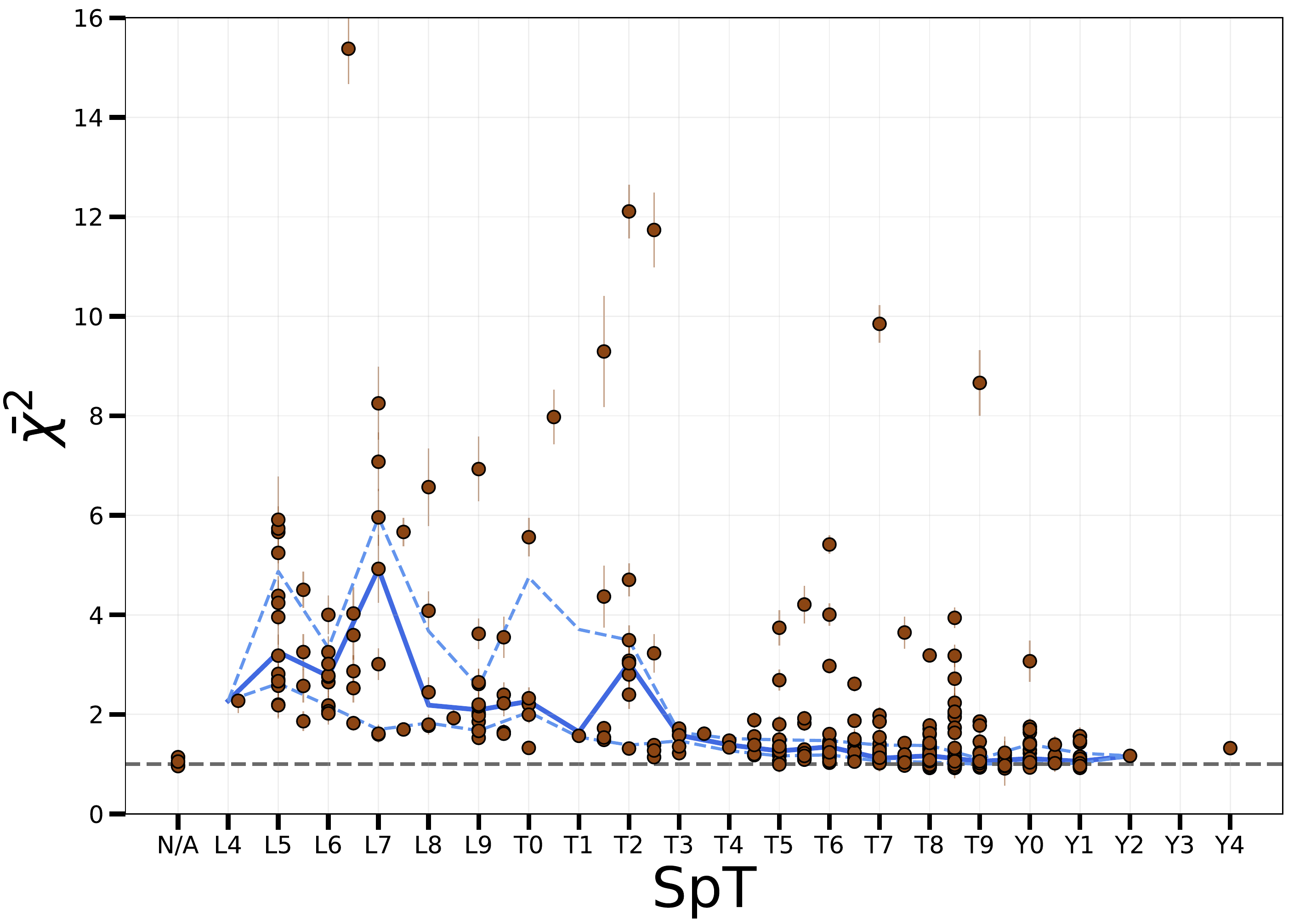}
        \caption{Comparison of the spectral type (Figure \ref{Figure 4} subplot a) to the averaged $\chi^2$ (for the ch2$_{PRF}$ fit) for each object. The solid blue line is the median for each spectral type, while the dashed light blue lines are the 25th and 75th percentile for each spectral type. Note the large outliers on this plot, most of which were identified as possible binary candidates (further discussed in \S\ref{binary}). Lastly, one object, 2MASS J22551861$-$5713056 (\citealt{Kendall et al.(2007)}), which has a $\bar{\chi^2}$ value of $\sim$68, is not shown on this plot. This is a known brown dwarf binary (\citealt{Reid et al.(2008)}).}
        \label{Figure 7}
\end{figure}

An important statistic to consider is the $\chi^2$ value for the \textit{Spitzer} PRF-fit. The cause of a high $\chi^2$ is that the MOPEX ch2 PRF was a poor fit to the data. In Figure \ref{Figure 7} we compare spectral type to the averaged $\chi^2$ value. The object's AOR $\chi^2$ value was calculated by performing a weighted average of all the dithers, and then the overall $\bar{\chi^2}$ was computed using a weighted average of all the AORs (the weight for these averages are the inverse variance for the photometry). The solid blue line in Figure \ref{Figure 7} is the median for each spectral type, with the dashed light blue lines showing the lower and upper quartiles Q1 (25$\%$) and Q3 (75$\%$). Outliers are discussed further in \S\ref{binaries}. 

\section{Variability Results}\label{variability}

Now that the photometry has been analyzed, we can discuss our definition of variability. To formally define variability, we require that there are points lying \textgreater3$\sigma$ away from the median value. A 3$\sigma$ requirement means that we would expect only one object in a sample of 361 independent objects to show such an excursion by chance. This \textgreater 3$\sigma$ requirement searches for variability for each photometry type separately: ch2$_{aperture}$, ch2$_{PRF}$, and W2$_{PSF}$. We found that 24 of our 361 objects have such variability. 

\begin{table*}[ht]
\centering{
\caption{Objects With Spurious Variability}\label{Table 1}
\begin{tabular}{lllcc}
\hline
\hline
Object Name & RA (Deg) & Dec (Deg) & Dataset & Note\\
\hline
\hline
WISE J003110.04+574936.3   & 7.79433   & 57.826773  & W2$_{PSF}$  &  Blended with Background Source  \\
WISE J003231.09$-$494651.4   & 8.128736  & $-$49.782059 & W2$_{PSF}$   &  Blended with Background Source \\
WISEP J015010.86+382724.3  & 27.546953 & 38.456528  & W2$_{PSF}$   & Blended with Background Source   \\
WISEPA J031325.96+780744.2 & 48.359129 & 78.129099  & ch2$_{PRF}$   &  Blended with Background Source  \\
WISE J032504.33$-$504400.3   & 51.268182 & $-$50.733501 & W2$_{PSF}$  &  Blended with Background Source    \\
SDSSp J033035.13$-$002534.5  & 52.64629  & $-$0.42628   & W2$_{PSF}$   &  Blended with Background Source   \\
 2MASS J04210718$-$6306022    & 65.281665 & $-$63.099437 & W2$_{PSF}$  & Blended with Background Source    \\
WISE J064723.23$-$623235.5   & 101.84721 & $-$62.542746 & ch2$_{PRF}$  & Cosmic Ray Affects 1 Epoch     \\
WISE J071322.55$-$291751.9   & 108.34441 & $-$29.298264 & W2$_{PSF}$    &  Blended with Background Source  \\
WISE J103907.73$-$160002.9   & 159.78173 & $-$16.001139 & ch2$_{PRF}$   & Cosmic Ray Affects 1 Epoch    \\
WISE J112438.12$-$042149.7   & 171.15781 & $-$4.363732  & W2$_{PSF}$  &  Blended with Background Source    \\
WISEA J114156.67$-$332635.5  & 175.48445 & $-$33.443475 & ch2$_{PRF}$/W2$_{PSF}$ &  Blended with Background Source \\
WISEPC J151906.64+700931.5 & 229.77879 & 70.157919  & ch2$_{PRF}$/W2$_{PSF}$ &  Blended with Background Source \\
CWISEP J160835.01$-$244244.7 & 242.14633 & $-$24.712494 & W2$_{PSF}$  &  Blended with Background Source   \\
WISEPA J171104.60+350036.8 & 257.76888 & 35.010238  & ch2$_{PRF}$  & Cosmic Ray Affects 1 Epoch    \\
WISE J174303.71+421150.0   & 265.76558 & 42.196369  & W2$_{PSF}$    &  Blended with Background Source  \\
WISE J192841.35+235604.9   & 292.17202 & 23.93489   & ch2$_{PRF}$  &  Blended with Background Source     \\
WISE J195500.42$-$254013.9   & 298.7526  & $-$25.671162 & ch2$_{PRF}$   & Cosmic Ray Affects 1 Epoch    \\
WISEPA J195905.66$-$333833.7 & 299.77339 & $-$33.642956 & ch2$_{aperture}$ & Cosmic Ray Affects 1 Epoch \\
WISE J201546.27+664645.1   & 303.94207 & 66.779688  & ch2$_{PRF}$   &  Blended with Background Source    \\
CWISEP J210007.87$-$293139.8 & 315.03349 & $-$29.52782  & W2$_{PSF}$ &  Blended with Background Source     \\
WISEPA J231336.40$-$803700.3 & 348.40273 & $-$80.616879 & ch2$_{PRF}$   &  Blended with Background Source    \\
WISE J235716.49+122741.8   & 359.31997 & 12.459123  & ch2$_{aperture}$ & Cosmic Ray Affects 1 Epoch \\
\hline
\end{tabular}}
\end{table*}

We found that the variability for all 24 objects is a result of a non-physical property of the brown dwarf. All 24 objects are listed in Table \ref{Table 1}. The most common cause of W2$_{PSF}$ variability is contamination from a background source that the object was blended with in one or more epochs. All objects with ch2$_{aperture}$ variability have a single AOR that is bombarded with cosmic ray hits in all but a few dithers. With all 24 objects being labeled as false, we find 0 out of 361 objects with true long-term variability. We further discuss this result in \S\ref{defvari}. The complete figure set of light curves (361 images) is available in the online journal. An example from this figure set can be seen in Figure \ref{Figure 8}. 

\figsetstart
\figsetnum{8}
\figsettitle{Light curves for the 361 objects}

\figsetgrpstart
\figsetgrpnum{8.1}
\figsetgrptitle{Light Curve for WISE0005p3737}
\figsetplot{WISE0005p3737_total_variable_SURF2022.pdf}
\figsetgrpnote{The light curve of WISE0005p3737 with \textit{Spitzer} ch2$_{aperture}$, ch2$_{PRF}$, and \textit{WISE} W2$_{PSF}$ photometry. The solid lines are the medians for each photometry type, while the dashed lines are $\pm$3$\sigma$. The spectral type is on the scale of 0 is M0, 10 is L0, 20 is T0, and 30 is Y0. }
\figsetgrpend

\figsetgrpstart
\figsetgrpnum{8.2}
\figsetgrptitle{Light Curve for WISE0015m4615}
\figsetplot{WISE0015m4615_total_variable_SURF2022.pdf}
\figsetgrpnote{The light curve of WISE0015m4615 with \textit{Spitzer} ch2$_{aperture}$, ch2$_{PRF}$, and \textit{WISE} W2$_{PSF}$ photometry. The solid lines are the medians for each photometry type, while the dashed lines are $\pm$3$\sigma$. The spectral type is on the scale of 0 is M0, 10 is L0, 20 is T0, and 30 is Y0. }
\figsetgrpend

\figsetgrpstart
\figsetgrpnum{8.3}
\figsetgrptitle{Light Curve for WISE0027m0121}
\figsetplot{WISE0027m0121_total_variable_SURF2022.pdf}
\figsetgrpnote{The light curve of WISE0027m0121 with \textit{Spitzer} ch2$_{aperture}$, ch2$_{PRF}$, and \textit{WISE} W2$_{PSF}$ photometry. The solid lines are the medians for each photometry type, while the dashed lines are $\pm$3$\sigma$. The spectral type is on the scale of 0 is M0, 10 is L0, 20 is T0, and 30 is Y0. }
\figsetgrpend

\figsetgrpstart
\figsetgrpnum{8.4}
\figsetgrptitle{Light Curve for WISE0031p5749}
\figsetplot{WISE0031p5749_total_variable_SURF2022.pdf}
\figsetgrpnote{The light curve of WISE0031p5749 with \textit{Spitzer} ch2$_{aperture}$, ch2$_{PRF}$, and \textit{WISE} W2$_{PSF}$ photometry. The solid lines are the medians for each photometry type, while the dashed lines are $\pm$3$\sigma$. The spectral type is on the scale of 0 is M0, 10 is L0, 20 is T0, and 30 is Y0. }
\figsetgrpend

\figsetgrpstart
\figsetgrpnum{8.5}
\figsetgrptitle{Light Curve for WISE0031p3335}
\figsetplot{WISE0031p3335_total_variable_SURF2022.pdf}
\figsetgrpnote{The light curve of WISE0031p3335 with \textit{Spitzer} ch2$_{aperture}$, ch2$_{PRF}$, and \textit{WISE} W2$_{PSF}$ photometry. The solid lines are the medians for each photometry type, while the dashed lines are $\pm$3$\sigma$. The spectral type is on the scale of 0 is M0, 10 is L0, 20 is T0, and 30 is Y0. }
\figsetgrpend

\figsetgrpstart
\figsetgrpnum{8.6}
\figsetgrptitle{Light Curve for WISE0032m4946}
\figsetplot{WISE0032m4946_total_variable_SURF2022.pdf}
\figsetgrpnote{The light curve of WISE0032m4946 with \textit{Spitzer} ch2$_{aperture}$, ch2$_{PRF}$, and \textit{WISE} W2$_{PSF}$ photometry. The solid lines are the medians for each photometry type, while the dashed lines are $\pm$3$\sigma$. The spectral type is on the scale of 0 is M0, 10 is L0, 20 is T0, and 30 is Y0. }
\figsetgrpend

\figsetgrpstart
\figsetgrpnum{8.7}
\figsetgrptitle{Light Curve for WISE0034p0523}
\figsetplot{WISE0034p0523_total_variable_SURF2022.pdf}
\figsetgrpnote{The light curve of WISE0034p0523 with \textit{Spitzer} ch2$_{aperture}$, ch2$_{PRF}$, and \textit{WISE} W2$_{PSF}$ photometry. The solid lines are the medians for each photometry type, while the dashed lines are $\pm$3$\sigma$. The spectral type is on the scale of 0 is M0, 10 is L0, 20 is T0, and 30 is Y0. }
\figsetgrpend

\figsetgrpstart
\figsetgrpnum{8.8}
\figsetgrptitle{Light Curve for WISE0038p2758}
\figsetplot{WISE0038p2758_total_variable_SURF2022.pdf}
\figsetgrpnote{The light curve of WISE0038p2758 with \textit{Spitzer} ch2$_{aperture}$, ch2$_{PRF}$, and \textit{WISE} W2$_{PSF}$ photometry. The solid lines are the medians for each photometry type, while the dashed lines are $\pm$3$\sigma$. The spectral type is on the scale of 0 is M0, 10 is L0, 20 is T0, and 30 is Y0. }
\figsetgrpend

\figsetgrpstart
\figsetgrpnum{8.9}
\figsetgrptitle{Light Curve for WISE0041m4019}
\figsetplot{WISE0041m4019_total_variable_SURF2022.pdf}
\figsetgrpnote{The light curve of WISE0041m4019 with \textit{Spitzer} ch2$_{aperture}$, ch2$_{PRF}$, and \textit{WISE} W2$_{PSF}$ photometry. The solid lines are the medians for each photometry type, while the dashed lines are $\pm$3$\sigma$. The spectral type is on the scale of 0 is M0, 10 is L0, 20 is T0, and 30 is Y0. }
\figsetgrpend

\figsetgrpstart
\figsetgrpnum{8.10}
\figsetgrptitle{Light Curve for WISE0043m3822}
\figsetplot{WISE0043m3822_total_variable_SURF2022.pdf}
\figsetgrpnote{The light curve of WISE0043m3822 with \textit{Spitzer} ch2$_{aperture}$, ch2$_{PRF}$, and \textit{WISE} W2$_{PSF}$ photometry. The solid lines are the medians for each photometry type, while the dashed lines are $\pm$3$\sigma$. The spectral type is on the scale of 0 is M0, 10 is L0, 20 is T0, and 30 is Y0. }
\figsetgrpend

\figsetgrpstart
\figsetgrpnum{8.11}
\figsetgrptitle{Light Curve for WISE0045p3611}
\figsetplot{WISE0045p3611_total_variable_SURF2022.pdf}
\figsetgrpnote{The light curve of WISE0045p3611 with \textit{Spitzer} ch2$_{aperture}$, ch2$_{PRF}$, and \textit{WISE} W2$_{PSF}$ photometry. The solid lines are the medians for each photometry type, while the dashed lines are $\pm$3$\sigma$. The spectral type is on the scale of 0 is M0, 10 is L0, 20 is T0, and 30 is Y0. }
\figsetgrpend

\figsetgrpstart
\figsetgrpnum{8.12}
\figsetgrptitle{Light Curve for WISE0048p2508}
\figsetplot{WISE0048p2508_total_variable_SURF2022.pdf}
\figsetgrpnote{The light curve of WISE0048p2508 with \textit{Spitzer} ch2$_{aperture}$, ch2$_{PRF}$, and \textit{WISE} W2$_{PSF}$ photometry. The solid lines are the medians for each photometry type, while the dashed lines are $\pm$3$\sigma$. The spectral type is on the scale of 0 is M0, 10 is L0, 20 is T0, and 30 is Y0. }
\figsetgrpend

\figsetgrpstart
\figsetgrpnum{8.13}
\figsetgrptitle{Light Curve for WISE0049p2151}
\figsetplot{WISE0049p2151_total_variable_SURF2022.pdf}
\figsetgrpnote{The light curve of WISE0049p2151 with \textit{Spitzer} ch2$_{aperture}$, ch2$_{PRF}$, and \textit{WISE} W2$_{PSF}$ photometry. The solid lines are the medians for each photometry type, while the dashed lines are $\pm$3$\sigma$. The spectral type is on the scale of 0 is M0, 10 is L0, 20 is T0, and 30 is Y0. }
\figsetgrpend

\figsetgrpstart
\figsetgrpnum{8.14}
\figsetgrptitle{Light Curve for WISE0051m1544}
\figsetplot{WISE0051m1544_total_variable_SURF2022.pdf}
\figsetgrpnote{The light curve of WISE0051m1544 with \textit{Spitzer} ch2$_{aperture}$, ch2$_{PRF}$, and \textit{WISE} W2$_{PSF}$ photometry. The solid lines are the medians for each photometry type, while the dashed lines are $\pm$3$\sigma$. The spectral type is on the scale of 0 is M0, 10 is L0, 20 is T0, and 30 is Y0. }
\figsetgrpend

\figsetgrpstart
\figsetgrpnum{8.15}
\figsetgrptitle{Light Curve for WISE0058m5653}
\figsetplot{WISE0058m5653_total_variable_SURF2022.pdf}
\figsetgrpnote{The light curve of WISE0058m5653 with \textit{Spitzer} ch2$_{aperture}$, ch2$_{PRF}$, and \textit{WISE} W2$_{PSF}$ photometry. The solid lines are the medians for each photometry type, while the dashed lines are $\pm$3$\sigma$. The spectral type is on the scale of 0 is M0, 10 is L0, 20 is T0, and 30 is Y0. }
\figsetgrpend

\figsetgrpstart
\figsetgrpnum{8.16}
\figsetgrptitle{Light Curve for WISE0103p1935}
\figsetplot{WISE0103p1935_total_variable_SURF2022.pdf}
\figsetgrpnote{The light curve of WISE0103p1935 with \textit{Spitzer} ch2$_{aperture}$, ch2$_{PRF}$, and \textit{WISE} W2$_{PSF}$ photometry. The solid lines are the medians for each photometry type, while the dashed lines are $\pm$3$\sigma$. The spectral type is on the scale of 0 is M0, 10 is L0, 20 is T0, and 30 is Y0. }
\figsetgrpend

\figsetgrpstart
\figsetgrpnum{8.17}
\figsetgrptitle{Light Curve for WISE0105m7834}
\figsetplot{WISE0105m7834_total_variable_SURF2022.pdf}
\figsetgrpnote{The light curve of WISE0105m7834 with \textit{Spitzer} ch2$_{aperture}$, ch2$_{PRF}$, and \textit{WISE} W2$_{PSF}$ photometry. The solid lines are the medians for each photometry type, while the dashed lines are $\pm$3$\sigma$. The spectral type is on the scale of 0 is M0, 10 is L0, 20 is T0, and 30 is Y0. }
\figsetgrpend

\figsetgrpstart
\figsetgrpnum{8.18}
\figsetgrptitle{Light Curve for WISE0111m5053}
\figsetplot{WISE0111m5053_total_variable_SURF2022.pdf}
\figsetgrpnote{The light curve of WISE0111m5053 with \textit{Spitzer} ch2$_{aperture}$, ch2$_{PRF}$, and \textit{WISE} W2$_{PSF}$ photometry. The solid lines are the medians for each photometry type, while the dashed lines are $\pm$3$\sigma$. The spectral type is on the scale of 0 is M0, 10 is L0, 20 is T0, and 30 is Y0. }
\figsetgrpend

\figsetgrpstart
\figsetgrpnum{8.19}
\figsetgrptitle{Light Curve for WISE0123p4142}
\figsetplot{WISE0123p4142_total_variable_SURF2022.pdf}
\figsetgrpnote{The light curve of WISE0123p4142 with \textit{Spitzer} ch2$_{aperture}$, ch2$_{PRF}$, and \textit{WISE} W2$_{PSF}$ photometry. The solid lines are the medians for each photometry type, while the dashed lines are $\pm$3$\sigma$. The spectral type is on the scale of 0 is M0, 10 is L0, 20 is T0, and 30 is Y0. }
\figsetgrpend

\figsetgrpstart
\figsetgrpnum{8.20}
\figsetgrptitle{Light Curve for WISE0132m5818}
\figsetplot{WISE0132m5818_total_variable_SURF2022.pdf}
\figsetgrpnote{The light curve of WISE0132m5818 with \textit{Spitzer} ch2$_{aperture}$, ch2$_{PRF}$, and \textit{WISE} W2$_{PSF}$ photometry. The solid lines are the medians for each photometry type, while the dashed lines are $\pm$3$\sigma$. The spectral type is on the scale of 0 is M0, 10 is L0, 20 is T0, and 30 is Y0. }
\figsetgrpend

\figsetgrpstart
\figsetgrpnum{8.21}
\figsetgrptitle{Light Curve for WISE0133p0231}
\figsetplot{WISE0133p0231_total_variable_SURF2022.pdf}
\figsetgrpnote{The light curve of WISE0133p0231 with \textit{Spitzer} ch2$_{aperture}$, ch2$_{PRF}$, and \textit{WISE} W2$_{PSF}$ photometry. The solid lines are the medians for each photometry type, while the dashed lines are $\pm$3$\sigma$. The spectral type is on the scale of 0 is M0, 10 is L0, 20 is T0, and 30 is Y0. }
\figsetgrpend

\figsetgrpstart
\figsetgrpnum{8.22}
\figsetgrptitle{Light Curve for WISE0135p1715}
\figsetplot{WISE0135p1715_total_variable_SURF2022.pdf}
\figsetgrpnote{The light curve of WISE0135p1715 with \textit{Spitzer} ch2$_{aperture}$, ch2$_{PRF}$, and \textit{WISE} W2$_{PSF}$ photometry. The solid lines are the medians for each photometry type, while the dashed lines are $\pm$3$\sigma$. The spectral type is on the scale of 0 is M0, 10 is L0, 20 is T0, and 30 is Y0. }
\figsetgrpend

\figsetgrpstart
\figsetgrpnum{8.23}
\figsetgrptitle{Light Curve for WISE0138m0322}
\figsetplot{WISE0138m0322_total_variable_SURF2022.pdf}
\figsetgrpnote{The light curve of WISE0138m0322 with \textit{Spitzer} ch2$_{aperture}$, ch2$_{PRF}$, and \textit{WISE} W2$_{PSF}$ photometry. The solid lines are the medians for each photometry type, while the dashed lines are $\pm$3$\sigma$. The spectral type is on the scale of 0 is M0, 10 is L0, 20 is T0, and 30 is Y0. }
\figsetgrpend

\figsetgrpstart
\figsetgrpnum{8.24}
\figsetgrptitle{Light Curve for WISE0146p4234}
\figsetplot{WISE0146p4234_total_variable_SURF2022.pdf}
\figsetgrpnote{The light curve of WISE0146p4234 with \textit{Spitzer} ch2$_{aperture}$, ch2$_{PRF}$, and \textit{WISE} W2$_{PSF}$ photometry. The solid lines are the medians for each photometry type, while the dashed lines are $\pm$3$\sigma$. The spectral type is on the scale of 0 is M0, 10 is L0, 20 is T0, and 30 is Y0. }
\figsetgrpend

\figsetgrpstart
\figsetgrpnum{8.25}
\figsetgrptitle{Light Curve for WISE0148m1048}
\figsetplot{WISE0148m1048_total_variable_SURF2022.pdf}
\figsetgrpnote{The light curve of WISE0148m1048 with \textit{Spitzer} ch2$_{aperture}$, ch2$_{PRF}$, and \textit{WISE} W2$_{PSF}$ photometry. The solid lines are the medians for each photometry type, while the dashed lines are $\pm$3$\sigma$. The spectral type is on the scale of 0 is M0, 10 is L0, 20 is T0, and 30 is Y0. }
\figsetgrpend

\figsetgrpstart
\figsetgrpnum{8.26}
\figsetgrptitle{Light Curve for WISE0150p3827}
\figsetplot{WISE0150p3827_total_variable_SURF2022.pdf}
\figsetgrpnote{The light curve of WISE0150p3827 with \textit{Spitzer} ch2$_{aperture}$, ch2$_{PRF}$, and \textit{WISE} W2$_{PSF}$ photometry. The solid lines are the medians for each photometry type, while the dashed lines are $\pm$3$\sigma$. The spectral type is on the scale of 0 is M0, 10 is L0, 20 is T0, and 30 is Y0. }
\figsetgrpend

\figsetgrpstart
\figsetgrpnum{8.27}
\figsetgrptitle{Light Curve for WISE0155p0950}
\figsetplot{WISE0155p0950_total_variable_SURF2022.pdf}
\figsetgrpnote{The light curve of WISE0155p0950 with \textit{Spitzer} ch2$_{aperture}$, ch2$_{PRF}$, and \textit{WISE} W2$_{PSF}$ photometry. The solid lines are the medians for each photometry type, while the dashed lines are $\pm$3$\sigma$. The spectral type is on the scale of 0 is M0, 10 is L0, 20 is T0, and 30 is Y0. }
\figsetgrpend

\figsetgrpstart
\figsetgrpnum{8.28}
\figsetgrptitle{Light Curve for WISE0200m5105}
\figsetplot{WISE0200m5105_total_variable_SURF2022.pdf}
\figsetgrpnote{The light curve of WISE0200m5105 with \textit{Spitzer} ch2$_{aperture}$, ch2$_{PRF}$, and \textit{WISE} W2$_{PSF}$ photometry. The solid lines are the medians for each photometry type, while the dashed lines are $\pm$3$\sigma$. The spectral type is on the scale of 0 is M0, 10 is L0, 20 is T0, and 30 is Y0. }
\figsetgrpend

\figsetgrpstart
\figsetgrpnum{8.29}
\figsetgrptitle{Light Curve for WISE0205p1251}
\figsetplot{WISE0205p1251_total_variable_SURF2022.pdf}
\figsetgrpnote{The light curve of WISE0205p1251 with \textit{Spitzer} ch2$_{aperture}$, ch2$_{PRF}$, and \textit{WISE} W2$_{PSF}$ photometry. The solid lines are the medians for each photometry type, while the dashed lines are $\pm$3$\sigma$. The spectral type is on the scale of 0 is M0, 10 is L0, 20 is T0, and 30 is Y0. }
\figsetgrpend

\figsetgrpstart
\figsetgrpnum{8.30}
\figsetgrptitle{Light Curve for WISE0212p0531}
\figsetplot{WISE0212p0531_total_variable_SURF2022.pdf}
\figsetgrpnote{The light curve of WISE0212p0531 with \textit{Spitzer} ch2$_{aperture}$, ch2$_{PRF}$, and \textit{WISE} W2$_{PSF}$ photometry. The solid lines are the medians for each photometry type, while the dashed lines are $\pm$3$\sigma$. The spectral type is on the scale of 0 is M0, 10 is L0, 20 is T0, and 30 is Y0. }
\figsetgrpend

\figsetgrpstart
\figsetgrpnum{8.31}
\figsetgrptitle{Light Curve for WISE0221p3842}
\figsetplot{WISE0221p3842_total_variable_SURF2022.pdf}
\figsetgrpnote{The light curve of WISE0221p3842 with \textit{Spitzer} ch2$_{aperture}$, ch2$_{PRF}$, and \textit{WISE} W2$_{PSF}$ photometry. The solid lines are the medians for each photometry type, while the dashed lines are $\pm$3$\sigma$. The spectral type is on the scale of 0 is M0, 10 is L0, 20 is T0, and 30 is Y0. }
\figsetgrpend

\figsetgrpstart
\figsetgrpnum{8.32}
\figsetgrptitle{Light Curve for WISE0226m0211}
\figsetplot{WISE0226m0211_total_variable_SURF2022.pdf}
\figsetgrpnote{The light curve of WISE0226m0211 with \textit{Spitzer} ch2$_{aperture}$, ch2$_{PRF}$, and \textit{WISE} W2$_{PSF}$ photometry. The solid lines are the medians for each photometry type, while the dashed lines are $\pm$3$\sigma$. The spectral type is on the scale of 0 is M0, 10 is L0, 20 is T0, and 30 is Y0. }
\figsetgrpend

\figsetgrpstart
\figsetgrpnum{8.33}
\figsetgrptitle{Light Curve for WISE0233p3030}
\figsetplot{WISE0233p3030_total_variable_SURF2022.pdf}
\figsetgrpnote{The light curve of WISE0233p3030 with \textit{Spitzer} ch2$_{aperture}$, ch2$_{PRF}$, and \textit{WISE} W2$_{PSF}$ photometry. The solid lines are the medians for each photometry type, while the dashed lines are $\pm$3$\sigma$. The spectral type is on the scale of 0 is M0, 10 is L0, 20 is T0, and 30 is Y0. }
\figsetgrpend

\figsetgrpstart
\figsetgrpnum{8.34}
\figsetgrptitle{Light Curve for WISE0238m1332}
\figsetplot{WISE0238m1332_total_variable_SURF2022.pdf}
\figsetgrpnote{The light curve of WISE0238m1332 with \textit{Spitzer} ch2$_{aperture}$, ch2$_{PRF}$, and \textit{WISE} W2$_{PSF}$ photometry. The solid lines are the medians for each photometry type, while the dashed lines are $\pm$3$\sigma$. The spectral type is on the scale of 0 is M0, 10 is L0, 20 is T0, and 30 is Y0. }
\figsetgrpend

\figsetgrpstart
\figsetgrpnum{8.35}
\figsetgrptitle{Light Curve for WISE0241m3653}
\figsetplot{WISE0241m3653_total_variable_SURF2022.pdf}
\figsetgrpnote{The light curve of WISE0241m3653 with \textit{Spitzer} ch2$_{aperture}$, ch2$_{PRF}$, and \textit{WISE} W2$_{PSF}$ photometry. The solid lines are the medians for each photometry type, while the dashed lines are $\pm$3$\sigma$. The spectral type is on the scale of 0 is M0, 10 is L0, 20 is T0, and 30 is Y0. }
\figsetgrpend

\figsetgrpstart
\figsetgrpnum{8.36}
\figsetgrptitle{Light Curve for WISE0245m3450}
\figsetplot{WISE0245m3450_total_variable_SURF2022.pdf}
\figsetgrpnote{The light curve of WISE0245m3450 with \textit{Spitzer} ch2$_{aperture}$, ch2$_{PRF}$, and \textit{WISE} W2$_{PSF}$ photometry. The solid lines are the medians for each photometry type, while the dashed lines are $\pm$3$\sigma$. The spectral type is on the scale of 0 is M0, 10 is L0, 20 is T0, and 30 is Y0. }
\figsetgrpend

\figsetgrpstart
\figsetgrpnum{8.37}
\figsetgrptitle{Light Curve for WISE0247p3725}
\figsetplot{WISE0247p3725_total_variable_SURF2022.pdf}
\figsetgrpnote{The light curve of WISE0247p3725 with \textit{Spitzer} ch2$_{aperture}$, ch2$_{PRF}$, and \textit{WISE} W2$_{PSF}$ photometry. The solid lines are the medians for each photometry type, while the dashed lines are $\pm$3$\sigma$. The spectral type is on the scale of 0 is M0, 10 is L0, 20 is T0, and 30 is Y0. }
\figsetgrpend

\figsetgrpstart
\figsetgrpnum{8.38}
\figsetgrptitle{Light Curve for WISE0257m2655}
\figsetplot{WISE0257m2655_total_variable_SURF2022.pdf}
\figsetgrpnote{The light curve of WISE0257m2655 with \textit{Spitzer} ch2$_{aperture}$, ch2$_{PRF}$, and \textit{WISE} W2$_{PSF}$ photometry. The solid lines are the medians for each photometry type, while the dashed lines are $\pm$3$\sigma$. The spectral type is on the scale of 0 is M0, 10 is L0, 20 is T0, and 30 is Y0. }
\figsetgrpend

\figsetgrpstart
\figsetgrpnum{8.39}
\figsetgrptitle{Light Curve for WISE0302m5817}
\figsetplot{WISE0302m5817_total_variable_SURF2022.pdf}
\figsetgrpnote{The light curve of WISE0302m5817 with \textit{Spitzer} ch2$_{aperture}$, ch2$_{PRF}$, and \textit{WISE} W2$_{PSF}$ photometry. The solid lines are the medians for each photometry type, while the dashed lines are $\pm$3$\sigma$. The spectral type is on the scale of 0 is M0, 10 is L0, 20 is T0, and 30 is Y0. }
\figsetgrpend

\figsetgrpstart
\figsetgrpnum{8.40}
\figsetgrptitle{Light Curve for WISE0304m2705}
\figsetplot{WISE0304m2705_total_variable_SURF2022.pdf}
\figsetgrpnote{The light curve of WISE0304m2705 with \textit{Spitzer} ch2$_{aperture}$, ch2$_{PRF}$, and \textit{WISE} W2$_{PSF}$ photometry. The solid lines are the medians for each photometry type, while the dashed lines are $\pm$3$\sigma$. The spectral type is on the scale of 0 is M0, 10 is L0, 20 is T0, and 30 is Y0. }
\figsetgrpend

\figsetgrpstart
\figsetgrpnum{8.41}
\figsetgrptitle{Light Curve for WISE0305p3954}
\figsetplot{WISE0305p3954_total_variable_SURF2022.pdf}
\figsetgrpnote{The light curve of WISE0305p3954 with \textit{Spitzer} ch2$_{aperture}$, ch2$_{PRF}$, and \textit{WISE} W2$_{PSF}$ photometry. The solid lines are the medians for each photometry type, while the dashed lines are $\pm$3$\sigma$. The spectral type is on the scale of 0 is M0, 10 is L0, 20 is T0, and 30 is Y0. }
\figsetgrpend

\figsetgrpstart
\figsetgrpnum{8.42}
\figsetgrptitle{Light Curve for WISE0309m5016}
\figsetplot{WISE0309m5016_total_variable_SURF2022.pdf}
\figsetgrpnote{The light curve of WISE0309m5016 with \textit{Spitzer} ch2$_{aperture}$, ch2$_{PRF}$, and \textit{WISE} W2$_{PSF}$ photometry. The solid lines are the medians for each photometry type, while the dashed lines are $\pm$3$\sigma$. The spectral type is on the scale of 0 is M0, 10 is L0, 20 is T0, and 30 is Y0. }
\figsetgrpend

\figsetgrpstart
\figsetgrpnum{8.43}
\figsetgrptitle{Light Curve for WISE0310m2756}
\figsetplot{WISE0310m2756_total_variable_SURF2022.pdf}
\figsetgrpnote{The light curve of WISE0310m2756 with \textit{Spitzer} ch2$_{aperture}$, ch2$_{PRF}$, and \textit{WISE} W2$_{PSF}$ photometry. The solid lines are the medians for each photometry type, while the dashed lines are $\pm$3$\sigma$. The spectral type is on the scale of 0 is M0, 10 is L0, 20 is T0, and 30 is Y0. }
\figsetgrpend

\figsetgrpstart
\figsetgrpnum{8.44}
\figsetgrptitle{Light Curve for WISE0313p7807}
\figsetplot{WISE0313p7807_total_variable_SURF2022.pdf}
\figsetgrpnote{The light curve of WISE0313p7807 with \textit{Spitzer} ch2$_{aperture}$, ch2$_{PRF}$, and \textit{WISE} W2$_{PSF}$ photometry. The solid lines are the medians for each photometry type, while the dashed lines are $\pm$3$\sigma$. The spectral type is on the scale of 0 is M0, 10 is L0, 20 is T0, and 30 is Y0. }
\figsetgrpend

\figsetgrpstart
\figsetgrpnum{8.45}
\figsetgrptitle{Light Curve for WISE0316p3820}
\figsetplot{WISE0316p3820_total_variable_SURF2022.pdf}
\figsetgrpnote{The light curve of WISE0316p3820 with \textit{Spitzer} ch2$_{aperture}$, ch2$_{PRF}$, and \textit{WISE} W2$_{PSF}$ photometry. The solid lines are the medians for each photometry type, while the dashed lines are $\pm$3$\sigma$. The spectral type is on the scale of 0 is M0, 10 is L0, 20 is T0, and 30 is Y0. }
\figsetgrpend

\figsetgrpstart
\figsetgrpnum{8.46}
\figsetgrptitle{Light Curve for WISE0316p4307}
\figsetplot{WISE0316p4307_total_variable_SURF2022.pdf}
\figsetgrpnote{The light curve of WISE0316p4307 with \textit{Spitzer} ch2$_{aperture}$, ch2$_{PRF}$, and \textit{WISE} W2$_{PSF}$ photometry. The solid lines are the medians for each photometry type, while the dashed lines are $\pm$3$\sigma$. The spectral type is on the scale of 0 is M0, 10 is L0, 20 is T0, and 30 is Y0. }
\figsetgrpend

\figsetgrpstart
\figsetgrpnum{8.47}
\figsetgrptitle{Light Curve for WISE0318m3421}
\figsetplot{WISE0318m3421_total_variable_SURF2022.pdf}
\figsetgrpnote{The light curve of WISE0318m3421 with \textit{Spitzer} ch2$_{aperture}$, ch2$_{PRF}$, and \textit{WISE} W2$_{PSF}$ photometry. The solid lines are the medians for each photometry type, while the dashed lines are $\pm$3$\sigma$. The spectral type is on the scale of 0 is M0, 10 is L0, 20 is T0, and 30 is Y0. }
\figsetgrpend

\figsetgrpstart
\figsetgrpnum{8.48}
\figsetgrptitle{Light Curve for WISE0321p6932}
\figsetplot{WISE0321p6932_total_variable_SURF2022.pdf}
\figsetgrpnote{The light curve of WISE0321p6932 with \textit{Spitzer} ch2$_{aperture}$, ch2$_{PRF}$, and \textit{WISE} W2$_{PSF}$ photometry. The solid lines are the medians for each photometry type, while the dashed lines are $\pm$3$\sigma$. The spectral type is on the scale of 0 is M0, 10 is L0, 20 is T0, and 30 is Y0. }
\figsetgrpend

\figsetgrpstart
\figsetgrpnum{8.49}
\figsetgrptitle{Light Curve for WISE0323p5625}
\figsetplot{WISE0323p5625_total_variable_SURF2022.pdf}
\figsetgrpnote{The light curve of WISE0323p5625 with \textit{Spitzer} ch2$_{aperture}$, ch2$_{PRF}$, and \textit{WISE} W2$_{PSF}$ photometry. The solid lines are the medians for each photometry type, while the dashed lines are $\pm$3$\sigma$. The spectral type is on the scale of 0 is M0, 10 is L0, 20 is T0, and 30 is Y0. }
\figsetgrpend

\figsetgrpstart
\figsetgrpnum{8.50}
\figsetgrptitle{Light Curve for WISE0323m5907}
\figsetplot{WISE0323m5907_total_variable_SURF2022.pdf}
\figsetgrpnote{The light curve of WISE0323m5907 with \textit{Spitzer} ch2$_{aperture}$, ch2$_{PRF}$, and \textit{WISE} W2$_{PSF}$ photometry. The solid lines are the medians for each photometry type, while the dashed lines are $\pm$3$\sigma$. The spectral type is on the scale of 0 is M0, 10 is L0, 20 is T0, and 30 is Y0. }
\figsetgrpend

\figsetgrpstart
\figsetgrpnum{8.51}
\figsetgrptitle{Light Curve for WISE0323m6025}
\figsetplot{WISE0323m6025_total_variable_SURF2022.pdf}
\figsetgrpnote{The light curve of WISE0323m6025 with \textit{Spitzer} ch2$_{aperture}$, ch2$_{PRF}$, and \textit{WISE} W2$_{PSF}$ photometry. The solid lines are the medians for each photometry type, while the dashed lines are $\pm$3$\sigma$. The spectral type is on the scale of 0 is M0, 10 is L0, 20 is T0, and 30 is Y0. }
\figsetgrpend

\figsetgrpstart
\figsetgrpnum{8.52}
\figsetgrptitle{Light Curve for WISE0325m5044}
\figsetplot{WISE0325m5044_total_variable_SURF2022.pdf}
\figsetgrpnote{The light curve of WISE0325m5044 with \textit{Spitzer} ch2$_{aperture}$, ch2$_{PRF}$, and \textit{WISE} W2$_{PSF}$ photometry. The solid lines are the medians for each photometry type, while the dashed lines are $\pm$3$\sigma$. The spectral type is on the scale of 0 is M0, 10 is L0, 20 is T0, and 30 is Y0. }
\figsetgrpend

\figsetgrpstart
\figsetgrpnum{8.53}
\figsetgrptitle{Light Curve for WISE0325m3854}
\figsetplot{WISE0325m3854_total_variable_SURF2022.pdf}
\figsetgrpnote{The light curve of WISE0325m3854 with \textit{Spitzer} ch2$_{aperture}$, ch2$_{PRF}$, and \textit{WISE} W2$_{PSF}$ photometry. The solid lines are the medians for each photometry type, while the dashed lines are $\pm$3$\sigma$. The spectral type is on the scale of 0 is M0, 10 is L0, 20 is T0, and 30 is Y0. }
\figsetgrpend

\figsetgrpstart
\figsetgrpnum{8.54}
\figsetgrptitle{Light Curve for WISE0325p0831}
\figsetplot{WISE0325p0831_total_variable_SURF2022.pdf}
\figsetgrpnote{The light curve of WISE0325p0831 with \textit{Spitzer} ch2$_{aperture}$, ch2$_{PRF}$, and \textit{WISE} W2$_{PSF}$ photometry. The solid lines are the medians for each photometry type, while the dashed lines are $\pm$3$\sigma$. The spectral type is on the scale of 0 is M0, 10 is L0, 20 is T0, and 30 is Y0. }
\figsetgrpend

\figsetgrpstart
\figsetgrpnum{8.55}
\figsetgrptitle{Light Curve for WISE0325p0425}
\figsetplot{WISE0325p0425_total_variable_SURF2022.pdf}
\figsetgrpnote{The light curve of WISE0325p0425 with \textit{Spitzer} ch2$_{aperture}$, ch2$_{PRF}$, and \textit{WISE} W2$_{PSF}$ photometry. The solid lines are the medians for each photometry type, while the dashed lines are $\pm$3$\sigma$. The spectral type is on the scale of 0 is M0, 10 is L0, 20 is T0, and 30 is Y0. }
\figsetgrpend

\figsetgrpstart
\figsetgrpnum{8.56}
\figsetgrptitle{Light Curve for WISE0330m0025}
\figsetplot{WISE0330m0025_total_variable_SURF2022.pdf}
\figsetgrpnote{The light curve of WISE0330m0025 with \textit{Spitzer} ch2$_{aperture}$, ch2$_{PRF}$, and \textit{WISE} W2$_{PSF}$ photometry. The solid lines are the medians for each photometry type, while the dashed lines are $\pm$3$\sigma$. The spectral type is on the scale of 0 is M0, 10 is L0, 20 is T0, and 30 is Y0. }
\figsetgrpend

\figsetgrpstart
\figsetgrpnum{8.57}
\figsetgrptitle{Light Curve for WISE0330m0350}
\figsetplot{WISE0330m0350_total_variable_SURF2022.pdf}
\figsetgrpnote{The light curve of WISE0330m0350 with \textit{Spitzer} ch2$_{aperture}$, ch2$_{PRF}$, and \textit{WISE} W2$_{PSF}$ photometry. The solid lines are the medians for each photometry type, while the dashed lines are $\pm$3$\sigma$. The spectral type is on the scale of 0 is M0, 10 is L0, 20 is T0, and 30 is Y0. }
\figsetgrpend

\figsetgrpstart
\figsetgrpnum{8.58}
\figsetgrptitle{Light Curve for WISE0333m5856}
\figsetplot{WISE0333m5856_total_variable_SURF2022.pdf}
\figsetgrpnote{The light curve of WISE0333m5856 with \textit{Spitzer} ch2$_{aperture}$, ch2$_{PRF}$, and \textit{WISE} W2$_{PSF}$ photometry. The solid lines are the medians for each photometry type, while the dashed lines are $\pm$3$\sigma$. The spectral type is on the scale of 0 is M0, 10 is L0, 20 is T0, and 30 is Y0. }
\figsetgrpend

\figsetgrpstart
\figsetgrpnum{8.59}
\figsetgrptitle{Light Curve for WISE0335p4310}
\figsetplot{WISE0335p4310_total_variable_SURF2022.pdf}
\figsetgrpnote{The light curve of WISE0335p4310 with \textit{Spitzer} ch2$_{aperture}$, ch2$_{PRF}$, and \textit{WISE} W2$_{PSF}$ photometry. The solid lines are the medians for each photometry type, while the dashed lines are $\pm$3$\sigma$. The spectral type is on the scale of 0 is M0, 10 is L0, 20 is T0, and 30 is Y0. }
\figsetgrpend

\figsetgrpstart
\figsetgrpnum{8.60}
\figsetgrptitle{Light Curve for WISE0336m0143}
\figsetplot{WISE0336m0143_total_variable_SURF2022.pdf}
\figsetgrpnote{The light curve of WISE0336m0143 with \textit{Spitzer} ch2$_{aperture}$, ch2$_{PRF}$, and \textit{WISE} W2$_{PSF}$ photometry. The solid lines are the medians for each photometry type, while the dashed lines are $\pm$3$\sigma$. The spectral type is on the scale of 0 is M0, 10 is L0, 20 is T0, and 30 is Y0. }
\figsetgrpend

\figsetgrpstart
\figsetgrpnum{8.61}
\figsetgrptitle{Light Curve for WISE0336p2826}
\figsetplot{WISE0336p2826_total_variable_SURF2022.pdf}
\figsetgrpnote{The light curve of WISE0336p2826 with \textit{Spitzer} ch2$_{aperture}$, ch2$_{PRF}$, and \textit{WISE} W2$_{PSF}$ photometry. The solid lines are the medians for each photometry type, while the dashed lines are $\pm$3$\sigma$. The spectral type is on the scale of 0 is M0, 10 is L0, 20 is T0, and 30 is Y0. }
\figsetgrpend

\figsetgrpstart
\figsetgrpnum{8.62}
\figsetgrptitle{Light Curve for WISE0337m1758}
\figsetplot{WISE0337m1758_total_variable_SURF2022.pdf}
\figsetgrpnote{The light curve of WISE0337m1758 with \textit{Spitzer} ch2$_{aperture}$, ch2$_{PRF}$, and \textit{WISE} W2$_{PSF}$ photometry. The solid lines are the medians for each photometry type, while the dashed lines are $\pm$3$\sigma$. The spectral type is on the scale of 0 is M0, 10 is L0, 20 is T0, and 30 is Y0. }
\figsetgrpend

\figsetgrpstart
\figsetgrpnum{8.63}
\figsetgrptitle{Light Curve for WISE0340m6724}
\figsetplot{WISE0340m6724_total_variable_SURF2022.pdf}
\figsetgrpnote{The light curve of WISE0340m6724 with \textit{Spitzer} ch2$_{aperture}$, ch2$_{PRF}$, and \textit{WISE} W2$_{PSF}$ photometry. The solid lines are the medians for each photometry type, while the dashed lines are $\pm$3$\sigma$. The spectral type is on the scale of 0 is M0, 10 is L0, 20 is T0, and 30 is Y0. }
\figsetgrpend

\figsetgrpstart
\figsetgrpnum{8.64}
\figsetgrptitle{Light Curve for WISE0350m5658}
\figsetplot{WISE0350m5658_total_variable_SURF2022.pdf}
\figsetgrpnote{The light curve of WISE0350m5658 with \textit{Spitzer} ch2$_{aperture}$, ch2$_{PRF}$, and \textit{WISE} W2$_{PSF}$ photometry. The solid lines are the medians for each photometry type, while the dashed lines are $\pm$3$\sigma$. The spectral type is on the scale of 0 is M0, 10 is L0, 20 is T0, and 30 is Y0. }
\figsetgrpend

\figsetgrpstart
\figsetgrpnum{8.65}
\figsetgrptitle{Light Curve for WISE0355p4743}
\figsetplot{WISE0355p4743_total_variable_SURF2022.pdf}
\figsetgrpnote{The light curve of WISE0355p4743 with \textit{Spitzer} ch2$_{aperture}$, ch2$_{PRF}$, and \textit{WISE} W2$_{PSF}$ photometry. The solid lines are the medians for each photometry type, while the dashed lines are $\pm$3$\sigma$. The spectral type is on the scale of 0 is M0, 10 is L0, 20 is T0, and 30 is Y0. }
\figsetgrpend

\figsetgrpstart
\figsetgrpnum{8.66}
\figsetgrptitle{Light Curve for WISE0358m4116}
\figsetplot{WISE0358m4116_total_variable_SURF2022.pdf}
\figsetgrpnote{The light curve of WISE0358m4116 with \textit{Spitzer} ch2$_{aperture}$, ch2$_{PRF}$, and \textit{WISE} W2$_{PSF}$ photometry. The solid lines are the medians for each photometry type, while the dashed lines are $\pm$3$\sigma$. The spectral type is on the scale of 0 is M0, 10 is L0, 20 is T0, and 30 is Y0. }
\figsetgrpend

\figsetgrpstart
\figsetgrpnum{8.67}
\figsetgrptitle{Light Curve for WISE0359m5401}
\figsetplot{WISE0359m5401_total_variable_SURF2022.pdf}
\figsetgrpnote{The light curve of WISE0359m5401 with \textit{Spitzer} ch2$_{aperture}$, ch2$_{PRF}$, and \textit{WISE} W2$_{PSF}$ photometry. The solid lines are the medians for each photometry type, while the dashed lines are $\pm$3$\sigma$. The spectral type is on the scale of 0 is M0, 10 is L0, 20 is T0, and 30 is Y0. }
\figsetgrpend

\figsetgrpstart
\figsetgrpnum{8.68}
\figsetgrptitle{Light Curve for WISE0402m2651}
\figsetplot{WISE0402m2651_total_variable_SURF2022.pdf}
\figsetgrpnote{The light curve of WISE0402m2651 with \textit{Spitzer} ch2$_{aperture}$, ch2$_{PRF}$, and \textit{WISE} W2$_{PSF}$ photometry. The solid lines are the medians for each photometry type, while the dashed lines are $\pm$3$\sigma$. The spectral type is on the scale of 0 is M0, 10 is L0, 20 is T0, and 30 is Y0. }
\figsetgrpend

\figsetgrpstart
\figsetgrpnum{8.69}
\figsetgrptitle{Light Curve for WISE0404m6420}
\figsetplot{WISE0404m6420_total_variable_SURF2022.pdf}
\figsetgrpnote{The light curve of WISE0404m6420 with \textit{Spitzer} ch2$_{aperture}$, ch2$_{PRF}$, and \textit{WISE} W2$_{PSF}$ photometry. The solid lines are the medians for each photometry type, while the dashed lines are $\pm$3$\sigma$. The spectral type is on the scale of 0 is M0, 10 is L0, 20 is T0, and 30 is Y0. }
\figsetgrpend

\figsetgrpstart
\figsetgrpnum{8.70}
\figsetgrptitle{Light Curve for WISE0410p1502}
\figsetplot{WISE0410p1502_total_variable_SURF2022.pdf}
\figsetgrpnote{The light curve of WISE0410p1502 with \textit{Spitzer} ch2$_{aperture}$, ch2$_{PRF}$, and \textit{WISE} W2$_{PSF}$ photometry. The solid lines are the medians for each photometry type, while the dashed lines are $\pm$3$\sigma$. The spectral type is on the scale of 0 is M0, 10 is L0, 20 is T0, and 30 is Y0. }
\figsetgrpend

\figsetgrpstart
\figsetgrpnum{8.71}
\figsetgrptitle{Light Curve for WISE0413m4750}
\figsetplot{WISE0413m4750_total_variable_SURF2022.pdf}
\figsetgrpnote{The light curve of WISE0413m4750 with \textit{Spitzer} ch2$_{aperture}$, ch2$_{PRF}$, and \textit{WISE} W2$_{PSF}$ photometry. The solid lines are the medians for each photometry type, while the dashed lines are $\pm$3$\sigma$. The spectral type is on the scale of 0 is M0, 10 is L0, 20 is T0, and 30 is Y0. }
\figsetgrpend

\figsetgrpstart
\figsetgrpnum{8.72}
\figsetgrptitle{Light Curve for WISE0421m6306}
\figsetplot{WISE0421m6306_total_variable_SURF2022.pdf}
\figsetgrpnote{The light curve of WISE0421m6306 with \textit{Spitzer} ch2$_{aperture}$, ch2$_{PRF}$, and \textit{WISE} W2$_{PSF}$ photometry. The solid lines are the medians for each photometry type, while the dashed lines are $\pm$3$\sigma$. The spectral type is on the scale of 0 is M0, 10 is L0, 20 is T0, and 30 is Y0. }
\figsetgrpend

\figsetgrpstart
\figsetgrpnum{8.73}
\figsetgrptitle{Light Curve for WISE0423m4019}
\figsetplot{WISE0423m4019_total_variable_SURF2022.pdf}
\figsetgrpnote{The light curve of WISE0423m4019 with \textit{Spitzer} ch2$_{aperture}$, ch2$_{PRF}$, and \textit{WISE} W2$_{PSF}$ photometry. The solid lines are the medians for each photometry type, while the dashed lines are $\pm$3$\sigma$. The spectral type is on the scale of 0 is M0, 10 is L0, 20 is T0, and 30 is Y0. }
\figsetgrpend

\figsetgrpstart
\figsetgrpnum{8.74}
\figsetgrptitle{Light Curve for WISE0424p0002}
\figsetplot{WISE0424p0002_total_variable_SURF2022.pdf}
\figsetgrpnote{The light curve of WISE0424p0002 with \textit{Spitzer} ch2$_{aperture}$, ch2$_{PRF}$, and \textit{WISE} W2$_{PSF}$ photometry. The solid lines are the medians for each photometry type, while the dashed lines are $\pm$3$\sigma$. The spectral type is on the scale of 0 is M0, 10 is L0, 20 is T0, and 30 is Y0. }
\figsetgrpend

\figsetgrpstart
\figsetgrpnum{8.75}
\figsetgrptitle{Light Curve for WISE0430p4633}
\figsetplot{WISE0430p4633_total_variable_SURF2022.pdf}
\figsetgrpnote{The light curve of WISE0430p4633 with \textit{Spitzer} ch2$_{aperture}$, ch2$_{PRF}$, and \textit{WISE} W2$_{PSF}$ photometry. The solid lines are the medians for each photometry type, while the dashed lines are $\pm$3$\sigma$. The spectral type is on the scale of 0 is M0, 10 is L0, 20 is T0, and 30 is Y0. }
\figsetgrpend

\figsetgrpstart
\figsetgrpnum{8.76}
\figsetgrptitle{Light Curve for WISE0442m3855}
\figsetplot{WISE0442m3855_total_variable_SURF2022.pdf}
\figsetgrpnote{The light curve of WISE0442m3855 with \textit{Spitzer} ch2$_{aperture}$, ch2$_{PRF}$, and \textit{WISE} W2$_{PSF}$ photometry. The solid lines are the medians for each photometry type, while the dashed lines are $\pm$3$\sigma$. The spectral type is on the scale of 0 is M0, 10 is L0, 20 is T0, and 30 is Y0. }
\figsetgrpend

\figsetgrpstart
\figsetgrpnum{8.77}
\figsetgrptitle{Light Curve for WISE0443m3202}
\figsetplot{WISE0443m3202_total_variable_SURF2022.pdf}
\figsetgrpnote{The light curve of WISE0443m3202 with \textit{Spitzer} ch2$_{aperture}$, ch2$_{PRF}$, and \textit{WISE} W2$_{PSF}$ photometry. The solid lines are the medians for each photometry type, while the dashed lines are $\pm$3$\sigma$. The spectral type is on the scale of 0 is M0, 10 is L0, 20 is T0, and 30 is Y0. }
\figsetgrpend

\figsetgrpstart
\figsetgrpnum{8.78}
\figsetgrptitle{Light Curve for WISE0448m1935}
\figsetplot{WISE0448m1935_total_variable_SURF2022.pdf}
\figsetgrpnote{The light curve of WISE0448m1935 with \textit{Spitzer} ch2$_{aperture}$, ch2$_{PRF}$, and \textit{WISE} W2$_{PSF}$ photometry. The solid lines are the medians for each photometry type, while the dashed lines are $\pm$3$\sigma$. The spectral type is on the scale of 0 is M0, 10 is L0, 20 is T0, and 30 is Y0. }
\figsetgrpend

\figsetgrpstart
\figsetgrpnum{8.79}
\figsetgrptitle{Light Curve for WISE0457m0207}
\figsetplot{WISE0457m0207_total_variable_SURF2022.pdf}
\figsetgrpnote{The light curve of WISE0457m0207 with \textit{Spitzer} ch2$_{aperture}$, ch2$_{PRF}$, and \textit{WISE} W2$_{PSF}$ photometry. The solid lines are the medians for each photometry type, while the dashed lines are $\pm$3$\sigma$. The spectral type is on the scale of 0 is M0, 10 is L0, 20 is T0, and 30 is Y0. }
\figsetgrpend

\figsetgrpstart
\figsetgrpnum{8.80}
\figsetgrptitle{Light Curve for WISE0458p6434}
\figsetplot{WISE0458p6434_total_variable_SURF2022.pdf}
\figsetgrpnote{The light curve of WISE0458p6434 with \textit{Spitzer} ch2$_{aperture}$, ch2$_{PRF}$, and \textit{WISE} W2$_{PSF}$ photometry. The solid lines are the medians for each photometry type, while the dashed lines are $\pm$3$\sigma$. The spectral type is on the scale of 0 is M0, 10 is L0, 20 is T0, and 30 is Y0. }
\figsetgrpend

\figsetgrpstart
\figsetgrpnum{8.81}
\figsetgrptitle{Light Curve for WISE0500m1223}
\figsetplot{WISE0500m1223_total_variable_SURF2022.pdf}
\figsetgrpnote{The light curve of WISE0500m1223 with \textit{Spitzer} ch2$_{aperture}$, ch2$_{PRF}$, and \textit{WISE} W2$_{PSF}$ photometry. The solid lines are the medians for each photometry type, while the dashed lines are $\pm$3$\sigma$. The spectral type is on the scale of 0 is M0, 10 is L0, 20 is T0, and 30 is Y0. }
\figsetgrpend

\figsetgrpstart
\figsetgrpnum{8.82}
\figsetgrptitle{Light Curve for WISE0502p1007}
\figsetplot{WISE0502p1007_total_variable_SURF2022.pdf}
\figsetgrpnote{The light curve of WISE0502p1007 with \textit{Spitzer} ch2$_{aperture}$, ch2$_{PRF}$, and \textit{WISE} W2$_{PSF}$ photometry. The solid lines are the medians for each photometry type, while the dashed lines are $\pm$3$\sigma$. The spectral type is on the scale of 0 is M0, 10 is L0, 20 is T0, and 30 is Y0. }
\figsetgrpend

\figsetgrpstart
\figsetgrpnum{8.83}
\figsetgrptitle{Light Curve for WISE0503m5648}
\figsetplot{WISE0503m5648_total_variable_SURF2022.pdf}
\figsetgrpnote{The light curve of WISE0503m5648 with \textit{Spitzer} ch2$_{aperture}$, ch2$_{PRF}$, and \textit{WISE} W2$_{PSF}$ photometry. The solid lines are the medians for each photometry type, while the dashed lines are $\pm$3$\sigma$. The spectral type is on the scale of 0 is M0, 10 is L0, 20 is T0, and 30 is Y0. }
\figsetgrpend

\figsetgrpstart
\figsetgrpnum{8.84}
\figsetgrptitle{Light Curve for WISE0506p5236}
\figsetplot{WISE0506p5236_total_variable_SURF2022.pdf}
\figsetgrpnote{The light curve of WISE0506p5236 with \textit{Spitzer} ch2$_{aperture}$, ch2$_{PRF}$, and \textit{WISE} W2$_{PSF}$ photometry. The solid lines are the medians for each photometry type, while the dashed lines are $\pm$3$\sigma$. The spectral type is on the scale of 0 is M0, 10 is L0, 20 is T0, and 30 is Y0. }
\figsetgrpend

\figsetgrpstart
\figsetgrpnum{8.85}
\figsetgrptitle{Light Curve for WISE0512m2949}
\figsetplot{WISE0512m2949_total_variable_SURF2022.pdf}
\figsetgrpnote{The light curve of WISE0512m2949 with \textit{Spitzer} ch2$_{aperture}$, ch2$_{PRF}$, and \textit{WISE} W2$_{PSF}$ photometry. The solid lines are the medians for each photometry type, while the dashed lines are $\pm$3$\sigma$. The spectral type is on the scale of 0 is M0, 10 is L0, 20 is T0, and 30 is Y0. }
\figsetgrpend

\figsetgrpstart
\figsetgrpnum{8.86}
\figsetgrptitle{Light Curve for WISE0512m3004}
\figsetplot{WISE0512m3004_total_variable_SURF2022.pdf}
\figsetgrpnote{The light curve of WISE0512m3004 with \textit{Spitzer} ch2$_{aperture}$, ch2$_{PRF}$, and \textit{WISE} W2$_{PSF}$ photometry. The solid lines are the medians for each photometry type, while the dashed lines are $\pm$3$\sigma$. The spectral type is on the scale of 0 is M0, 10 is L0, 20 is T0, and 30 is Y0. }
\figsetgrpend

\figsetgrpstart
\figsetgrpnum{8.87}
\figsetgrptitle{Light Curve for WISE0521p1025}
\figsetplot{WISE0521p1025_total_variable_SURF2022.pdf}
\figsetgrpnote{The light curve of WISE0521p1025 with \textit{Spitzer} ch2$_{aperture}$, ch2$_{PRF}$, and \textit{WISE} W2$_{PSF}$ photometry. The solid lines are the medians for each photometry type, while the dashed lines are $\pm$3$\sigma$. The spectral type is on the scale of 0 is M0, 10 is L0, 20 is T0, and 30 is Y0. }
\figsetgrpend

\figsetgrpstart
\figsetgrpnum{8.88}
\figsetgrptitle{Light Curve for WISE0535m7500}
\figsetplot{WISE0535m7500_total_variable_SURF2022.pdf}
\figsetgrpnote{The light curve of WISE0535m7500 with \textit{Spitzer} ch2$_{aperture}$, ch2$_{PRF}$, and \textit{WISE} W2$_{PSF}$ photometry. The solid lines are the medians for each photometry type, while the dashed lines are $\pm$3$\sigma$. The spectral type is on the scale of 0 is M0, 10 is L0, 20 is T0, and 30 is Y0. }
\figsetgrpend

\figsetgrpstart
\figsetgrpnum{8.89}
\figsetgrptitle{Light Curve for WISE0536m3055}
\figsetplot{WISE0536m3055_total_variable_SURF2022.pdf}
\figsetgrpnote{The light curve of WISE0536m3055 with \textit{Spitzer} ch2$_{aperture}$, ch2$_{PRF}$, and \textit{WISE} W2$_{PSF}$ photometry. The solid lines are the medians for each photometry type, while the dashed lines are $\pm$3$\sigma$. The spectral type is on the scale of 0 is M0, 10 is L0, 20 is T0, and 30 is Y0. }
\figsetgrpend

\figsetgrpstart
\figsetgrpnum{8.90}
\figsetgrptitle{Light Curve for WISE0540m1802}
\figsetplot{WISE0540m1802_total_variable_SURF2022.pdf}
\figsetgrpnote{The light curve of WISE0540m1802 with \textit{Spitzer} ch2$_{aperture}$, ch2$_{PRF}$, and \textit{WISE} W2$_{PSF}$ photometry. The solid lines are the medians for each photometry type, while the dashed lines are $\pm$3$\sigma$. The spectral type is on the scale of 0 is M0, 10 is L0, 20 is T0, and 30 is Y0. }
\figsetgrpend

\figsetgrpstart
\figsetgrpnum{8.91}
\figsetgrptitle{Light Curve for WISE0540p4832}
\figsetplot{WISE0540p4832_total_variable_SURF2022.pdf}
\figsetgrpnote{The light curve of WISE0540p4832 with \textit{Spitzer} ch2$_{aperture}$, ch2$_{PRF}$, and \textit{WISE} W2$_{PSF}$ photometry. The solid lines are the medians for each photometry type, while the dashed lines are $\pm$3$\sigma$. The spectral type is on the scale of 0 is M0, 10 is L0, 20 is T0, and 30 is Y0. }
\figsetgrpend

\figsetgrpstart
\figsetgrpnum{8.92}
\figsetgrptitle{Light Curve for WISE0546m0959}
\figsetplot{WISE0546m0959_total_variable_SURF2022.pdf}
\figsetgrpnote{The light curve of WISE0546m0959 with \textit{Spitzer} ch2$_{aperture}$, ch2$_{PRF}$, and \textit{WISE} W2$_{PSF}$ photometry. The solid lines are the medians for each photometry type, while the dashed lines are $\pm$3$\sigma$. The spectral type is on the scale of 0 is M0, 10 is L0, 20 is T0, and 30 is Y0. }
\figsetgrpend

\figsetgrpstart
\figsetgrpnum{8.93}
\figsetgrptitle{Light Curve for WISE0602p4043}
\figsetplot{WISE0602p4043_total_variable_SURF2022.pdf}
\figsetgrpnote{The light curve of WISE0602p4043 with \textit{Spitzer} ch2$_{aperture}$, ch2$_{PRF}$, and \textit{WISE} W2$_{PSF}$ photometry. The solid lines are the medians for each photometry type, while the dashed lines are $\pm$3$\sigma$. The spectral type is on the scale of 0 is M0, 10 is L0, 20 is T0, and 30 is Y0. }
\figsetgrpend

\figsetgrpstart
\figsetgrpnum{8.94}
\figsetgrptitle{Light Curve for WISE0613p4808}
\figsetplot{WISE0613p4808_total_variable_SURF2022.pdf}
\figsetgrpnote{The light curve of WISE0613p4808 with \textit{Spitzer} ch2$_{aperture}$, ch2$_{PRF}$, and \textit{WISE} W2$_{PSF}$ photometry. The solid lines are the medians for each photometry type, while the dashed lines are $\pm$3$\sigma$. The spectral type is on the scale of 0 is M0, 10 is L0, 20 is T0, and 30 is Y0. }
\figsetgrpend

\figsetgrpstart
\figsetgrpnum{8.95}
\figsetgrptitle{Light Curve for WISE0614p0951}
\figsetplot{WISE0614p0951_total_variable_SURF2022.pdf}
\figsetgrpnote{The light curve of WISE0614p0951 with \textit{Spitzer} ch2$_{aperture}$, ch2$_{PRF}$, and \textit{WISE} W2$_{PSF}$ photometry. The solid lines are the medians for each photometry type, while the dashed lines are $\pm$3$\sigma$. The spectral type is on the scale of 0 is M0, 10 is L0, 20 is T0, and 30 is Y0. }
\figsetgrpend

\figsetgrpstart
\figsetgrpnum{8.96}
\figsetgrptitle{Light Curve for WISE0615p1526}
\figsetplot{WISE0615p1526_total_variable_SURF2022.pdf}
\figsetgrpnote{The light curve of WISE0615p1526 with \textit{Spitzer} ch2$_{aperture}$, ch2$_{PRF}$, and \textit{WISE} W2$_{PSF}$ photometry. The solid lines are the medians for each photometry type, while the dashed lines are $\pm$3$\sigma$. The spectral type is on the scale of 0 is M0, 10 is L0, 20 is T0, and 30 is Y0. }
\figsetgrpend

\figsetgrpstart
\figsetgrpnum{8.97}
\figsetgrptitle{Light Curve for WISE0628m8057}
\figsetplot{WISE0628m8057_total_variable_SURF2022.pdf}
\figsetgrpnote{The light curve of WISE0628m8057 with \textit{Spitzer} ch2$_{aperture}$, ch2$_{PRF}$, and \textit{WISE} W2$_{PSF}$ photometry. The solid lines are the medians for each photometry type, while the dashed lines are $\pm$3$\sigma$. The spectral type is on the scale of 0 is M0, 10 is L0, 20 is T0, and 30 is Y0. }
\figsetgrpend

\figsetgrpstart
\figsetgrpnum{8.98}
\figsetgrptitle{Light Curve for WISE0629p2418}
\figsetplot{WISE0629p2418_total_variable_SURF2022.pdf}
\figsetgrpnote{The light curve of WISE0629p2418 with \textit{Spitzer} ch2$_{aperture}$, ch2$_{PRF}$, and \textit{WISE} W2$_{PSF}$ photometry. The solid lines are the medians for each photometry type, while the dashed lines are $\pm$3$\sigma$. The spectral type is on the scale of 0 is M0, 10 is L0, 20 is T0, and 30 is Y0. }
\figsetgrpend

\figsetgrpstart
\figsetgrpnum{8.99}
\figsetgrptitle{Light Curve for WISE0634p5049}
\figsetplot{WISE0634p5049_total_variable_SURF2022.pdf}
\figsetgrpnote{The light curve of WISE0634p5049 with \textit{Spitzer} ch2$_{aperture}$, ch2$_{PRF}$, and \textit{WISE} W2$_{PSF}$ photometry. The solid lines are the medians for each photometry type, while the dashed lines are $\pm$3$\sigma$. The spectral type is on the scale of 0 is M0, 10 is L0, 20 is T0, and 30 is Y0. }
\figsetgrpend

\figsetgrpstart
\figsetgrpnum{8.100}
\figsetgrptitle{Light Curve for WISE0642p4101}
\figsetplot{WISE0642p4101_total_variable_SURF2022.pdf}
\figsetgrpnote{The light curve of WISE0642p4101 with \textit{Spitzer} ch2$_{aperture}$, ch2$_{PRF}$, and \textit{WISE} W2$_{PSF}$ photometry. The solid lines are the medians for each photometry type, while the dashed lines are $\pm$3$\sigma$. The spectral type is on the scale of 0 is M0, 10 is L0, 20 is T0, and 30 is Y0. }
\figsetgrpend

\figsetgrpstart
\figsetgrpnum{8.101}
\figsetgrptitle{Light Curve for WISE0642p0423}
\figsetplot{WISE0642p0423_total_variable_SURF2022.pdf}
\figsetgrpnote{The light curve of WISE0642p0423 with \textit{Spitzer} ch2$_{aperture}$, ch2$_{PRF}$, and \textit{WISE} W2$_{PSF}$ photometry. The solid lines are the medians for each photometry type, while the dashed lines are $\pm$3$\sigma$. The spectral type is on the scale of 0 is M0, 10 is L0, 20 is T0, and 30 is Y0. }
\figsetgrpend

\figsetgrpstart
\figsetgrpnum{8.102}
\figsetgrptitle{Light Curve for WISE0645p5240}
\figsetplot{WISE0645p5240_total_variable_SURF2022.pdf}
\figsetgrpnote{The light curve of WISE0645p5240 with \textit{Spitzer} ch2$_{aperture}$, ch2$_{PRF}$, and \textit{WISE} W2$_{PSF}$ photometry. The solid lines are the medians for each photometry type, while the dashed lines are $\pm$3$\sigma$. The spectral type is on the scale of 0 is M0, 10 is L0, 20 is T0, and 30 is Y0. }
\figsetgrpend

\figsetgrpstart
\figsetgrpnum{8.103}
\figsetgrptitle{Light Curve for WISE0645m6645}
\figsetplot{WISE0645m6645_total_variable_SURF2022.pdf}
\figsetgrpnote{The light curve of WISE0645m6645 with \textit{Spitzer} ch2$_{aperture}$, ch2$_{PRF}$, and \textit{WISE} W2$_{PSF}$ photometry. The solid lines are the medians for each photometry type, while the dashed lines are $\pm$3$\sigma$. The spectral type is on the scale of 0 is M0, 10 is L0, 20 is T0, and 30 is Y0. }
\figsetgrpend

\figsetgrpstart
\figsetgrpnum{8.104}
\figsetgrptitle{Light Curve for WISE0645m0302}
\figsetplot{WISE0645m0302_total_variable_SURF2022.pdf}
\figsetgrpnote{The light curve of WISE0645m0302 with \textit{Spitzer} ch2$_{aperture}$, ch2$_{PRF}$, and \textit{WISE} W2$_{PSF}$ photometry. The solid lines are the medians for each photometry type, while the dashed lines are $\pm$3$\sigma$. The spectral type is on the scale of 0 is M0, 10 is L0, 20 is T0, and 30 is Y0. }
\figsetgrpend

\figsetgrpstart
\figsetgrpnum{8.105}
\figsetgrptitle{Light Curve for WISE0647m6232}
\figsetplot{WISE0647m6232_total_variable_SURF2022.pdf}
\figsetgrpnote{The light curve of WISE0647m6232 with \textit{Spitzer} ch2$_{aperture}$, ch2$_{PRF}$, and \textit{WISE} W2$_{PSF}$ photometry. The solid lines are the medians for each photometry type, while the dashed lines are $\pm$3$\sigma$. The spectral type is on the scale of 0 is M0, 10 is L0, 20 is T0, and 30 is Y0. }
\figsetgrpend

\figsetgrpstart
\figsetgrpnum{8.106}
\figsetgrptitle{Light Curve for WISE0647m1546}
\figsetplot{WISE0647m1546_total_variable_SURF2022.pdf}
\figsetgrpnote{The light curve of WISE0647m1546 with \textit{Spitzer} ch2$_{aperture}$, ch2$_{PRF}$, and \textit{WISE} W2$_{PSF}$ photometry. The solid lines are the medians for each photometry type, while the dashed lines are $\pm$3$\sigma$. The spectral type is on the scale of 0 is M0, 10 is L0, 20 is T0, and 30 is Y0. }
\figsetgrpend

\figsetgrpstart
\figsetgrpnum{8.107}
\figsetgrptitle{Light Curve for WISE0652p4127}
\figsetplot{WISE0652p4127_total_variable_SURF2022.pdf}
\figsetgrpnote{The light curve of WISE0652p4127 with \textit{Spitzer} ch2$_{aperture}$, ch2$_{PRF}$, and \textit{WISE} W2$_{PSF}$ photometry. The solid lines are the medians for each photometry type, while the dashed lines are $\pm$3$\sigma$. The spectral type is on the scale of 0 is M0, 10 is L0, 20 is T0, and 30 is Y0. }
\figsetgrpend

\figsetgrpstart
\figsetgrpnum{8.108}
\figsetgrptitle{Light Curve for WISE0701p6321}
\figsetplot{WISE0701p6321_total_variable_SURF2022.pdf}
\figsetgrpnote{The light curve of WISE0701p6321 with \textit{Spitzer} ch2$_{aperture}$, ch2$_{PRF}$, and \textit{WISE} W2$_{PSF}$ photometry. The solid lines are the medians for each photometry type, while the dashed lines are $\pm$3$\sigma$. The spectral type is on the scale of 0 is M0, 10 is L0, 20 is T0, and 30 is Y0. }
\figsetgrpend

\figsetgrpstart
\figsetgrpnum{8.109}
\figsetgrptitle{Light Curve for WISE0713m5854}
\figsetplot{WISE0713m5854_total_variable_SURF2022.pdf}
\figsetgrpnote{The light curve of WISE0713m5854 with \textit{Spitzer} ch2$_{aperture}$, ch2$_{PRF}$, and \textit{WISE} W2$_{PSF}$ photometry. The solid lines are the medians for each photometry type, while the dashed lines are $\pm$3$\sigma$. The spectral type is on the scale of 0 is M0, 10 is L0, 20 is T0, and 30 is Y0. }
\figsetgrpend

\figsetgrpstart
\figsetgrpnum{8.110}
\figsetgrptitle{Light Curve for WISE0713m2917}
\figsetplot{WISE0713m2917_total_variable_SURF2022.pdf}
\figsetgrpnote{The light curve of WISE0713m2917 with \textit{Spitzer} ch2$_{aperture}$, ch2$_{PRF}$, and \textit{WISE} W2$_{PSF}$ photometry. The solid lines are the medians for each photometry type, while the dashed lines are $\pm$3$\sigma$. The spectral type is on the scale of 0 is M0, 10 is L0, 20 is T0, and 30 is Y0. }
\figsetgrpend

\figsetgrpstart
\figsetgrpnum{8.111}
\figsetgrptitle{Light Curve for WISE0723p3403}
\figsetplot{WISE0723p3403_total_variable_SURF2022.pdf}
\figsetgrpnote{The light curve of WISE0723p3403 with \textit{Spitzer} ch2$_{aperture}$, ch2$_{PRF}$, and \textit{WISE} W2$_{PSF}$ photometry. The solid lines are the medians for each photometry type, while the dashed lines are $\pm$3$\sigma$. The spectral type is on the scale of 0 is M0, 10 is L0, 20 is T0, and 30 is Y0. }
\figsetgrpend

\figsetgrpstart
\figsetgrpnum{8.112}
\figsetgrptitle{Light Curve for WISE0734m7157}
\figsetplot{WISE0734m7157_total_variable_SURF2022.pdf}
\figsetgrpnote{The light curve of WISE0734m7157 with \textit{Spitzer} ch2$_{aperture}$, ch2$_{PRF}$, and \textit{WISE} W2$_{PSF}$ photometry. The solid lines are the medians for each photometry type, while the dashed lines are $\pm$3$\sigma$. The spectral type is on the scale of 0 is M0, 10 is L0, 20 is T0, and 30 is Y0. }
\figsetgrpend

\figsetgrpstart
\figsetgrpnum{8.113}
\figsetgrptitle{Light Curve for WISE0741m0506}
\figsetplot{WISE0741m0506_total_variable_SURF2022.pdf}
\figsetgrpnote{The light curve of WISE0741m0506 with \textit{Spitzer} ch2$_{aperture}$, ch2$_{PRF}$, and \textit{WISE} W2$_{PSF}$ photometry. The solid lines are the medians for each photometry type, while the dashed lines are $\pm$3$\sigma$. The spectral type is on the scale of 0 is M0, 10 is L0, 20 is T0, and 30 is Y0. }
\figsetgrpend

\figsetgrpstart
\figsetgrpnum{8.114}
\figsetgrptitle{Light Curve for WISE0741p2351}
\figsetplot{WISE0741p2351_total_variable_SURF2022.pdf}
\figsetgrpnote{The light curve of WISE0741p2351 with \textit{Spitzer} ch2$_{aperture}$, ch2$_{PRF}$, and \textit{WISE} W2$_{PSF}$ photometry. The solid lines are the medians for each photometry type, while the dashed lines are $\pm$3$\sigma$. The spectral type is on the scale of 0 is M0, 10 is L0, 20 is T0, and 30 is Y0. }
\figsetgrpend

\figsetgrpstart
\figsetgrpnum{8.115}
\figsetgrptitle{Light Curve for WISE0742p2055}
\figsetplot{WISE0742p2055_total_variable_SURF2022.pdf}
\figsetgrpnote{The light curve of WISE0742p2055 with \textit{Spitzer} ch2$_{aperture}$, ch2$_{PRF}$, and \textit{WISE} W2$_{PSF}$ photometry. The solid lines are the medians for each photometry type, while the dashed lines are $\pm$3$\sigma$. The spectral type is on the scale of 0 is M0, 10 is L0, 20 is T0, and 30 is Y0. }
\figsetgrpend

\figsetgrpstart
\figsetgrpnum{8.116}
\figsetgrptitle{Light Curve for WISE0744p5628}
\figsetplot{WISE0744p5628_total_variable_SURF2022.pdf}
\figsetgrpnote{The light curve of WISE0744p5628 with \textit{Spitzer} ch2$_{aperture}$, ch2$_{PRF}$, and \textit{WISE} W2$_{PSF}$ photometry. The solid lines are the medians for each photometry type, while the dashed lines are $\pm$3$\sigma$. The spectral type is on the scale of 0 is M0, 10 is L0, 20 is T0, and 30 is Y0. }
\figsetgrpend

\figsetgrpstart
\figsetgrpnum{8.117}
\figsetgrptitle{Light Curve for WISE0745p2332}
\figsetplot{WISE0745p2332_total_variable_SURF2022.pdf}
\figsetgrpnote{The light curve of WISE0745p2332 with \textit{Spitzer} ch2$_{aperture}$, ch2$_{PRF}$, and \textit{WISE} W2$_{PSF}$ photometry. The solid lines are the medians for each photometry type, while the dashed lines are $\pm$3$\sigma$. The spectral type is on the scale of 0 is M0, 10 is L0, 20 is T0, and 30 is Y0. }
\figsetgrpend

\figsetgrpstart
\figsetgrpnum{8.118}
\figsetgrptitle{Light Curve for WISE0755p2212}
\figsetplot{WISE0755p2212_total_variable_SURF2022.pdf}
\figsetgrpnote{The light curve of WISE0755p2212 with \textit{Spitzer} ch2$_{aperture}$, ch2$_{PRF}$, and \textit{WISE} W2$_{PSF}$ photometry. The solid lines are the medians for each photometry type, while the dashed lines are $\pm$3$\sigma$. The spectral type is on the scale of 0 is M0, 10 is L0, 20 is T0, and 30 is Y0. }
\figsetgrpend

\figsetgrpstart
\figsetgrpnum{8.119}
\figsetgrptitle{Light Curve for WISE0755m3259}
\figsetplot{WISE0755m3259_total_variable_SURF2022.pdf}
\figsetgrpnote{The light curve of WISE0755m3259 with \textit{Spitzer} ch2$_{aperture}$, ch2$_{PRF}$, and \textit{WISE} W2$_{PSF}$ photometry. The solid lines are the medians for each photometry type, while the dashed lines are $\pm$3$\sigma$. The spectral type is on the scale of 0 is M0, 10 is L0, 20 is T0, and 30 is Y0. }
\figsetgrpend

\figsetgrpstart
\figsetgrpnum{8.120}
\figsetgrptitle{Light Curve for WISE0758p3247}
\figsetplot{WISE0758p3247_total_variable_SURF2022.pdf}
\figsetgrpnote{The light curve of WISE0758p3247 with \textit{Spitzer} ch2$_{aperture}$, ch2$_{PRF}$, and \textit{WISE} W2$_{PSF}$ photometry. The solid lines are the medians for each photometry type, while the dashed lines are $\pm$3$\sigma$. The spectral type is on the scale of 0 is M0, 10 is L0, 20 is T0, and 30 is Y0. }
\figsetgrpend

\figsetgrpstart
\figsetgrpnum{8.121}
\figsetgrptitle{Light Curve for WISE0759m4904}
\figsetplot{WISE0759m4904_total_variable_SURF2022.pdf}
\figsetgrpnote{The light curve of WISE0759m4904 with \textit{Spitzer} ch2$_{aperture}$, ch2$_{PRF}$, and \textit{WISE} W2$_{PSF}$ photometry. The solid lines are the medians for each photometry type, while the dashed lines are $\pm$3$\sigma$. The spectral type is on the scale of 0 is M0, 10 is L0, 20 is T0, and 30 is Y0. }
\figsetgrpend

\figsetgrpstart
\figsetgrpnum{8.122}
\figsetgrptitle{Light Curve for WISE0806m0820}
\figsetplot{WISE0806m0820_total_variable_SURF2022.pdf}
\figsetgrpnote{The light curve of WISE0806m0820 with \textit{Spitzer} ch2$_{aperture}$, ch2$_{PRF}$, and \textit{WISE} W2$_{PSF}$ photometry. The solid lines are the medians for each photometry type, while the dashed lines are $\pm$3$\sigma$. The spectral type is on the scale of 0 is M0, 10 is L0, 20 is T0, and 30 is Y0. }
\figsetgrpend

\figsetgrpstart
\figsetgrpnum{8.123}
\figsetgrptitle{Light Curve for WISE0807p4130}
\figsetplot{WISE0807p4130_total_variable_SURF2022.pdf}
\figsetgrpnote{The light curve of WISE0807p4130 with \textit{Spitzer} ch2$_{aperture}$, ch2$_{PRF}$, and \textit{WISE} W2$_{PSF}$ photometry. The solid lines are the medians for each photometry type, while the dashed lines are $\pm$3$\sigma$. The spectral type is on the scale of 0 is M0, 10 is L0, 20 is T0, and 30 is Y0. }
\figsetgrpend

\figsetgrpstart
\figsetgrpnum{8.124}
\figsetgrptitle{Light Curve for WISE0809p4434}
\figsetplot{WISE0809p4434_total_variable_SURF2022.pdf}
\figsetgrpnote{The light curve of WISE0809p4434 with \textit{Spitzer} ch2$_{aperture}$, ch2$_{PRF}$, and \textit{WISE} W2$_{PSF}$ photometry. The solid lines are the medians for each photometry type, while the dashed lines are $\pm$3$\sigma$. The spectral type is on the scale of 0 is M0, 10 is L0, 20 is T0, and 30 is Y0. }
\figsetgrpend

\figsetgrpstart
\figsetgrpnum{8.125}
\figsetgrptitle{Light Curve for WISE0812p4021}
\figsetplot{WISE0812p4021_total_variable_SURF2022.pdf}
\figsetgrpnote{The light curve of WISE0812p4021 with \textit{Spitzer} ch2$_{aperture}$, ch2$_{PRF}$, and \textit{WISE} W2$_{PSF}$ photometry. The solid lines are the medians for each photometry type, while the dashed lines are $\pm$3$\sigma$. The spectral type is on the scale of 0 is M0, 10 is L0, 20 is T0, and 30 is Y0. }
\figsetgrpend

\figsetgrpstart
\figsetgrpnum{8.126}
\figsetgrptitle{Light Curve for WISE0820m6622}
\figsetplot{WISE0820m6622_total_variable_SURF2022.pdf}
\figsetgrpnote{The light curve of WISE0820m6622 with \textit{Spitzer} ch2$_{aperture}$, ch2$_{PRF}$, and \textit{WISE} W2$_{PSF}$ photometry. The solid lines are the medians for each photometry type, while the dashed lines are $\pm$3$\sigma$. The spectral type is on the scale of 0 is M0, 10 is L0, 20 is T0, and 30 is Y0. }
\figsetgrpend

\figsetgrpstart
\figsetgrpnum{8.127}
\figsetgrptitle{Light Curve for WISE0825p2805}
\figsetplot{WISE0825p2805_total_variable_SURF2022.pdf}
\figsetgrpnote{The light curve of WISE0825p2805 with \textit{Spitzer} ch2$_{aperture}$, ch2$_{PRF}$, and \textit{WISE} W2$_{PSF}$ photometry. The solid lines are the medians for each photometry type, while the dashed lines are $\pm$3$\sigma$. The spectral type is on the scale of 0 is M0, 10 is L0, 20 is T0, and 30 is Y0. }
\figsetgrpend

\figsetgrpstart
\figsetgrpnum{8.128}
\figsetgrptitle{Light Curve for WISE0826m1640}
\figsetplot{WISE0826m1640_total_variable_SURF2022.pdf}
\figsetgrpnote{The light curve of WISE0826m1640 with \textit{Spitzer} ch2$_{aperture}$, ch2$_{PRF}$, and \textit{WISE} W2$_{PSF}$ photometry. The solid lines are the medians for each photometry type, while the dashed lines are $\pm$3$\sigma$. The spectral type is on the scale of 0 is M0, 10 is L0, 20 is T0, and 30 is Y0. }
\figsetgrpend

\figsetgrpstart
\figsetgrpnum{8.129}
\figsetgrptitle{Light Curve for WISE0830p2837}
\figsetplot{WISE0830p2837_total_variable_SURF2022.pdf}
\figsetgrpnote{The light curve of WISE0830p2837 with \textit{Spitzer} ch2$_{aperture}$, ch2$_{PRF}$, and \textit{WISE} W2$_{PSF}$ photometry. The solid lines are the medians for each photometry type, while the dashed lines are $\pm$3$\sigma$. The spectral type is on the scale of 0 is M0, 10 is L0, 20 is T0, and 30 is Y0. }
\figsetgrpend

\figsetgrpstart
\figsetgrpnum{8.130}
\figsetgrptitle{Light Curve for WISE0833p0052}
\figsetplot{WISE0833p0052_total_variable_SURF2022.pdf}
\figsetgrpnote{The light curve of WISE0833p0052 with \textit{Spitzer} ch2$_{aperture}$, ch2$_{PRF}$, and \textit{WISE} W2$_{PSF}$ photometry. The solid lines are the medians for each photometry type, while the dashed lines are $\pm$3$\sigma$. The spectral type is on the scale of 0 is M0, 10 is L0, 20 is T0, and 30 is Y0. }
\figsetgrpend

\figsetgrpstart
\figsetgrpnum{8.131}
\figsetgrptitle{Light Curve for WISE0836m1859}
\figsetplot{WISE0836m1859_total_variable_SURF2022.pdf}
\figsetgrpnote{The light curve of WISE0836m1859 with \textit{Spitzer} ch2$_{aperture}$, ch2$_{PRF}$, and \textit{WISE} W2$_{PSF}$ photometry. The solid lines are the medians for each photometry type, while the dashed lines are $\pm$3$\sigma$. The spectral type is on the scale of 0 is M0, 10 is L0, 20 is T0, and 30 is Y0. }
\figsetgrpend

\figsetgrpstart
\figsetgrpnum{8.132}
\figsetgrptitle{Light Curve for WISE0852p4720}
\figsetplot{WISE0852p4720_total_variable_SURF2022.pdf}
\figsetgrpnote{The light curve of WISE0852p4720 with \textit{Spitzer} ch2$_{aperture}$, ch2$_{PRF}$, and \textit{WISE} W2$_{PSF}$ photometry. The solid lines are the medians for each photometry type, while the dashed lines are $\pm$3$\sigma$. The spectral type is on the scale of 0 is M0, 10 is L0, 20 is T0, and 30 is Y0. }
\figsetgrpend

\figsetgrpstart
\figsetgrpnum{8.133}
\figsetgrptitle{Light Curve for WISE0855m0714}
\figsetplot{WISE0855m0714_total_variable_SURF2022.pdf}
\figsetgrpnote{The light curve of WISE0855m0714 with \textit{Spitzer} ch2$_{aperture}$, ch2$_{PRF}$, and \textit{WISE} W2$_{PSF}$ photometry. The solid lines are the medians for each photometry type, while the dashed lines are $\pm$3$\sigma$. The spectral type is on the scale of 0 is M0, 10 is L0, 20 is T0, and 30 is Y0. }
\figsetgrpend

\figsetgrpstart
\figsetgrpnum{8.134}
\figsetgrptitle{Light Curve for WISE0857p5604}
\figsetplot{WISE0857p5604_total_variable_SURF2022.pdf}
\figsetgrpnote{The light curve of WISE0857p5604 with \textit{Spitzer} ch2$_{aperture}$, ch2$_{PRF}$, and \textit{WISE} W2$_{PSF}$ photometry. The solid lines are the medians for each photometry type, while the dashed lines are $\pm$3$\sigma$. The spectral type is on the scale of 0 is M0, 10 is L0, 20 is T0, and 30 is Y0. }
\figsetgrpend

\figsetgrpstart
\figsetgrpnum{8.135}
\figsetgrptitle{Light Curve for WISE0857p5708}
\figsetplot{WISE0857p5708_total_variable_SURF2022.pdf}
\figsetgrpnote{The light curve of WISE0857p5708 with \textit{Spitzer} ch2$_{aperture}$, ch2$_{PRF}$, and \textit{WISE} W2$_{PSF}$ photometry. The solid lines are the medians for each photometry type, while the dashed lines are $\pm$3$\sigma$. The spectral type is on the scale of 0 is M0, 10 is L0, 20 is T0, and 30 is Y0. }
\figsetgrpend

\figsetgrpstart
\figsetgrpnum{8.136}
\figsetgrptitle{Light Curve for WISE0858p3256}
\figsetplot{WISE0858p3256_total_variable_SURF2022.pdf}
\figsetgrpnote{The light curve of WISE0858p3256 with \textit{Spitzer} ch2$_{aperture}$, ch2$_{PRF}$, and \textit{WISE} W2$_{PSF}$ photometry. The solid lines are the medians for each photometry type, while the dashed lines are $\pm$3$\sigma$. The spectral type is on the scale of 0 is M0, 10 is L0, 20 is T0, and 30 is Y0. }
\figsetgrpend

\figsetgrpstart
\figsetgrpnum{8.137}
\figsetgrptitle{Light Curve for WISE0859p5349}
\figsetplot{WISE0859p5349_total_variable_SURF2022.pdf}
\figsetgrpnote{The light curve of WISE0859p5349 with \textit{Spitzer} ch2$_{aperture}$, ch2$_{PRF}$, and \textit{WISE} W2$_{PSF}$ photometry. The solid lines are the medians for each photometry type, while the dashed lines are $\pm$3$\sigma$. The spectral type is on the scale of 0 is M0, 10 is L0, 20 is T0, and 30 is Y0. }
\figsetgrpend

\figsetgrpstart
\figsetgrpnum{8.138}
\figsetgrptitle{Light Curve for WISE0905p5623}
\figsetplot{WISE0905p5623_total_variable_SURF2022.pdf}
\figsetgrpnote{The light curve of WISE0905p5623 with \textit{Spitzer} ch2$_{aperture}$, ch2$_{PRF}$, and \textit{WISE} W2$_{PSF}$ photometry. The solid lines are the medians for each photometry type, while the dashed lines are $\pm$3$\sigma$. The spectral type is on the scale of 0 is M0, 10 is L0, 20 is T0, and 30 is Y0. }
\figsetgrpend

\figsetgrpstart
\figsetgrpnum{8.139}
\figsetgrptitle{Light Curve for WISE0906p4735}
\figsetplot{WISE0906p4735_total_variable_SURF2022.pdf}
\figsetgrpnote{The light curve of WISE0906p4735 with \textit{Spitzer} ch2$_{aperture}$, ch2$_{PRF}$, and \textit{WISE} W2$_{PSF}$ photometry. The solid lines are the medians for each photometry type, while the dashed lines are $\pm$3$\sigma$. The spectral type is on the scale of 0 is M0, 10 is L0, 20 is T0, and 30 is Y0. }
\figsetgrpend

\figsetgrpstart
\figsetgrpnum{8.140}
\figsetgrptitle{Light Curve for WISE0909p6525}
\figsetplot{WISE0909p6525_total_variable_SURF2022.pdf}
\figsetgrpnote{The light curve of WISE0909p6525 with \textit{Spitzer} ch2$_{aperture}$, ch2$_{PRF}$, and \textit{WISE} W2$_{PSF}$ photometry. The solid lines are the medians for each photometry type, while the dashed lines are $\pm$3$\sigma$. The spectral type is on the scale of 0 is M0, 10 is L0, 20 is T0, and 30 is Y0. }
\figsetgrpend

\figsetgrpstart
\figsetgrpnum{8.141}
\figsetgrptitle{Light Curve for WISE0911p2146}
\figsetplot{WISE0911p2146_total_variable_SURF2022.pdf}
\figsetgrpnote{The light curve of WISE0911p2146 with \textit{Spitzer} ch2$_{aperture}$, ch2$_{PRF}$, and \textit{WISE} W2$_{PSF}$ photometry. The solid lines are the medians for each photometry type, while the dashed lines are $\pm$3$\sigma$. The spectral type is on the scale of 0 is M0, 10 is L0, 20 is T0, and 30 is Y0. }
\figsetgrpend

\figsetgrpstart
\figsetgrpnum{8.142}
\figsetgrptitle{Light Curve for WISE0914m3459}
\figsetplot{WISE0914m3459_total_variable_SURF2022.pdf}
\figsetgrpnote{The light curve of WISE0914m3459 with \textit{Spitzer} ch2$_{aperture}$, ch2$_{PRF}$, and \textit{WISE} W2$_{PSF}$ photometry. The solid lines are the medians for each photometry type, while the dashed lines are $\pm$3$\sigma$. The spectral type is on the scale of 0 is M0, 10 is L0, 20 is T0, and 30 is Y0. }
\figsetgrpend

\figsetgrpstart
\figsetgrpnum{8.143}
\figsetgrptitle{Light Curve for WISE0920p4538}
\figsetplot{WISE0920p4538_total_variable_SURF2022.pdf}
\figsetgrpnote{The light curve of WISE0920p4538 with \textit{Spitzer} ch2$_{aperture}$, ch2$_{PRF}$, and \textit{WISE} W2$_{PSF}$ photometry. The solid lines are the medians for each photometry type, while the dashed lines are $\pm$3$\sigma$. The spectral type is on the scale of 0 is M0, 10 is L0, 20 is T0, and 30 is Y0. }
\figsetgrpend

\figsetgrpstart
\figsetgrpnum{8.144}
\figsetgrptitle{Light Curve for WISE0938p0634}
\figsetplot{WISE0938p0634_total_variable_SURF2022.pdf}
\figsetgrpnote{The light curve of WISE0938p0634 with \textit{Spitzer} ch2$_{aperture}$, ch2$_{PRF}$, and \textit{WISE} W2$_{PSF}$ photometry. The solid lines are the medians for each photometry type, while the dashed lines are $\pm$3$\sigma$. The spectral type is on the scale of 0 is M0, 10 is L0, 20 is T0, and 30 is Y0. }
\figsetgrpend

\figsetgrpstart
\figsetgrpnum{8.145}
\figsetgrptitle{Light Curve for WISE0940p5233}
\figsetplot{WISE0940p5233_total_variable_SURF2022.pdf}
\figsetgrpnote{The light curve of WISE0940p5233 with \textit{Spitzer} ch2$_{aperture}$, ch2$_{PRF}$, and \textit{WISE} W2$_{PSF}$ photometry. The solid lines are the medians for each photometry type, while the dashed lines are $\pm$3$\sigma$. The spectral type is on the scale of 0 is M0, 10 is L0, 20 is T0, and 30 is Y0. }
\figsetgrpend

\figsetgrpstart
\figsetgrpnum{8.146}
\figsetgrptitle{Light Curve for WISE0940m2208}
\figsetplot{WISE0940m2208_total_variable_SURF2022.pdf}
\figsetgrpnote{The light curve of WISE0940m2208 with \textit{Spitzer} ch2$_{aperture}$, ch2$_{PRF}$, and \textit{WISE} W2$_{PSF}$ photometry. The solid lines are the medians for each photometry type, while the dashed lines are $\pm$3$\sigma$. The spectral type is on the scale of 0 is M0, 10 is L0, 20 is T0, and 30 is Y0. }
\figsetgrpend

\figsetgrpstart
\figsetgrpnum{8.147}
\figsetgrptitle{Light Curve for WISE0943p3607}
\figsetplot{WISE0943p3607_total_variable_SURF2022.pdf}
\figsetgrpnote{The light curve of WISE0943p3607 with \textit{Spitzer} ch2$_{aperture}$, ch2$_{PRF}$, and \textit{WISE} W2$_{PSF}$ photometry. The solid lines are the medians for each photometry type, while the dashed lines are $\pm$3$\sigma$. The spectral type is on the scale of 0 is M0, 10 is L0, 20 is T0, and 30 is Y0. }
\figsetgrpend

\figsetgrpstart
\figsetgrpnum{8.148}
\figsetgrptitle{Light Curve for WISE0952p1955}
\figsetplot{WISE0952p1955_total_variable_SURF2022.pdf}
\figsetgrpnote{The light curve of WISE0952p1955 with \textit{Spitzer} ch2$_{aperture}$, ch2$_{PRF}$, and \textit{WISE} W2$_{PSF}$ photometry. The solid lines are the medians for each photometry type, while the dashed lines are $\pm$3$\sigma$. The spectral type is on the scale of 0 is M0, 10 is L0, 20 is T0, and 30 is Y0. }
\figsetgrpend

\figsetgrpstart
\figsetgrpnum{8.149}
\figsetgrptitle{Light Curve for WISE0956m1447}
\figsetplot{WISE0956m1447_total_variable_SURF2022.pdf}
\figsetgrpnote{The light curve of WISE0956m1447 with \textit{Spitzer} ch2$_{aperture}$, ch2$_{PRF}$, and \textit{WISE} W2$_{PSF}$ photometry. The solid lines are the medians for each photometry type, while the dashed lines are $\pm$3$\sigma$. The spectral type is on the scale of 0 is M0, 10 is L0, 20 is T0, and 30 is Y0. }
\figsetgrpend

\figsetgrpstart
\figsetgrpnum{8.150}
\figsetgrptitle{Light Curve for WISE1008p2031}
\figsetplot{WISE1008p2031_total_variable_SURF2022.pdf}
\figsetgrpnote{The light curve of WISE1008p2031 with \textit{Spitzer} ch2$_{aperture}$, ch2$_{PRF}$, and \textit{WISE} W2$_{PSF}$ photometry. The solid lines are the medians for each photometry type, while the dashed lines are $\pm$3$\sigma$. The spectral type is on the scale of 0 is M0, 10 is L0, 20 is T0, and 30 is Y0. }
\figsetgrpend

\figsetgrpstart
\figsetgrpnum{8.151}
\figsetgrptitle{Light Curve for WISE1010m0406}
\figsetplot{WISE1010m0406_total_variable_SURF2022.pdf}
\figsetgrpnote{The light curve of WISE1010m0406 with \textit{Spitzer} ch2$_{aperture}$, ch2$_{PRF}$, and \textit{WISE} W2$_{PSF}$ photometry. The solid lines are the medians for each photometry type, while the dashed lines are $\pm$3$\sigma$. The spectral type is on the scale of 0 is M0, 10 is L0, 20 is T0, and 30 is Y0. }
\figsetgrpend

\figsetgrpstart
\figsetgrpnum{8.152}
\figsetgrptitle{Light Curve for WISE1012p1021}
\figsetplot{WISE1012p1021_total_variable_SURF2022.pdf}
\figsetgrpnote{The light curve of WISE1012p1021 with \textit{Spitzer} ch2$_{aperture}$, ch2$_{PRF}$, and \textit{WISE} W2$_{PSF}$ photometry. The solid lines are the medians for each photometry type, while the dashed lines are $\pm$3$\sigma$. The spectral type is on the scale of 0 is M0, 10 is L0, 20 is T0, and 30 is Y0. }
\figsetgrpend

\figsetgrpstart
\figsetgrpnum{8.153}
\figsetgrptitle{Light Curve for WISE1018m2445}
\figsetplot{WISE1018m2445_total_variable_SURF2022.pdf}
\figsetgrpnote{The light curve of WISE1018m2445 with \textit{Spitzer} ch2$_{aperture}$, ch2$_{PRF}$, and \textit{WISE} W2$_{PSF}$ photometry. The solid lines are the medians for each photometry type, while the dashed lines are $\pm$3$\sigma$. The spectral type is on the scale of 0 is M0, 10 is L0, 20 is T0, and 30 is Y0. }
\figsetgrpend

\figsetgrpstart
\figsetgrpnum{8.154}
\figsetgrptitle{Light Curve for WISE1022p1455}
\figsetplot{WISE1022p1455_total_variable_SURF2022.pdf}
\figsetgrpnote{The light curve of WISE1022p1455 with \textit{Spitzer} ch2$_{aperture}$, ch2$_{PRF}$, and \textit{WISE} W2$_{PSF}$ photometry. The solid lines are the medians for each photometry type, while the dashed lines are $\pm$3$\sigma$. The spectral type is on the scale of 0 is M0, 10 is L0, 20 is T0, and 30 is Y0. }
\figsetgrpend

\figsetgrpstart
\figsetgrpnum{8.155}
\figsetgrptitle{Light Curve for WISE1025p0307}
\figsetplot{WISE1025p0307_total_variable_SURF2022.pdf}
\figsetgrpnote{The light curve of WISE1025p0307 with \textit{Spitzer} ch2$_{aperture}$, ch2$_{PRF}$, and \textit{WISE} W2$_{PSF}$ photometry. The solid lines are the medians for each photometry type, while the dashed lines are $\pm$3$\sigma$. The spectral type is on the scale of 0 is M0, 10 is L0, 20 is T0, and 30 is Y0. }
\figsetgrpend

\figsetgrpstart
\figsetgrpnum{8.156}
\figsetgrptitle{Light Curve for WISE1028p5654}
\figsetplot{WISE1028p5654_total_variable_SURF2022.pdf}
\figsetgrpnote{The light curve of WISE1028p5654 with \textit{Spitzer} ch2$_{aperture}$, ch2$_{PRF}$, and \textit{WISE} W2$_{PSF}$ photometry. The solid lines are the medians for each photometry type, while the dashed lines are $\pm$3$\sigma$. The spectral type is on the scale of 0 is M0, 10 is L0, 20 is T0, and 30 is Y0. }
\figsetgrpend

\figsetgrpstart
\figsetgrpnum{8.157}
\figsetgrptitle{Light Curve for WISE1036m3441}
\figsetplot{WISE1036m3441_total_variable_SURF2022.pdf}
\figsetgrpnote{The light curve of WISE1036m3441 with \textit{Spitzer} ch2$_{aperture}$, ch2$_{PRF}$, and \textit{WISE} W2$_{PSF}$ photometry. The solid lines are the medians for each photometry type, while the dashed lines are $\pm$3$\sigma$. The spectral type is on the scale of 0 is M0, 10 is L0, 20 is T0, and 30 is Y0. }
\figsetgrpend

\figsetgrpstart
\figsetgrpnum{8.158}
\figsetgrptitle{Light Curve for WISE1039m1600}
\figsetplot{WISE1039m1600_total_variable_SURF2022.pdf}
\figsetgrpnote{The light curve of WISE1039m1600 with \textit{Spitzer} ch2$_{aperture}$, ch2$_{PRF}$, and \textit{WISE} W2$_{PSF}$ photometry. The solid lines are the medians for each photometry type, while the dashed lines are $\pm$3$\sigma$. The spectral type is on the scale of 0 is M0, 10 is L0, 20 is T0, and 30 is Y0. }
\figsetgrpend

\figsetgrpstart
\figsetgrpnum{8.159}
\figsetgrptitle{Light Curve for WISE1040p4503}
\figsetplot{WISE1040p4503_total_variable_SURF2022.pdf}
\figsetgrpnote{The light curve of WISE1040p4503 with \textit{Spitzer} ch2$_{aperture}$, ch2$_{PRF}$, and \textit{WISE} W2$_{PSF}$ photometry. The solid lines are the medians for each photometry type, while the dashed lines are $\pm$3$\sigma$. The spectral type is on the scale of 0 is M0, 10 is L0, 20 is T0, and 30 is Y0. }
\figsetgrpend

\figsetgrpstart
\figsetgrpnum{8.160}
\figsetgrptitle{Light Curve for WISE1043p2225}
\figsetplot{WISE1043p2225_total_variable_SURF2022.pdf}
\figsetgrpnote{The light curve of WISE1043p2225 with \textit{Spitzer} ch2$_{aperture}$, ch2$_{PRF}$, and \textit{WISE} W2$_{PSF}$ photometry. The solid lines are the medians for each photometry type, while the dashed lines are $\pm$3$\sigma$. The spectral type is on the scale of 0 is M0, 10 is L0, 20 is T0, and 30 is Y0. }
\figsetgrpend

\figsetgrpstart
\figsetgrpnum{8.161}
\figsetgrptitle{Light Curve for WISE1043p1213}
\figsetplot{WISE1043p1213_total_variable_SURF2022.pdf}
\figsetgrpnote{The light curve of WISE1043p1213 with \textit{Spitzer} ch2$_{aperture}$, ch2$_{PRF}$, and \textit{WISE} W2$_{PSF}$ photometry. The solid lines are the medians for each photometry type, while the dashed lines are $\pm$3$\sigma$. The spectral type is on the scale of 0 is M0, 10 is L0, 20 is T0, and 30 is Y0. }
\figsetgrpend

\figsetgrpstart
\figsetgrpnum{8.162}
\figsetgrptitle{Light Curve for WISE1043p1048}
\figsetplot{WISE1043p1048_total_variable_SURF2022.pdf}
\figsetgrpnote{The light curve of WISE1043p1048 with \textit{Spitzer} ch2$_{aperture}$, ch2$_{PRF}$, and \textit{WISE} W2$_{PSF}$ photometry. The solid lines are the medians for each photometry type, while the dashed lines are $\pm$3$\sigma$. The spectral type is on the scale of 0 is M0, 10 is L0, 20 is T0, and 30 is Y0. }
\figsetgrpend

\figsetgrpstart
\figsetgrpnum{8.163}
\figsetgrptitle{Light Curve for WISE1044p0429}
\figsetplot{WISE1044p0429_total_variable_SURF2022.pdf}
\figsetgrpnote{The light curve of WISE1044p0429 with \textit{Spitzer} ch2$_{aperture}$, ch2$_{PRF}$, and \textit{WISE} W2$_{PSF}$ photometry. The solid lines are the medians for each photometry type, while the dashed lines are $\pm$3$\sigma$. The spectral type is on the scale of 0 is M0, 10 is L0, 20 is T0, and 30 is Y0. }
\figsetgrpend

\figsetgrpstart
\figsetgrpnum{8.164}
\figsetgrptitle{Light Curve for WISE1047p5457}
\figsetplot{WISE1047p5457_total_variable_SURF2022.pdf}
\figsetgrpnote{The light curve of WISE1047p5457 with \textit{Spitzer} ch2$_{aperture}$, ch2$_{PRF}$, and \textit{WISE} W2$_{PSF}$ photometry. The solid lines are the medians for each photometry type, while the dashed lines are $\pm$3$\sigma$. The spectral type is on the scale of 0 is M0, 10 is L0, 20 is T0, and 30 is Y0. }
\figsetgrpend

\figsetgrpstart
\figsetgrpnum{8.165}
\figsetgrptitle{Light Curve for WISE1050p5056}
\figsetplot{WISE1050p5056_total_variable_SURF2022.pdf}
\figsetgrpnote{The light curve of WISE1050p5056 with \textit{Spitzer} ch2$_{aperture}$, ch2$_{PRF}$, and \textit{WISE} W2$_{PSF}$ photometry. The solid lines are the medians for each photometry type, while the dashed lines are $\pm$3$\sigma$. The spectral type is on the scale of 0 is M0, 10 is L0, 20 is T0, and 30 is Y0. }
\figsetgrpend

\figsetgrpstart
\figsetgrpnum{8.166}
\figsetgrptitle{Light Curve for WISE1051m2138}
\figsetplot{WISE1051m2138_total_variable_SURF2022.pdf}
\figsetgrpnote{The light curve of WISE1051m2138 with \textit{Spitzer} ch2$_{aperture}$, ch2$_{PRF}$, and \textit{WISE} W2$_{PSF}$ photometry. The solid lines are the medians for each photometry type, while the dashed lines are $\pm$3$\sigma$. The spectral type is on the scale of 0 is M0, 10 is L0, 20 is T0, and 30 is Y0. }
\figsetgrpend

\figsetgrpstart
\figsetgrpnum{8.167}
\figsetgrptitle{Light Curve for WISE1052m1942}
\figsetplot{WISE1052m1942_total_variable_SURF2022.pdf}
\figsetgrpnote{The light curve of WISE1052m1942 with \textit{Spitzer} ch2$_{aperture}$, ch2$_{PRF}$, and \textit{WISE} W2$_{PSF}$ photometry. The solid lines are the medians for each photometry type, while the dashed lines are $\pm$3$\sigma$. The spectral type is on the scale of 0 is M0, 10 is L0, 20 is T0, and 30 is Y0. }
\figsetgrpend

\figsetgrpstart
\figsetgrpnum{8.168}
\figsetgrptitle{Light Curve for WISE1055p5443}
\figsetplot{WISE1055p5443_total_variable_SURF2022.pdf}
\figsetgrpnote{The light curve of WISE1055p5443 with \textit{Spitzer} ch2$_{aperture}$, ch2$_{PRF}$, and \textit{WISE} W2$_{PSF}$ photometry. The solid lines are the medians for each photometry type, while the dashed lines are $\pm$3$\sigma$. The spectral type is on the scale of 0 is M0, 10 is L0, 20 is T0, and 30 is Y0. }
\figsetgrpend

\figsetgrpstart
\figsetgrpnum{8.169}
\figsetgrptitle{Light Curve for WISE1055m1652}
\figsetplot{WISE1055m1652_total_variable_SURF2022.pdf}
\figsetgrpnote{The light curve of WISE1055m1652 with \textit{Spitzer} ch2$_{aperture}$, ch2$_{PRF}$, and \textit{WISE} W2$_{PSF}$ photometry. The solid lines are the medians for each photometry type, while the dashed lines are $\pm$3$\sigma$. The spectral type is on the scale of 0 is M0, 10 is L0, 20 is T0, and 30 is Y0. }
\figsetgrpend

\figsetgrpstart
\figsetgrpnum{8.170}
\figsetgrptitle{Light Curve for WISE1104p1959}
\figsetplot{WISE1104p1959_total_variable_SURF2022.pdf}
\figsetgrpnote{The light curve of WISE1104p1959 with \textit{Spitzer} ch2$_{aperture}$, ch2$_{PRF}$, and \textit{WISE} W2$_{PSF}$ photometry. The solid lines are the medians for each photometry type, while the dashed lines are $\pm$3$\sigma$. The spectral type is on the scale of 0 is M0, 10 is L0, 20 is T0, and 30 is Y0. }
\figsetgrpend

\figsetgrpstart
\figsetgrpnum{8.171}
\figsetgrptitle{Light Curve for WISE1112m3857}
\figsetplot{WISE1112m3857_total_variable_SURF2022.pdf}
\figsetgrpnote{The light curve of WISE1112m3857 with \textit{Spitzer} ch2$_{aperture}$, ch2$_{PRF}$, and \textit{WISE} W2$_{PSF}$ photometry. The solid lines are the medians for each photometry type, while the dashed lines are $\pm$3$\sigma$. The spectral type is on the scale of 0 is M0, 10 is L0, 20 is T0, and 30 is Y0. }
\figsetgrpend

\figsetgrpstart
\figsetgrpnum{8.172}
\figsetgrptitle{Light Curve for WISE1118m0640}
\figsetplot{WISE1118m0640_total_variable_SURF2022.pdf}
\figsetgrpnote{The light curve of WISE1118m0640 with \textit{Spitzer} ch2$_{aperture}$, ch2$_{PRF}$, and \textit{WISE} W2$_{PSF}$ photometry. The solid lines are the medians for each photometry type, while the dashed lines are $\pm$3$\sigma$. The spectral type is on the scale of 0 is M0, 10 is L0, 20 is T0, and 30 is Y0. }
\figsetgrpend

\figsetgrpstart
\figsetgrpnum{8.173}
\figsetgrptitle{Light Curve for WISE1124m0421}
\figsetplot{WISE1124m0421_total_variable_SURF2022.pdf}
\figsetgrpnote{The light curve of WISE1124m0421 with \textit{Spitzer} ch2$_{aperture}$, ch2$_{PRF}$, and \textit{WISE} W2$_{PSF}$ photometry. The solid lines are the medians for each photometry type, while the dashed lines are $\pm$3$\sigma$. The spectral type is on the scale of 0 is M0, 10 is L0, 20 is T0, and 30 is Y0. }
\figsetgrpend

\figsetgrpstart
\figsetgrpnum{8.174}
\figsetgrptitle{Light Curve for WISE1130m1158}
\figsetplot{WISE1130m1158_total_variable_SURF2022.pdf}
\figsetgrpnote{The light curve of WISE1130m1158 with \textit{Spitzer} ch2$_{aperture}$, ch2$_{PRF}$, and \textit{WISE} W2$_{PSF}$ photometry. The solid lines are the medians for each photometry type, while the dashed lines are $\pm$3$\sigma$. The spectral type is on the scale of 0 is M0, 10 is L0, 20 is T0, and 30 is Y0. }
\figsetgrpend

\figsetgrpstart
\figsetgrpnum{8.175}
\figsetgrptitle{Light Curve for WISE1132m3809}
\figsetplot{WISE1132m3809_total_variable_SURF2022.pdf}
\figsetgrpnote{The light curve of WISE1132m3809 with \textit{Spitzer} ch2$_{aperture}$, ch2$_{PRF}$, and \textit{WISE} W2$_{PSF}$ photometry. The solid lines are the medians for each photometry type, while the dashed lines are $\pm$3$\sigma$. The spectral type is on the scale of 0 is M0, 10 is L0, 20 is T0, and 30 is Y0. }
\figsetgrpend

\figsetgrpstart
\figsetgrpnum{8.176}
\figsetgrptitle{Light Curve for WISE1137m5320}
\figsetplot{WISE1137m5320_total_variable_SURF2022.pdf}
\figsetgrpnote{The light curve of WISE1137m5320 with \textit{Spitzer} ch2$_{aperture}$, ch2$_{PRF}$, and \textit{WISE} W2$_{PSF}$ photometry. The solid lines are the medians for each photometry type, while the dashed lines are $\pm$3$\sigma$. The spectral type is on the scale of 0 is M0, 10 is L0, 20 is T0, and 30 is Y0. }
\figsetgrpend

\figsetgrpstart
\figsetgrpnum{8.177}
\figsetgrptitle{Light Curve for WISE1138p7212}
\figsetplot{WISE1138p7212_total_variable_SURF2022.pdf}
\figsetgrpnote{The light curve of WISE1138p7212 with \textit{Spitzer} ch2$_{aperture}$, ch2$_{PRF}$, and \textit{WISE} W2$_{PSF}$ photometry. The solid lines are the medians for each photometry type, while the dashed lines are $\pm$3$\sigma$. The spectral type is on the scale of 0 is M0, 10 is L0, 20 is T0, and 30 is Y0. }
\figsetgrpend

\figsetgrpstart
\figsetgrpnum{8.178}
\figsetgrptitle{Light Curve for WISE1139m3324}
\figsetplot{WISE1139m3324_total_variable_SURF2022.pdf}
\figsetgrpnote{The light curve of WISE1139m3324 with \textit{Spitzer} ch2$_{aperture}$, ch2$_{PRF}$, and \textit{WISE} W2$_{PSF}$ photometry. The solid lines are the medians for each photometry type, while the dashed lines are $\pm$3$\sigma$. The spectral type is on the scale of 0 is M0, 10 is L0, 20 is T0, and 30 is Y0. }
\figsetgrpend

\figsetgrpstart
\figsetgrpnum{8.179}
\figsetgrptitle{Light Curve for WISE1141m2110}
\figsetplot{WISE1141m2110_total_variable_SURF2022.pdf}
\figsetgrpnote{The light curve of WISE1141m2110 with \textit{Spitzer} ch2$_{aperture}$, ch2$_{PRF}$, and \textit{WISE} W2$_{PSF}$ photometry. The solid lines are the medians for each photometry type, while the dashed lines are $\pm$3$\sigma$. The spectral type is on the scale of 0 is M0, 10 is L0, 20 is T0, and 30 is Y0. }
\figsetgrpend

\figsetgrpstart
\figsetgrpnum{8.180}
\figsetgrptitle{Light Curve for WISE1141m3326}
\figsetplot{WISE1141m3326_total_variable_SURF2022.pdf}
\figsetgrpnote{The light curve of WISE1141m3326 with \textit{Spitzer} ch2$_{aperture}$, ch2$_{PRF}$, and \textit{WISE} W2$_{PSF}$ photometry. The solid lines are the medians for each photometry type, while the dashed lines are $\pm$3$\sigma$. The spectral type is on the scale of 0 is M0, 10 is L0, 20 is T0, and 30 is Y0. }
\figsetgrpend

\figsetgrpstart
\figsetgrpnum{8.181}
\figsetgrptitle{Light Curve for WISE1143p4431}
\figsetplot{WISE1143p4431_total_variable_SURF2022.pdf}
\figsetgrpnote{The light curve of WISE1143p4431 with \textit{Spitzer} ch2$_{aperture}$, ch2$_{PRF}$, and \textit{WISE} W2$_{PSF}$ photometry. The solid lines are the medians for each photometry type, while the dashed lines are $\pm$3$\sigma$. The spectral type is on the scale of 0 is M0, 10 is L0, 20 is T0, and 30 is Y0. }
\figsetgrpend

\figsetgrpstart
\figsetgrpnum{8.182}
\figsetgrptitle{Light Curve for WISE1150p6302}
\figsetplot{WISE1150p6302_total_variable_SURF2022.pdf}
\figsetgrpnote{The light curve of WISE1150p6302 with \textit{Spitzer} ch2$_{aperture}$, ch2$_{PRF}$, and \textit{WISE} W2$_{PSF}$ photometry. The solid lines are the medians for each photometry type, while the dashed lines are $\pm$3$\sigma$. The spectral type is on the scale of 0 is M0, 10 is L0, 20 is T0, and 30 is Y0. }
\figsetgrpend

\figsetgrpstart
\figsetgrpnum{8.183}
\figsetgrptitle{Light Curve for WISE1152p1134}
\figsetplot{WISE1152p1134_total_variable_SURF2022.pdf}
\figsetgrpnote{The light curve of WISE1152p1134 with \textit{Spitzer} ch2$_{aperture}$, ch2$_{PRF}$, and \textit{WISE} W2$_{PSF}$ photometry. The solid lines are the medians for each photometry type, while the dashed lines are $\pm$3$\sigma$. The spectral type is on the scale of 0 is M0, 10 is L0, 20 is T0, and 30 is Y0. }
\figsetgrpend

\figsetgrpstart
\figsetgrpnum{8.184}
\figsetgrptitle{Light Curve for WISE1155p0559}
\figsetplot{WISE1155p0559_total_variable_SURF2022.pdf}
\figsetgrpnote{The light curve of WISE1155p0559 with \textit{Spitzer} ch2$_{aperture}$, ch2$_{PRF}$, and \textit{WISE} W2$_{PSF}$ photometry. The solid lines are the medians for each photometry type, while the dashed lines are $\pm$3$\sigma$. The spectral type is on the scale of 0 is M0, 10 is L0, 20 is T0, and 30 is Y0. }
\figsetgrpend

\figsetgrpstart
\figsetgrpnum{8.185}
\figsetgrptitle{Light Curve for WISE1158p0435}
\figsetplot{WISE1158p0435_total_variable_SURF2022.pdf}
\figsetgrpnote{The light curve of WISE1158p0435 with \textit{Spitzer} ch2$_{aperture}$, ch2$_{PRF}$, and \textit{WISE} W2$_{PSF}$ photometry. The solid lines are the medians for each photometry type, while the dashed lines are $\pm$3$\sigma$. The spectral type is on the scale of 0 is M0, 10 is L0, 20 is T0, and 30 is Y0. }
\figsetgrpend

\figsetgrpstart
\figsetgrpnum{8.186}
\figsetgrptitle{Light Curve for WISE1203p0015}
\figsetplot{WISE1203p0015_total_variable_SURF2022.pdf}
\figsetgrpnote{The light curve of WISE1203p0015 with \textit{Spitzer} ch2$_{aperture}$, ch2$_{PRF}$, and \textit{WISE} W2$_{PSF}$ photometry. The solid lines are the medians for each photometry type, while the dashed lines are $\pm$3$\sigma$. The spectral type is on the scale of 0 is M0, 10 is L0, 20 is T0, and 30 is Y0. }
\figsetgrpend

\figsetgrpstart
\figsetgrpnum{8.187}
\figsetgrptitle{Light Curve for WISE1205m1802}
\figsetplot{WISE1205m1802_total_variable_SURF2022.pdf}
\figsetgrpnote{The light curve of WISE1205m1802 with \textit{Spitzer} ch2$_{aperture}$, ch2$_{PRF}$, and \textit{WISE} W2$_{PSF}$ photometry. The solid lines are the medians for each photometry type, while the dashed lines are $\pm$3$\sigma$. The spectral type is on the scale of 0 is M0, 10 is L0, 20 is T0, and 30 is Y0. }
\figsetgrpend

\figsetgrpstart
\figsetgrpnum{8.188}
\figsetgrptitle{Light Curve for WISE1206p8401}
\figsetplot{WISE1206p8401_total_variable_SURF2022.pdf}
\figsetgrpnote{The light curve of WISE1206p8401 with \textit{Spitzer} ch2$_{aperture}$, ch2$_{PRF}$, and \textit{WISE} W2$_{PSF}$ photometry. The solid lines are the medians for each photometry type, while the dashed lines are $\pm$3$\sigma$. The spectral type is on the scale of 0 is M0, 10 is L0, 20 is T0, and 30 is Y0. }
\figsetgrpend

\figsetgrpstart
\figsetgrpnum{8.189}
\figsetgrptitle{Light Curve for WISE1213m0432}
\figsetplot{WISE1213m0432_total_variable_SURF2022.pdf}
\figsetgrpnote{The light curve of WISE1213m0432 with \textit{Spitzer} ch2$_{aperture}$, ch2$_{PRF}$, and \textit{WISE} W2$_{PSF}$ photometry. The solid lines are the medians for each photometry type, while the dashed lines are $\pm$3$\sigma$. The spectral type is on the scale of 0 is M0, 10 is L0, 20 is T0, and 30 is Y0. }
\figsetgrpend

\figsetgrpstart
\figsetgrpnum{8.190}
\figsetgrptitle{Light Curve for WISE1214p6316}
\figsetplot{WISE1214p6316_total_variable_SURF2022.pdf}
\figsetgrpnote{The light curve of WISE1214p6316 with \textit{Spitzer} ch2$_{aperture}$, ch2$_{PRF}$, and \textit{WISE} W2$_{PSF}$ photometry. The solid lines are the medians for each photometry type, while the dashed lines are $\pm$3$\sigma$. The spectral type is on the scale of 0 is M0, 10 is L0, 20 is T0, and 30 is Y0. }
\figsetgrpend

\figsetgrpstart
\figsetgrpnum{8.191}
\figsetgrptitle{Light Curve for WISE1217p1626}
\figsetplot{WISE1217p1626_total_variable_SURF2022.pdf}
\figsetgrpnote{The light curve of WISE1217p1626 with \textit{Spitzer} ch2$_{aperture}$, ch2$_{PRF}$, and \textit{WISE} W2$_{PSF}$ photometry. The solid lines are the medians for each photometry type, while the dashed lines are $\pm$3$\sigma$. The spectral type is on the scale of 0 is M0, 10 is L0, 20 is T0, and 30 is Y0. }
\figsetgrpend

\figsetgrpstart
\figsetgrpnum{8.192}
\figsetgrptitle{Light Curve for WISE1219p3128}
\figsetplot{WISE1219p3128_total_variable_SURF2022.pdf}
\figsetgrpnote{The light curve of WISE1219p3128 with \textit{Spitzer} ch2$_{aperture}$, ch2$_{PRF}$, and \textit{WISE} W2$_{PSF}$ photometry. The solid lines are the medians for each photometry type, while the dashed lines are $\pm$3$\sigma$. The spectral type is on the scale of 0 is M0, 10 is L0, 20 is T0, and 30 is Y0. }
\figsetgrpend

\figsetgrpstart
\figsetgrpnum{8.193}
\figsetgrptitle{Light Curve for WISE1220p5407}
\figsetplot{WISE1220p5407_total_variable_SURF2022.pdf}
\figsetgrpnote{The light curve of WISE1220p5407 with \textit{Spitzer} ch2$_{aperture}$, ch2$_{PRF}$, and \textit{WISE} W2$_{PSF}$ photometry. The solid lines are the medians for each photometry type, while the dashed lines are $\pm$3$\sigma$. The spectral type is on the scale of 0 is M0, 10 is L0, 20 is T0, and 30 is Y0. }
\figsetgrpend

\figsetgrpstart
\figsetgrpnum{8.194}
\figsetgrptitle{Light Curve for WISE1221m3136}
\figsetplot{WISE1221m3136_total_variable_SURF2022.pdf}
\figsetgrpnote{The light curve of WISE1221m3136 with \textit{Spitzer} ch2$_{aperture}$, ch2$_{PRF}$, and \textit{WISE} W2$_{PSF}$ photometry. The solid lines are the medians for each photometry type, while the dashed lines are $\pm$3$\sigma$. The spectral type is on the scale of 0 is M0, 10 is L0, 20 is T0, and 30 is Y0. }
\figsetgrpend

\figsetgrpstart
\figsetgrpnum{8.195}
\figsetgrptitle{Light Curve for WISE1225m1013}
\figsetplot{WISE1225m1013_total_variable_SURF2022.pdf}
\figsetgrpnote{The light curve of WISE1225m1013 with \textit{Spitzer} ch2$_{aperture}$, ch2$_{PRF}$, and \textit{WISE} W2$_{PSF}$ photometry. The solid lines are the medians for each photometry type, while the dashed lines are $\pm$3$\sigma$. The spectral type is on the scale of 0 is M0, 10 is L0, 20 is T0, and 30 is Y0. }
\figsetgrpend

\figsetgrpstart
\figsetgrpnum{8.196}
\figsetgrptitle{Light Curve for WISE1231p0847}
\figsetplot{WISE1231p0847_total_variable_SURF2022.pdf}
\figsetgrpnote{The light curve of WISE1231p0847 with \textit{Spitzer} ch2$_{aperture}$, ch2$_{PRF}$, and \textit{WISE} W2$_{PSF}$ photometry. The solid lines are the medians for each photometry type, while the dashed lines are $\pm$3$\sigma$. The spectral type is on the scale of 0 is M0, 10 is L0, 20 is T0, and 30 is Y0. }
\figsetgrpend

\figsetgrpstart
\figsetgrpnum{8.197}
\figsetgrptitle{Light Curve for WISE1241m8200}
\figsetplot{WISE1241m8200_total_variable_SURF2022.pdf}
\figsetgrpnote{The light curve of WISE1241m8200 with \textit{Spitzer} ch2$_{aperture}$, ch2$_{PRF}$, and \textit{WISE} W2$_{PSF}$ photometry. The solid lines are the medians for each photometry type, while the dashed lines are $\pm$3$\sigma$. The spectral type is on the scale of 0 is M0, 10 is L0, 20 is T0, and 30 is Y0. }
\figsetgrpend

\figsetgrpstart
\figsetgrpnum{8.198}
\figsetgrptitle{Light Curve for WISE1243p8445}
\figsetplot{WISE1243p8445_total_variable_SURF2022.pdf}
\figsetgrpnote{The light curve of WISE1243p8445 with \textit{Spitzer} ch2$_{aperture}$, ch2$_{PRF}$, and \textit{WISE} W2$_{PSF}$ photometry. The solid lines are the medians for each photometry type, while the dashed lines are $\pm$3$\sigma$. The spectral type is on the scale of 0 is M0, 10 is L0, 20 is T0, and 30 is Y0. }
\figsetgrpend

\figsetgrpstart
\figsetgrpnum{8.199}
\figsetgrptitle{Light Curve for WISE1250p2628}
\figsetplot{WISE1250p2628_total_variable_SURF2022.pdf}
\figsetgrpnote{The light curve of WISE1250p2628 with \textit{Spitzer} ch2$_{aperture}$, ch2$_{PRF}$, and \textit{WISE} W2$_{PSF}$ photometry. The solid lines are the medians for each photometry type, while the dashed lines are $\pm$3$\sigma$. The spectral type is on the scale of 0 is M0, 10 is L0, 20 is T0, and 30 is Y0. }
\figsetgrpend

\figsetgrpstart
\figsetgrpnum{8.200}
\figsetgrptitle{Light Curve for WISE1254m0728}
\figsetplot{WISE1254m0728_total_variable_SURF2022.pdf}
\figsetgrpnote{The light curve of WISE1254m0728 with \textit{Spitzer} ch2$_{aperture}$, ch2$_{PRF}$, and \textit{WISE} W2$_{PSF}$ photometry. The solid lines are the medians for each photometry type, while the dashed lines are $\pm$3$\sigma$. The spectral type is on the scale of 0 is M0, 10 is L0, 20 is T0, and 30 is Y0. }
\figsetgrpend

\figsetgrpstart
\figsetgrpnum{8.201}
\figsetgrptitle{Light Curve for WISE1257p4008}
\figsetplot{WISE1257p4008_total_variable_SURF2022.pdf}
\figsetgrpnote{The light curve of WISE1257p4008 with \textit{Spitzer} ch2$_{aperture}$, ch2$_{PRF}$, and \textit{WISE} W2$_{PSF}$ photometry. The solid lines are the medians for each photometry type, while the dashed lines are $\pm$3$\sigma$. The spectral type is on the scale of 0 is M0, 10 is L0, 20 is T0, and 30 is Y0. }
\figsetgrpend

\figsetgrpstart
\figsetgrpnum{8.202}
\figsetgrptitle{Light Curve for WISE1257p7153}
\figsetplot{WISE1257p7153_total_variable_SURF2022.pdf}
\figsetgrpnote{The light curve of WISE1257p7153 with \textit{Spitzer} ch2$_{aperture}$, ch2$_{PRF}$, and \textit{WISE} W2$_{PSF}$ photometry. The solid lines are the medians for each photometry type, while the dashed lines are $\pm$3$\sigma$. The spectral type is on the scale of 0 is M0, 10 is L0, 20 is T0, and 30 is Y0. }
\figsetgrpend

\figsetgrpstart
\figsetgrpnum{8.203}
\figsetgrptitle{Light Curve for WISE1258m4412}
\figsetplot{WISE1258m4412_total_variable_SURF2022.pdf}
\figsetgrpnote{The light curve of WISE1258m4412 with \textit{Spitzer} ch2$_{aperture}$, ch2$_{PRF}$, and \textit{WISE} W2$_{PSF}$ photometry. The solid lines are the medians for each photometry type, while the dashed lines are $\pm$3$\sigma$. The spectral type is on the scale of 0 is M0, 10 is L0, 20 is T0, and 30 is Y0. }
\figsetgrpend

\figsetgrpstart
\figsetgrpnum{8.204}
\figsetgrptitle{Light Curve for WISE1301m0302}
\figsetplot{WISE1301m0302_total_variable_SURF2022.pdf}
\figsetgrpnote{The light curve of WISE1301m0302 with \textit{Spitzer} ch2$_{aperture}$, ch2$_{PRF}$, and \textit{WISE} W2$_{PSF}$ photometry. The solid lines are the medians for each photometry type, while the dashed lines are $\pm$3$\sigma$. The spectral type is on the scale of 0 is M0, 10 is L0, 20 is T0, and 30 is Y0. }
\figsetgrpend

\figsetgrpstart
\figsetgrpnum{8.205}
\figsetgrptitle{Light Curve for WISE1318m1758}
\figsetplot{WISE1318m1758_total_variable_SURF2022.pdf}
\figsetgrpnote{The light curve of WISE1318m1758 with \textit{Spitzer} ch2$_{aperture}$, ch2$_{PRF}$, and \textit{WISE} W2$_{PSF}$ photometry. The solid lines are the medians for each photometry type, while the dashed lines are $\pm$3$\sigma$. The spectral type is on the scale of 0 is M0, 10 is L0, 20 is T0, and 30 is Y0. }
\figsetgrpend

\figsetgrpstart
\figsetgrpnum{8.206}
\figsetgrptitle{Light Curve for WISE1319p1209}
\figsetplot{WISE1319p1209_total_variable_SURF2022.pdf}
\figsetgrpnote{The light curve of WISE1319p1209 with \textit{Spitzer} ch2$_{aperture}$, ch2$_{PRF}$, and \textit{WISE} W2$_{PSF}$ photometry. The solid lines are the medians for each photometry type, while the dashed lines are $\pm$3$\sigma$. The spectral type is on the scale of 0 is M0, 10 is L0, 20 is T0, and 30 is Y0. }
\figsetgrpend

\figsetgrpstart
\figsetgrpnum{8.207}
\figsetgrptitle{Light Curve for WISE1324p1906}
\figsetplot{WISE1324p1906_total_variable_SURF2022.pdf}
\figsetgrpnote{The light curve of WISE1324p1906 with \textit{Spitzer} ch2$_{aperture}$, ch2$_{PRF}$, and \textit{WISE} W2$_{PSF}$ photometry. The solid lines are the medians for each photometry type, while the dashed lines are $\pm$3$\sigma$. The spectral type is on the scale of 0 is M0, 10 is L0, 20 is T0, and 30 is Y0. }
\figsetgrpend

\figsetgrpstart
\figsetgrpnum{8.208}
\figsetgrptitle{Light Curve for WISE1324p6358}
\figsetplot{WISE1324p6358_total_variable_SURF2022.pdf}
\figsetgrpnote{The light curve of WISE1324p6358 with \textit{Spitzer} ch2$_{aperture}$, ch2$_{PRF}$, and \textit{WISE} W2$_{PSF}$ photometry. The solid lines are the medians for each photometry type, while the dashed lines are $\pm$3$\sigma$. The spectral type is on the scale of 0 is M0, 10 is L0, 20 is T0, and 30 is Y0. }
\figsetgrpend

\figsetgrpstart
\figsetgrpnum{8.209}
\figsetgrptitle{Light Curve for WISE1326m0038}
\figsetplot{WISE1326m0038_total_variable_SURF2022.pdf}
\figsetgrpnote{The light curve of WISE1326m0038 with \textit{Spitzer} ch2$_{aperture}$, ch2$_{PRF}$, and \textit{WISE} W2$_{PSF}$ photometry. The solid lines are the medians for each photometry type, while the dashed lines are $\pm$3$\sigma$. The spectral type is on the scale of 0 is M0, 10 is L0, 20 is T0, and 30 is Y0. }
\figsetgrpend

\figsetgrpstart
\figsetgrpnum{8.210}
\figsetgrptitle{Light Curve for WISE1333m1607}
\figsetplot{WISE1333m1607_total_variable_SURF2022.pdf}
\figsetgrpnote{The light curve of WISE1333m1607 with \textit{Spitzer} ch2$_{aperture}$, ch2$_{PRF}$, and \textit{WISE} W2$_{PSF}$ photometry. The solid lines are the medians for each photometry type, while the dashed lines are $\pm$3$\sigma$. The spectral type is on the scale of 0 is M0, 10 is L0, 20 is T0, and 30 is Y0. }
\figsetgrpend

\figsetgrpstart
\figsetgrpnum{8.211}
\figsetgrptitle{Light Curve for WISE1358p3747}
\figsetplot{WISE1358p3747_total_variable_SURF2022.pdf}
\figsetgrpnote{The light curve of WISE1358p3747 with \textit{Spitzer} ch2$_{aperture}$, ch2$_{PRF}$, and \textit{WISE} W2$_{PSF}$ photometry. The solid lines are the medians for each photometry type, while the dashed lines are $\pm$3$\sigma$. The spectral type is on the scale of 0 is M0, 10 is L0, 20 is T0, and 30 is Y0. }
\figsetgrpend

\figsetgrpstart
\figsetgrpnum{8.212}
\figsetgrptitle{Light Curve for WISE1359m4352}
\figsetplot{WISE1359m4352_total_variable_SURF2022.pdf}
\figsetgrpnote{The light curve of WISE1359m4352 with \textit{Spitzer} ch2$_{aperture}$, ch2$_{PRF}$, and \textit{WISE} W2$_{PSF}$ photometry. The solid lines are the medians for each photometry type, while the dashed lines are $\pm$3$\sigma$. The spectral type is on the scale of 0 is M0, 10 is L0, 20 is T0, and 30 is Y0. }
\figsetgrpend

\figsetgrpstart
\figsetgrpnum{8.213}
\figsetgrptitle{Light Curve for WISE1400m3850}
\figsetplot{WISE1400m3850_total_variable_SURF2022.pdf}
\figsetgrpnote{The light curve of WISE1400m3850 with \textit{Spitzer} ch2$_{aperture}$, ch2$_{PRF}$, and \textit{WISE} W2$_{PSF}$ photometry. The solid lines are the medians for each photometry type, while the dashed lines are $\pm$3$\sigma$. The spectral type is on the scale of 0 is M0, 10 is L0, 20 is T0, and 30 is Y0. }
\figsetgrpend

\figsetgrpstart
\figsetgrpnum{8.214}
\figsetgrptitle{Light Curve for WISE1405p5534}
\figsetplot{WISE1405p5534_total_variable_SURF2022.pdf}
\figsetgrpnote{The light curve of WISE1405p5534 with \textit{Spitzer} ch2$_{aperture}$, ch2$_{PRF}$, and \textit{WISE} W2$_{PSF}$ photometry. The solid lines are the medians for each photometry type, while the dashed lines are $\pm$3$\sigma$. The spectral type is on the scale of 0 is M0, 10 is L0, 20 is T0, and 30 is Y0. }
\figsetgrpend

\figsetgrpstart
\figsetgrpnum{8.215}
\figsetgrptitle{Light Curve for WISE1407p1241}
\figsetplot{WISE1407p1241_total_variable_SURF2022.pdf}
\figsetgrpnote{The light curve of WISE1407p1241 with \textit{Spitzer} ch2$_{aperture}$, ch2$_{PRF}$, and \textit{WISE} W2$_{PSF}$ photometry. The solid lines are the medians for each photometry type, while the dashed lines are $\pm$3$\sigma$. The spectral type is on the scale of 0 is M0, 10 is L0, 20 is T0, and 30 is Y0. }
\figsetgrpend

\figsetgrpstart
\figsetgrpnum{8.216}
\figsetgrptitle{Light Curve for WISE1411m4811}
\figsetplot{WISE1411m4811_total_variable_SURF2022.pdf}
\figsetgrpnote{The light curve of WISE1411m4811 with \textit{Spitzer} ch2$_{aperture}$, ch2$_{PRF}$, and \textit{WISE} W2$_{PSF}$ photometry. The solid lines are the medians for each photometry type, while the dashed lines are $\pm$3$\sigma$. The spectral type is on the scale of 0 is M0, 10 is L0, 20 is T0, and 30 is Y0. }
\figsetgrpend

\figsetgrpstart
\figsetgrpnum{8.217}
\figsetgrptitle{Light Curve for WISE1433m0837}
\figsetplot{WISE1433m0837_total_variable_SURF2022.pdf}
\figsetgrpnote{The light curve of WISE1433m0837 with \textit{Spitzer} ch2$_{aperture}$, ch2$_{PRF}$, and \textit{WISE} W2$_{PSF}$ photometry. The solid lines are the medians for each photometry type, while the dashed lines are $\pm$3$\sigma$. The spectral type is on the scale of 0 is M0, 10 is L0, 20 is T0, and 30 is Y0. }
\figsetgrpend

\figsetgrpstart
\figsetgrpnum{8.218}
\figsetgrptitle{Light Curve for WISE1436m1814}
\figsetplot{WISE1436m1814_total_variable_SURF2022.pdf}
\figsetgrpnote{The light curve of WISE1436m1814 with \textit{Spitzer} ch2$_{aperture}$, ch2$_{PRF}$, and \textit{WISE} W2$_{PSF}$ photometry. The solid lines are the medians for each photometry type, while the dashed lines are $\pm$3$\sigma$. The spectral type is on the scale of 0 is M0, 10 is L0, 20 is T0, and 30 is Y0. }
\figsetgrpend

\figsetgrpstart
\figsetgrpnum{8.219}
\figsetgrptitle{Light Curve for WISE1446m2317}
\figsetplot{WISE1446m2317_total_variable_SURF2022.pdf}
\figsetgrpnote{The light curve of WISE1446m2317 with \textit{Spitzer} ch2$_{aperture}$, ch2$_{PRF}$, and \textit{WISE} W2$_{PSF}$ photometry. The solid lines are the medians for each photometry type, while the dashed lines are $\pm$3$\sigma$. The spectral type is on the scale of 0 is M0, 10 is L0, 20 is T0, and 30 is Y0. }
\figsetgrpend

\figsetgrpstart
\figsetgrpnum{8.220}
\figsetgrptitle{Light Curve for WISE1448m2534}
\figsetplot{WISE1448m2534_total_variable_SURF2022.pdf}
\figsetgrpnote{The light curve of WISE1448m2534 with \textit{Spitzer} ch2$_{aperture}$, ch2$_{PRF}$, and \textit{WISE} W2$_{PSF}$ photometry. The solid lines are the medians for each photometry type, while the dashed lines are $\pm$3$\sigma$. The spectral type is on the scale of 0 is M0, 10 is L0, 20 is T0, and 30 is Y0. }
\figsetgrpend

\figsetgrpstart
\figsetgrpnum{8.221}
\figsetgrptitle{Light Curve for WISE1457p4724}
\figsetplot{WISE1457p4724_total_variable_SURF2022.pdf}
\figsetgrpnote{The light curve of WISE1457p4724 with \textit{Spitzer} ch2$_{aperture}$, ch2$_{PRF}$, and \textit{WISE} W2$_{PSF}$ photometry. The solid lines are the medians for each photometry type, while the dashed lines are $\pm$3$\sigma$. The spectral type is on the scale of 0 is M0, 10 is L0, 20 is T0, and 30 is Y0. }
\figsetgrpend

\figsetgrpstart
\figsetgrpnum{8.222}
\figsetgrptitle{Light Curve for WISE1458p1734}
\figsetplot{WISE1458p1734_total_variable_SURF2022.pdf}
\figsetgrpnote{The light curve of WISE1458p1734 with \textit{Spitzer} ch2$_{aperture}$, ch2$_{PRF}$, and \textit{WISE} W2$_{PSF}$ photometry. The solid lines are the medians for each photometry type, while the dashed lines are $\pm$3$\sigma$. The spectral type is on the scale of 0 is M0, 10 is L0, 20 is T0, and 30 is Y0. }
\figsetgrpend

\figsetgrpstart
\figsetgrpnum{8.223}
\figsetgrptitle{Light Curve for WISE1501m4004}
\figsetplot{WISE1501m4004_total_variable_SURF2022.pdf}
\figsetgrpnote{The light curve of WISE1501m4004 with \textit{Spitzer} ch2$_{aperture}$, ch2$_{PRF}$, and \textit{WISE} W2$_{PSF}$ photometry. The solid lines are the medians for each photometry type, while the dashed lines are $\pm$3$\sigma$. The spectral type is on the scale of 0 is M0, 10 is L0, 20 is T0, and 30 is Y0. }
\figsetgrpend

\figsetgrpstart
\figsetgrpnum{8.224}
\figsetgrptitle{Light Curve for WISE1505m2853}
\figsetplot{WISE1505m2853_total_variable_SURF2022.pdf}
\figsetgrpnote{The light curve of WISE1505m2853 with \textit{Spitzer} ch2$_{aperture}$, ch2$_{PRF}$, and \textit{WISE} W2$_{PSF}$ photometry. The solid lines are the medians for each photometry type, while the dashed lines are $\pm$3$\sigma$. The spectral type is on the scale of 0 is M0, 10 is L0, 20 is T0, and 30 is Y0. }
\figsetgrpend

\figsetgrpstart
\figsetgrpnum{8.225}
\figsetgrptitle{Light Curve for WISE1517p0529}
\figsetplot{WISE1517p0529_total_variable_SURF2022.pdf}
\figsetgrpnote{The light curve of WISE1517p0529 with \textit{Spitzer} ch2$_{aperture}$, ch2$_{PRF}$, and \textit{WISE} W2$_{PSF}$ photometry. The solid lines are the medians for each photometry type, while the dashed lines are $\pm$3$\sigma$. The spectral type is on the scale of 0 is M0, 10 is L0, 20 is T0, and 30 is Y0. }
\figsetgrpend

\figsetgrpstart
\figsetgrpnum{8.226}
\figsetgrptitle{Light Curve for WISE1519p7009}
\figsetplot{WISE1519p7009_total_variable_SURF2022.pdf}
\figsetgrpnote{The light curve of WISE1519p7009 with \textit{Spitzer} ch2$_{aperture}$, ch2$_{PRF}$, and \textit{WISE} W2$_{PSF}$ photometry. The solid lines are the medians for each photometry type, while the dashed lines are $\pm$3$\sigma$. The spectral type is on the scale of 0 is M0, 10 is L0, 20 is T0, and 30 is Y0. }
\figsetgrpend

\figsetgrpstart
\figsetgrpnum{8.227}
\figsetgrptitle{Light Curve for WISE1520p3546}
\figsetplot{WISE1520p3546_total_variable_SURF2022.pdf}
\figsetgrpnote{The light curve of WISE1520p3546 with \textit{Spitzer} ch2$_{aperture}$, ch2$_{PRF}$, and \textit{WISE} W2$_{PSF}$ photometry. The solid lines are the medians for each photometry type, while the dashed lines are $\pm$3$\sigma$. The spectral type is on the scale of 0 is M0, 10 is L0, 20 is T0, and 30 is Y0. }
\figsetgrpend

\figsetgrpstart
\figsetgrpnum{8.228}
\figsetgrptitle{Light Curve for WISE1523p3125}
\figsetplot{WISE1523p3125_total_variable_SURF2022.pdf}
\figsetgrpnote{The light curve of WISE1523p3125 with \textit{Spitzer} ch2$_{aperture}$, ch2$_{PRF}$, and \textit{WISE} W2$_{PSF}$ photometry. The solid lines are the medians for each photometry type, while the dashed lines are $\pm$3$\sigma$. The spectral type is on the scale of 0 is M0, 10 is L0, 20 is T0, and 30 is Y0. }
\figsetgrpend

\figsetgrpstart
\figsetgrpnum{8.229}
\figsetgrptitle{Light Curve for WISE1526p2043}
\figsetplot{WISE1526p2043_total_variable_SURF2022.pdf}
\figsetgrpnote{The light curve of WISE1526p2043 with \textit{Spitzer} ch2$_{aperture}$, ch2$_{PRF}$, and \textit{WISE} W2$_{PSF}$ photometry. The solid lines are the medians for each photometry type, while the dashed lines are $\pm$3$\sigma$. The spectral type is on the scale of 0 is M0, 10 is L0, 20 is T0, and 30 is Y0. }
\figsetgrpend

\figsetgrpstart
\figsetgrpnum{8.230}
\figsetgrptitle{Light Curve for WISE1533p1753}
\figsetplot{WISE1533p1753_total_variable_SURF2022.pdf}
\figsetgrpnote{The light curve of WISE1533p1753 with \textit{Spitzer} ch2$_{aperture}$, ch2$_{PRF}$, and \textit{WISE} W2$_{PSF}$ photometry. The solid lines are the medians for each photometry type, while the dashed lines are $\pm$3$\sigma$. The spectral type is on the scale of 0 is M0, 10 is L0, 20 is T0, and 30 is Y0. }
\figsetgrpend

\figsetgrpstart
\figsetgrpnum{8.231}
\figsetgrptitle{Light Curve for WISE1534m1043}
\figsetplot{WISE1534m1043_total_variable_SURF2022.pdf}
\figsetgrpnote{The light curve of WISE1534m1043 with \textit{Spitzer} ch2$_{aperture}$, ch2$_{PRF}$, and \textit{WISE} W2$_{PSF}$ photometry. The solid lines are the medians for each photometry type, while the dashed lines are $\pm$3$\sigma$. The spectral type is on the scale of 0 is M0, 10 is L0, 20 is T0, and 30 is Y0. }
\figsetgrpend

\figsetgrpstart
\figsetgrpnum{8.232}
\figsetgrptitle{Light Curve for WISE1534p1219}
\figsetplot{WISE1534p1219_total_variable_SURF2022.pdf}
\figsetgrpnote{The light curve of WISE1534p1219 with \textit{Spitzer} ch2$_{aperture}$, ch2$_{PRF}$, and \textit{WISE} W2$_{PSF}$ photometry. The solid lines are the medians for each photometry type, while the dashed lines are $\pm$3$\sigma$. The spectral type is on the scale of 0 is M0, 10 is L0, 20 is T0, and 30 is Y0. }
\figsetgrpend

\figsetgrpstart
\figsetgrpnum{8.233}
\figsetgrptitle{Light Curve for WISE1538p4826}
\figsetplot{WISE1538p4826_total_variable_SURF2022.pdf}
\figsetgrpnote{The light curve of WISE1538p4826 with \textit{Spitzer} ch2$_{aperture}$, ch2$_{PRF}$, and \textit{WISE} W2$_{PSF}$ photometry. The solid lines are the medians for each photometry type, while the dashed lines are $\pm$3$\sigma$. The spectral type is on the scale of 0 is M0, 10 is L0, 20 is T0, and 30 is Y0. }
\figsetgrpend

\figsetgrpstart
\figsetgrpnum{8.234}
\figsetgrptitle{Light Curve for WISE1541m2250}
\figsetplot{WISE1541m2250_total_variable_SURF2022.pdf}
\figsetgrpnote{The light curve of WISE1541m2250 with \textit{Spitzer} ch2$_{aperture}$, ch2$_{PRF}$, and \textit{WISE} W2$_{PSF}$ photometry. The solid lines are the medians for each photometry type, while the dashed lines are $\pm$3$\sigma$. The spectral type is on the scale of 0 is M0, 10 is L0, 20 is T0, and 30 is Y0. }
\figsetgrpend

\figsetgrpstart
\figsetgrpnum{8.235}
\figsetgrptitle{Light Curve for WISE1542p2230}
\figsetplot{WISE1542p2230_total_variable_SURF2022.pdf}
\figsetgrpnote{The light curve of WISE1542p2230 with \textit{Spitzer} ch2$_{aperture}$, ch2$_{PRF}$, and \textit{WISE} W2$_{PSF}$ photometry. The solid lines are the medians for each photometry type, while the dashed lines are $\pm$3$\sigma$. The spectral type is on the scale of 0 is M0, 10 is L0, 20 is T0, and 30 is Y0. }
\figsetgrpend

\figsetgrpstart
\figsetgrpnum{8.236}
\figsetgrptitle{Light Curve for WISE1544p5842}
\figsetplot{WISE1544p5842_total_variable_SURF2022.pdf}
\figsetgrpnote{The light curve of WISE1544p5842 with \textit{Spitzer} ch2$_{aperture}$, ch2$_{PRF}$, and \textit{WISE} W2$_{PSF}$ photometry. The solid lines are the medians for each photometry type, while the dashed lines are $\pm$3$\sigma$. The spectral type is on the scale of 0 is M0, 10 is L0, 20 is T0, and 30 is Y0. }
\figsetgrpend

\figsetgrpstart
\figsetgrpnum{8.237}
\figsetgrptitle{Light Curve for WISE1546p4932}
\figsetplot{WISE1546p4932_total_variable_SURF2022.pdf}
\figsetgrpnote{The light curve of WISE1546p4932 with \textit{Spitzer} ch2$_{aperture}$, ch2$_{PRF}$, and \textit{WISE} W2$_{PSF}$ photometry. The solid lines are the medians for each photometry type, while the dashed lines are $\pm$3$\sigma$. The spectral type is on the scale of 0 is M0, 10 is L0, 20 is T0, and 30 is Y0. }
\figsetgrpend

\figsetgrpstart
\figsetgrpnum{8.238}
\figsetgrptitle{Light Curve for WISE1608m2442}
\figsetplot{WISE1608m2442_total_variable_SURF2022.pdf}
\figsetgrpnote{The light curve of WISE1608m2442 with \textit{Spitzer} ch2$_{aperture}$, ch2$_{PRF}$, and \textit{WISE} W2$_{PSF}$ photometry. The solid lines are the medians for each photometry type, while the dashed lines are $\pm$3$\sigma$. The spectral type is on the scale of 0 is M0, 10 is L0, 20 is T0, and 30 is Y0. }
\figsetgrpend

\figsetgrpstart
\figsetgrpnum{8.239}
\figsetgrptitle{Light Curve for WISE1612m3420}
\figsetplot{WISE1612m3420_total_variable_SURF2022.pdf}
\figsetgrpnote{The light curve of WISE1612m3420 with \textit{Spitzer} ch2$_{aperture}$, ch2$_{PRF}$, and \textit{WISE} W2$_{PSF}$ photometry. The solid lines are the medians for each photometry type, while the dashed lines are $\pm$3$\sigma$. The spectral type is on the scale of 0 is M0, 10 is L0, 20 is T0, and 30 is Y0. }
\figsetgrpend

\figsetgrpstart
\figsetgrpnum{8.240}
\figsetgrptitle{Light Curve for WISE1614p1739}
\figsetplot{WISE1614p1739_total_variable_SURF2022.pdf}
\figsetgrpnote{The light curve of WISE1614p1739 with \textit{Spitzer} ch2$_{aperture}$, ch2$_{PRF}$, and \textit{WISE} W2$_{PSF}$ photometry. The solid lines are the medians for each photometry type, while the dashed lines are $\pm$3$\sigma$. The spectral type is on the scale of 0 is M0, 10 is L0, 20 is T0, and 30 is Y0. }
\figsetgrpend

\figsetgrpstart
\figsetgrpnum{8.241}
\figsetgrptitle{Light Curve for WISE1615p1340}
\figsetplot{WISE1615p1340_total_variable_SURF2022.pdf}
\figsetgrpnote{The light curve of WISE1615p1340 with \textit{Spitzer} ch2$_{aperture}$, ch2$_{PRF}$, and \textit{WISE} W2$_{PSF}$ photometry. The solid lines are the medians for each photometry type, while the dashed lines are $\pm$3$\sigma$. The spectral type is on the scale of 0 is M0, 10 is L0, 20 is T0, and 30 is Y0. }
\figsetgrpend

\figsetgrpstart
\figsetgrpnum{8.242}
\figsetgrptitle{Light Curve for WISE1619p0313}
\figsetplot{WISE1619p0313_total_variable_SURF2022.pdf}
\figsetgrpnote{The light curve of WISE1619p0313 with \textit{Spitzer} ch2$_{aperture}$, ch2$_{PRF}$, and \textit{WISE} W2$_{PSF}$ photometry. The solid lines are the medians for each photometry type, while the dashed lines are $\pm$3$\sigma$. The spectral type is on the scale of 0 is M0, 10 is L0, 20 is T0, and 30 is Y0. }
\figsetgrpend

\figsetgrpstart
\figsetgrpnum{8.243}
\figsetgrptitle{Light Curve for WISE1619p1347}
\figsetplot{WISE1619p1347_total_variable_SURF2022.pdf}
\figsetgrpnote{The light curve of WISE1619p1347 with \textit{Spitzer} ch2$_{aperture}$, ch2$_{PRF}$, and \textit{WISE} W2$_{PSF}$ photometry. The solid lines are the medians for each photometry type, while the dashed lines are $\pm$3$\sigma$. The spectral type is on the scale of 0 is M0, 10 is L0, 20 is T0, and 30 is Y0. }
\figsetgrpend

\figsetgrpstart
\figsetgrpnum{8.244}
\figsetgrptitle{Light Curve for WISE1622m0959}
\figsetplot{WISE1622m0959_total_variable_SURF2022.pdf}
\figsetgrpnote{The light curve of WISE1622m0959 with \textit{Spitzer} ch2$_{aperture}$, ch2$_{PRF}$, and \textit{WISE} W2$_{PSF}$ photometry. The solid lines are the medians for each photometry type, while the dashed lines are $\pm$3$\sigma$. The spectral type is on the scale of 0 is M0, 10 is L0, 20 is T0, and 30 is Y0. }
\figsetgrpend

\figsetgrpstart
\figsetgrpnum{8.245}
\figsetgrptitle{Light Curve for WISE1623m7402}
\figsetplot{WISE1623m7402_total_variable_SURF2022.pdf}
\figsetgrpnote{The light curve of WISE1623m7402 with \textit{Spitzer} ch2$_{aperture}$, ch2$_{PRF}$, and \textit{WISE} W2$_{PSF}$ photometry. The solid lines are the medians for each photometry type, while the dashed lines are $\pm$3$\sigma$. The spectral type is on the scale of 0 is M0, 10 is L0, 20 is T0, and 30 is Y0. }
\figsetgrpend

\figsetgrpstart
\figsetgrpnum{8.246}
\figsetgrptitle{Light Curve for WISE1629p0335}
\figsetplot{WISE1629p0335_total_variable_SURF2022.pdf}
\figsetgrpnote{The light curve of WISE1629p0335 with \textit{Spitzer} ch2$_{aperture}$, ch2$_{PRF}$, and \textit{WISE} W2$_{PSF}$ photometry. The solid lines are the medians for each photometry type, while the dashed lines are $\pm$3$\sigma$. The spectral type is on the scale of 0 is M0, 10 is L0, 20 is T0, and 30 is Y0. }
\figsetgrpend

\figsetgrpstart
\figsetgrpnum{8.247}
\figsetgrptitle{Light Curve for WISE1630p0818}
\figsetplot{WISE1630p0818_total_variable_SURF2022.pdf}
\figsetgrpnote{The light curve of WISE1630p0818 with \textit{Spitzer} ch2$_{aperture}$, ch2$_{PRF}$, and \textit{WISE} W2$_{PSF}$ photometry. The solid lines are the medians for each photometry type, while the dashed lines are $\pm$3$\sigma$. The spectral type is on the scale of 0 is M0, 10 is L0, 20 is T0, and 30 is Y0. }
\figsetgrpend

\figsetgrpstart
\figsetgrpnum{8.248}
\figsetgrptitle{Light Curve for WISE1639p1840}
\figsetplot{WISE1639p1840_total_variable_SURF2022.pdf}
\figsetgrpnote{The light curve of WISE1639p1840 with \textit{Spitzer} ch2$_{aperture}$, ch2$_{PRF}$, and \textit{WISE} W2$_{PSF}$ photometry. The solid lines are the medians for each photometry type, while the dashed lines are $\pm$3$\sigma$. The spectral type is on the scale of 0 is M0, 10 is L0, 20 is T0, and 30 is Y0. }
\figsetgrpend

\figsetgrpstart
\figsetgrpnum{8.249}
\figsetgrptitle{Light Curve for WISE1639m6847}
\figsetplot{WISE1639m6847_total_variable_SURF2022.pdf}
\figsetgrpnote{The light curve of WISE1639m6847 with \textit{Spitzer} ch2$_{aperture}$, ch2$_{PRF}$, and \textit{WISE} W2$_{PSF}$ photometry. The solid lines are the medians for each photometry type, while the dashed lines are $\pm$3$\sigma$. The spectral type is on the scale of 0 is M0, 10 is L0, 20 is T0, and 30 is Y0. }
\figsetgrpend

\figsetgrpstart
\figsetgrpnum{8.250}
\figsetgrptitle{Light Curve for WISE1653p4444}
\figsetplot{WISE1653p4444_total_variable_SURF2022.pdf}
\figsetgrpnote{The light curve of WISE1653p4444 with \textit{Spitzer} ch2$_{aperture}$, ch2$_{PRF}$, and \textit{WISE} W2$_{PSF}$ photometry. The solid lines are the medians for each photometry type, while the dashed lines are $\pm$3$\sigma$. The spectral type is on the scale of 0 is M0, 10 is L0, 20 is T0, and 30 is Y0. }
\figsetgrpend

\figsetgrpstart
\figsetgrpnum{8.251}
\figsetgrptitle{Light Curve for WISE1658p5103}
\figsetplot{WISE1658p5103_total_variable_SURF2022.pdf}
\figsetgrpnote{The light curve of WISE1658p5103 with \textit{Spitzer} ch2$_{aperture}$, ch2$_{PRF}$, and \textit{WISE} W2$_{PSF}$ photometry. The solid lines are the medians for each photometry type, while the dashed lines are $\pm$3$\sigma$. The spectral type is on the scale of 0 is M0, 10 is L0, 20 is T0, and 30 is Y0. }
\figsetgrpend

\figsetgrpstart
\figsetgrpnum{8.252}
\figsetgrptitle{Light Curve for WISE1701p4158}
\figsetplot{WISE1701p4158_total_variable_SURF2022.pdf}
\figsetgrpnote{The light curve of WISE1701p4158 with \textit{Spitzer} ch2$_{aperture}$, ch2$_{PRF}$, and \textit{WISE} W2$_{PSF}$ photometry. The solid lines are the medians for each photometry type, while the dashed lines are $\pm$3$\sigma$. The spectral type is on the scale of 0 is M0, 10 is L0, 20 is T0, and 30 is Y0. }
\figsetgrpend

\figsetgrpstart
\figsetgrpnum{8.253}
\figsetgrptitle{Light Curve for WISE1707m1744}
\figsetplot{WISE1707m1744_total_variable_SURF2022.pdf}
\figsetgrpnote{The light curve of WISE1707m1744 with \textit{Spitzer} ch2$_{aperture}$, ch2$_{PRF}$, and \textit{WISE} W2$_{PSF}$ photometry. The solid lines are the medians for each photometry type, while the dashed lines are $\pm$3$\sigma$. The spectral type is on the scale of 0 is M0, 10 is L0, 20 is T0, and 30 is Y0. }
\figsetgrpend

\figsetgrpstart
\figsetgrpnum{8.254}
\figsetgrptitle{Light Curve for WISE1711p3500}
\figsetplot{WISE1711p3500_total_variable_SURF2022.pdf}
\figsetgrpnote{The light curve of WISE1711p3500 with \textit{Spitzer} ch2$_{aperture}$, ch2$_{PRF}$, and \textit{WISE} W2$_{PSF}$ photometry. The solid lines are the medians for each photometry type, while the dashed lines are $\pm$3$\sigma$. The spectral type is on the scale of 0 is M0, 10 is L0, 20 is T0, and 30 is Y0. }
\figsetgrpend

\figsetgrpstart
\figsetgrpnum{8.255}
\figsetgrptitle{Light Curve for WISE1712p0645}
\figsetplot{WISE1712p0645_total_variable_SURF2022.pdf}
\figsetgrpnote{The light curve of WISE1712p0645 with \textit{Spitzer} ch2$_{aperture}$, ch2$_{PRF}$, and \textit{WISE} W2$_{PSF}$ photometry. The solid lines are the medians for each photometry type, while the dashed lines are $\pm$3$\sigma$. The spectral type is on the scale of 0 is M0, 10 is L0, 20 is T0, and 30 is Y0. }
\figsetgrpend

\figsetgrpstart
\figsetgrpnum{8.256}
\figsetgrptitle{Light Curve for WISE1717p6128}
\figsetplot{WISE1717p6128_total_variable_SURF2022.pdf}
\figsetgrpnote{The light curve of WISE1717p6128 with \textit{Spitzer} ch2$_{aperture}$, ch2$_{PRF}$, and \textit{WISE} W2$_{PSF}$ photometry. The solid lines are the medians for each photometry type, while the dashed lines are $\pm$3$\sigma$. The spectral type is on the scale of 0 is M0, 10 is L0, 20 is T0, and 30 is Y0. }
\figsetgrpend

\figsetgrpstart
\figsetgrpnum{8.257}
\figsetgrptitle{Light Curve for WISE1721p1117}
\figsetplot{WISE1721p1117_total_variable_SURF2022.pdf}
\figsetgrpnote{The light curve of WISE1721p1117 with \textit{Spitzer} ch2$_{aperture}$, ch2$_{PRF}$, and \textit{WISE} W2$_{PSF}$ photometry. The solid lines are the medians for each photometry type, while the dashed lines are $\pm$3$\sigma$. The spectral type is on the scale of 0 is M0, 10 is L0, 20 is T0, and 30 is Y0. }
\figsetgrpend

\figsetgrpstart
\figsetgrpnum{8.258}
\figsetgrptitle{Light Curve for WISE1729m7530}
\figsetplot{WISE1729m7530_total_variable_SURF2022.pdf}
\figsetgrpnote{The light curve of WISE1729m7530 with \textit{Spitzer} ch2$_{aperture}$, ch2$_{PRF}$, and \textit{WISE} W2$_{PSF}$ photometry. The solid lines are the medians for each photometry type, while the dashed lines are $\pm$3$\sigma$. The spectral type is on the scale of 0 is M0, 10 is L0, 20 is T0, and 30 is Y0. }
\figsetgrpend

\figsetgrpstart
\figsetgrpnum{8.259}
\figsetgrptitle{Light Curve for WISE1734m4813}
\figsetplot{WISE1734m4813_total_variable_SURF2022.pdf}
\figsetgrpnote{The light curve of WISE1734m4813 with \textit{Spitzer} ch2$_{aperture}$, ch2$_{PRF}$, and \textit{WISE} W2$_{PSF}$ photometry. The solid lines are the medians for each photometry type, while the dashed lines are $\pm$3$\sigma$. The spectral type is on the scale of 0 is M0, 10 is L0, 20 is T0, and 30 is Y0. }
\figsetgrpend

\figsetgrpstart
\figsetgrpnum{8.260}
\figsetgrptitle{Light Curve for WISE1735m8209}
\figsetplot{WISE1735m8209_total_variable_SURF2022.pdf}
\figsetgrpnote{The light curve of WISE1735m8209 with \textit{Spitzer} ch2$_{aperture}$, ch2$_{PRF}$, and \textit{WISE} W2$_{PSF}$ photometry. The solid lines are the medians for each photometry type, while the dashed lines are $\pm$3$\sigma$. The spectral type is on the scale of 0 is M0, 10 is L0, 20 is T0, and 30 is Y0. }
\figsetgrpend

\figsetgrpstart
\figsetgrpnum{8.261}
\figsetgrptitle{Light Curve for WISE1738p2732}
\figsetplot{WISE1738p2732_total_variable_SURF2022.pdf}
\figsetgrpnote{The light curve of WISE1738p2732 with \textit{Spitzer} ch2$_{aperture}$, ch2$_{PRF}$, and \textit{WISE} W2$_{PSF}$ photometry. The solid lines are the medians for each photometry type, while the dashed lines are $\pm$3$\sigma$. The spectral type is on the scale of 0 is M0, 10 is L0, 20 is T0, and 30 is Y0. }
\figsetgrpend

\figsetgrpstart
\figsetgrpnum{8.262}
\figsetgrptitle{Light Curve for WISE1738p6142}
\figsetplot{WISE1738p6142_total_variable_SURF2022.pdf}
\figsetgrpnote{The light curve of WISE1738p6142 with \textit{Spitzer} ch2$_{aperture}$, ch2$_{PRF}$, and \textit{WISE} W2$_{PSF}$ photometry. The solid lines are the medians for each photometry type, while the dashed lines are $\pm$3$\sigma$. The spectral type is on the scale of 0 is M0, 10 is L0, 20 is T0, and 30 is Y0. }
\figsetgrpend

\figsetgrpstart
\figsetgrpnum{8.263}
\figsetgrptitle{Light Curve for WISE1741m4642}
\figsetplot{WISE1741m4642_total_variable_SURF2022.pdf}
\figsetgrpnote{The light curve of WISE1741m4642 with \textit{Spitzer} ch2$_{aperture}$, ch2$_{PRF}$, and \textit{WISE} W2$_{PSF}$ photometry. The solid lines are the medians for each photometry type, while the dashed lines are $\pm$3$\sigma$. The spectral type is on the scale of 0 is M0, 10 is L0, 20 is T0, and 30 is Y0. }
\figsetgrpend

\figsetgrpstart
\figsetgrpnum{8.264}
\figsetgrptitle{Light Curve for WISE1743p4211}
\figsetplot{WISE1743p4211_total_variable_SURF2022.pdf}
\figsetgrpnote{The light curve of WISE1743p4211 with \textit{Spitzer} ch2$_{aperture}$, ch2$_{PRF}$, and \textit{WISE} W2$_{PSF}$ photometry. The solid lines are the medians for each photometry type, while the dashed lines are $\pm$3$\sigma$. The spectral type is on the scale of 0 is M0, 10 is L0, 20 is T0, and 30 is Y0. }
\figsetgrpend

\figsetgrpstart
\figsetgrpnum{8.265}
\figsetgrptitle{Light Curve for WISE1746p5034}
\figsetplot{WISE1746p5034_total_variable_SURF2022.pdf}
\figsetgrpnote{The light curve of WISE1746p5034 with \textit{Spitzer} ch2$_{aperture}$, ch2$_{PRF}$, and \textit{WISE} W2$_{PSF}$ photometry. The solid lines are the medians for each photometry type, while the dashed lines are $\pm$3$\sigma$. The spectral type is on the scale of 0 is M0, 10 is L0, 20 is T0, and 30 is Y0. }
\figsetgrpend

\figsetgrpstart
\figsetgrpnum{8.266}
\figsetgrptitle{Light Curve for WISE1746m0338}
\figsetplot{WISE1746m0338_total_variable_SURF2022.pdf}
\figsetgrpnote{The light curve of WISE1746m0338 with \textit{Spitzer} ch2$_{aperture}$, ch2$_{PRF}$, and \textit{WISE} W2$_{PSF}$ photometry. The solid lines are the medians for each photometry type, while the dashed lines are $\pm$3$\sigma$. The spectral type is on the scale of 0 is M0, 10 is L0, 20 is T0, and 30 is Y0. }
\figsetgrpend

\figsetgrpstart
\figsetgrpnum{8.267}
\figsetgrptitle{Light Curve for WISE1753m5904}
\figsetplot{WISE1753m5904_total_variable_SURF2022.pdf}
\figsetgrpnote{The light curve of WISE1753m5904 with \textit{Spitzer} ch2$_{aperture}$, ch2$_{PRF}$, and \textit{WISE} W2$_{PSF}$ photometry. The solid lines are the medians for each photometry type, while the dashed lines are $\pm$3$\sigma$. The spectral type is on the scale of 0 is M0, 10 is L0, 20 is T0, and 30 is Y0. }
\figsetgrpend

\figsetgrpstart
\figsetgrpnum{8.268}
\figsetgrptitle{Light Curve for WISE1754p1649}
\figsetplot{WISE1754p1649_total_variable_SURF2022.pdf}
\figsetgrpnote{The light curve of WISE1754p1649 with \textit{Spitzer} ch2$_{aperture}$, ch2$_{PRF}$, and \textit{WISE} W2$_{PSF}$ photometry. The solid lines are the medians for each photometry type, while the dashed lines are $\pm$3$\sigma$. The spectral type is on the scale of 0 is M0, 10 is L0, 20 is T0, and 30 is Y0. }
\figsetgrpend

\figsetgrpstart
\figsetgrpnum{8.269}
\figsetgrptitle{Light Curve for WISE1755p1803}
\figsetplot{WISE1755p1803_total_variable_SURF2022.pdf}
\figsetgrpnote{The light curve of WISE1755p1803 with \textit{Spitzer} ch2$_{aperture}$, ch2$_{PRF}$, and \textit{WISE} W2$_{PSF}$ photometry. The solid lines are the medians for each photometry type, while the dashed lines are $\pm$3$\sigma$. The spectral type is on the scale of 0 is M0, 10 is L0, 20 is T0, and 30 is Y0. }
\figsetgrpend

\figsetgrpstart
\figsetgrpnum{8.270}
\figsetgrptitle{Light Curve for WISE1804p3117}
\figsetplot{WISE1804p3117_total_variable_SURF2022.pdf}
\figsetgrpnote{The light curve of WISE1804p3117 with \textit{Spitzer} ch2$_{aperture}$, ch2$_{PRF}$, and \textit{WISE} W2$_{PSF}$ photometry. The solid lines are the medians for each photometry type, while the dashed lines are $\pm$3$\sigma$. The spectral type is on the scale of 0 is M0, 10 is L0, 20 is T0, and 30 is Y0. }
\figsetgrpend

\figsetgrpstart
\figsetgrpnum{8.271}
\figsetgrptitle{Light Curve for WISE1809m0448}
\figsetplot{WISE1809m0448_total_variable_SURF2022.pdf}
\figsetgrpnote{The light curve of WISE1809m0448 with \textit{Spitzer} ch2$_{aperture}$, ch2$_{PRF}$, and \textit{WISE} W2$_{PSF}$ photometry. The solid lines are the medians for each photometry type, while the dashed lines are $\pm$3$\sigma$. The spectral type is on the scale of 0 is M0, 10 is L0, 20 is T0, and 30 is Y0. }
\figsetgrpend

\figsetgrpstart
\figsetgrpnum{8.272}
\figsetgrptitle{Light Curve for WISE1812p2007}
\figsetplot{WISE1812p2007_total_variable_SURF2022.pdf}
\figsetgrpnote{The light curve of WISE1812p2007 with \textit{Spitzer} ch2$_{aperture}$, ch2$_{PRF}$, and \textit{WISE} W2$_{PSF}$ photometry. The solid lines are the medians for each photometry type, while the dashed lines are $\pm$3$\sigma$. The spectral type is on the scale of 0 is M0, 10 is L0, 20 is T0, and 30 is Y0. }
\figsetgrpend

\figsetgrpstart
\figsetgrpnum{8.273}
\figsetgrptitle{Light Curve for WISE1813p2835}
\figsetplot{WISE1813p2835_total_variable_SURF2022.pdf}
\figsetgrpnote{The light curve of WISE1813p2835 with \textit{Spitzer} ch2$_{aperture}$, ch2$_{PRF}$, and \textit{WISE} W2$_{PSF}$ photometry. The solid lines are the medians for each photometry type, while the dashed lines are $\pm$3$\sigma$. The spectral type is on the scale of 0 is M0, 10 is L0, 20 is T0, and 30 is Y0. }
\figsetgrpend

\figsetgrpstart
\figsetgrpnum{8.274}
\figsetgrptitle{Light Curve for WISE1818m4701}
\figsetplot{WISE1818m4701_total_variable_SURF2022.pdf}
\figsetgrpnote{The light curve of WISE1818m4701 with \textit{Spitzer} ch2$_{aperture}$, ch2$_{PRF}$, and \textit{WISE} W2$_{PSF}$ photometry. The solid lines are the medians for each photometry type, while the dashed lines are $\pm$3$\sigma$. The spectral type is on the scale of 0 is M0, 10 is L0, 20 is T0, and 30 is Y0. }
\figsetgrpend

\figsetgrpstart
\figsetgrpnum{8.275}
\figsetgrptitle{Light Curve for WISE1828p2650}
\figsetplot{WISE1828p2650_total_variable_SURF2022.pdf}
\figsetgrpnote{The light curve of WISE1828p2650 with \textit{Spitzer} ch2$_{aperture}$, ch2$_{PRF}$, and \textit{WISE} W2$_{PSF}$ photometry. The solid lines are the medians for each photometry type, while the dashed lines are $\pm$3$\sigma$. The spectral type is on the scale of 0 is M0, 10 is L0, 20 is T0, and 30 is Y0. }
\figsetgrpend

\figsetgrpstart
\figsetgrpnum{8.276}
\figsetgrptitle{Light Curve for WISE1832m5409}
\figsetplot{WISE1832m5409_total_variable_SURF2022.pdf}
\figsetgrpnote{The light curve of WISE1832m5409 with \textit{Spitzer} ch2$_{aperture}$, ch2$_{PRF}$, and \textit{WISE} W2$_{PSF}$ photometry. The solid lines are the medians for each photometry type, while the dashed lines are $\pm$3$\sigma$. The spectral type is on the scale of 0 is M0, 10 is L0, 20 is T0, and 30 is Y0. }
\figsetgrpend

\figsetgrpstart
\figsetgrpnum{8.277}
\figsetgrptitle{Light Curve for WISE1841p7000}
\figsetplot{WISE1841p7000_total_variable_SURF2022.pdf}
\figsetgrpnote{The light curve of WISE1841p7000 with \textit{Spitzer} ch2$_{aperture}$, ch2$_{PRF}$, and \textit{WISE} W2$_{PSF}$ photometry. The solid lines are the medians for each photometry type, while the dashed lines are $\pm$3$\sigma$. The spectral type is on the scale of 0 is M0, 10 is L0, 20 is T0, and 30 is Y0. }
\figsetgrpend

\figsetgrpstart
\figsetgrpnum{8.278}
\figsetgrptitle{Light Curve for WISE1851p5935}
\figsetplot{WISE1851p5935_total_variable_SURF2022.pdf}
\figsetgrpnote{The light curve of WISE1851p5935 with \textit{Spitzer} ch2$_{aperture}$, ch2$_{PRF}$, and \textit{WISE} W2$_{PSF}$ photometry. The solid lines are the medians for each photometry type, while the dashed lines are $\pm$3$\sigma$. The spectral type is on the scale of 0 is M0, 10 is L0, 20 is T0, and 30 is Y0. }
\figsetgrpend

\figsetgrpstart
\figsetgrpnum{8.279}
\figsetgrptitle{Light Curve for WISE1900m3108}
\figsetplot{WISE1900m3108_total_variable_SURF2022.pdf}
\figsetgrpnote{The light curve of WISE1900m3108 with \textit{Spitzer} ch2$_{aperture}$, ch2$_{PRF}$, and \textit{WISE} W2$_{PSF}$ photometry. The solid lines are the medians for each photometry type, while the dashed lines are $\pm$3$\sigma$. The spectral type is on the scale of 0 is M0, 10 is L0, 20 is T0, and 30 is Y0. }
\figsetgrpend

\figsetgrpstart
\figsetgrpnum{8.280}
\figsetgrptitle{Light Curve for WISE1901p4718}
\figsetplot{WISE1901p4718_total_variable_SURF2022.pdf}
\figsetgrpnote{The light curve of WISE1901p4718 with \textit{Spitzer} ch2$_{aperture}$, ch2$_{PRF}$, and \textit{WISE} W2$_{PSF}$ photometry. The solid lines are the medians for each photometry type, while the dashed lines are $\pm$3$\sigma$. The spectral type is on the scale of 0 is M0, 10 is L0, 20 is T0, and 30 is Y0. }
\figsetgrpend

\figsetgrpstart
\figsetgrpnum{8.281}
\figsetgrptitle{Light Curve for WISE1919p3045}
\figsetplot{WISE1919p3045_total_variable_SURF2022.pdf}
\figsetgrpnote{The light curve of WISE1919p3045 with \textit{Spitzer} ch2$_{aperture}$, ch2$_{PRF}$, and \textit{WISE} W2$_{PSF}$ photometry. The solid lines are the medians for each photometry type, while the dashed lines are $\pm$3$\sigma$. The spectral type is on the scale of 0 is M0, 10 is L0, 20 is T0, and 30 is Y0. }
\figsetgrpend

\figsetgrpstart
\figsetgrpnum{8.282}
\figsetgrptitle{Light Curve for WISE1925p0700}
\figsetplot{WISE1925p0700_total_variable_SURF2022.pdf}
\figsetgrpnote{The light curve of WISE1925p0700 with \textit{Spitzer} ch2$_{aperture}$, ch2$_{PRF}$, and \textit{WISE} W2$_{PSF}$ photometry. The solid lines are the medians for each photometry type, while the dashed lines are $\pm$3$\sigma$. The spectral type is on the scale of 0 is M0, 10 is L0, 20 is T0, and 30 is Y0. }
\figsetgrpend

\figsetgrpstart
\figsetgrpnum{8.283}
\figsetgrptitle{Light Curve for WISE1926m3429}
\figsetplot{WISE1926m3429_total_variable_SURF2022.pdf}
\figsetgrpnote{The light curve of WISE1926m3429 with \textit{Spitzer} ch2$_{aperture}$, ch2$_{PRF}$, and \textit{WISE} W2$_{PSF}$ photometry. The solid lines are the medians for each photometry type, while the dashed lines are $\pm$3$\sigma$. The spectral type is on the scale of 0 is M0, 10 is L0, 20 is T0, and 30 is Y0. }
\figsetgrpend

\figsetgrpstart
\figsetgrpnum{8.284}
\figsetgrptitle{Light Curve for WISE1928p2356}
\figsetplot{WISE1928p2356_total_variable_SURF2022.pdf}
\figsetgrpnote{The light curve of WISE1928p2356 with \textit{Spitzer} ch2$_{aperture}$, ch2$_{PRF}$, and \textit{WISE} W2$_{PSF}$ photometry. The solid lines are the medians for each photometry type, while the dashed lines are $\pm$3$\sigma$. The spectral type is on the scale of 0 is M0, 10 is L0, 20 is T0, and 30 is Y0. }
\figsetgrpend

\figsetgrpstart
\figsetgrpnum{8.285}
\figsetgrptitle{Light Curve for WISE1930m2059}
\figsetplot{WISE1930m2059_total_variable_SURF2022.pdf}
\figsetgrpnote{The light curve of WISE1930m2059 with \textit{Spitzer} ch2$_{aperture}$, ch2$_{PRF}$, and \textit{WISE} W2$_{PSF}$ photometry. The solid lines are the medians for each photometry type, while the dashed lines are $\pm$3$\sigma$. The spectral type is on the scale of 0 is M0, 10 is L0, 20 is T0, and 30 is Y0. }
\figsetgrpend

\figsetgrpstart
\figsetgrpnum{8.286}
\figsetgrptitle{Light Curve for WISE1935m1546}
\figsetplot{WISE1935m1546_total_variable_SURF2022.pdf}
\figsetgrpnote{The light curve of WISE1935m1546 with \textit{Spitzer} ch2$_{aperture}$, ch2$_{PRF}$, and \textit{WISE} W2$_{PSF}$ photometry. The solid lines are the medians for each photometry type, while the dashed lines are $\pm$3$\sigma$. The spectral type is on the scale of 0 is M0, 10 is L0, 20 is T0, and 30 is Y0. }
\figsetgrpend

\figsetgrpstart
\figsetgrpnum{8.287}
\figsetgrptitle{Light Curve for WISE1936p0407}
\figsetplot{WISE1936p0407_total_variable_SURF2022.pdf}
\figsetgrpnote{The light curve of WISE1936p0407 with \textit{Spitzer} ch2$_{aperture}$, ch2$_{PRF}$, and \textit{WISE} W2$_{PSF}$ photometry. The solid lines are the medians for each photometry type, while the dashed lines are $\pm$3$\sigma$. The spectral type is on the scale of 0 is M0, 10 is L0, 20 is T0, and 30 is Y0. }
\figsetgrpend

\figsetgrpstart
\figsetgrpnum{8.288}
\figsetgrptitle{Light Curve for WISE1955m2540}
\figsetplot{WISE1955m2540_total_variable_SURF2022.pdf}
\figsetgrpnote{The light curve of WISE1955m2540 with \textit{Spitzer} ch2$_{aperture}$, ch2$_{PRF}$, and \textit{WISE} W2$_{PSF}$ photometry. The solid lines are the medians for each photometry type, while the dashed lines are $\pm$3$\sigma$. The spectral type is on the scale of 0 is M0, 10 is L0, 20 is T0, and 30 is Y0. }
\figsetgrpend

\figsetgrpstart
\figsetgrpnum{8.289}
\figsetgrptitle{Light Curve for WISE1959m3338}
\figsetplot{WISE1959m3338_total_variable_SURF2022.pdf}
\figsetgrpnote{The light curve of WISE1959m3338 with \textit{Spitzer} ch2$_{aperture}$, ch2$_{PRF}$, and \textit{WISE} W2$_{PSF}$ photometry. The solid lines are the medians for each photometry type, while the dashed lines are $\pm$3$\sigma$. The spectral type is on the scale of 0 is M0, 10 is L0, 20 is T0, and 30 is Y0. }
\figsetgrpend

\figsetgrpstart
\figsetgrpnum{8.290}
\figsetgrptitle{Light Curve for WISE2000p3629}
\figsetplot{WISE2000p3629_total_variable_SURF2022.pdf}
\figsetgrpnote{The light curve of WISE2000p3629 with \textit{Spitzer} ch2$_{aperture}$, ch2$_{PRF}$, and \textit{WISE} W2$_{PSF}$ photometry. The solid lines are the medians for each photometry type, while the dashed lines are $\pm$3$\sigma$. The spectral type is on the scale of 0 is M0, 10 is L0, 20 is T0, and 30 is Y0. }
\figsetgrpend

\figsetgrpstart
\figsetgrpnum{8.291}
\figsetgrptitle{Light Curve for WISE2005p5424}
\figsetplot{WISE2005p5424_total_variable_SURF2022.pdf}
\figsetgrpnote{The light curve of WISE2005p5424 with \textit{Spitzer} ch2$_{aperture}$, ch2$_{PRF}$, and \textit{WISE} W2$_{PSF}$ photometry. The solid lines are the medians for each photometry type, while the dashed lines are $\pm$3$\sigma$. The spectral type is on the scale of 0 is M0, 10 is L0, 20 is T0, and 30 is Y0. }
\figsetgrpend

\figsetgrpstart
\figsetgrpnum{8.292}
\figsetgrptitle{Light Curve for WISE2008m0834}
\figsetplot{WISE2008m0834_total_variable_SURF2022.pdf}
\figsetgrpnote{The light curve of WISE2008m0834 with \textit{Spitzer} ch2$_{aperture}$, ch2$_{PRF}$, and \textit{WISE} W2$_{PSF}$ photometry. The solid lines are the medians for each photometry type, while the dashed lines are $\pm$3$\sigma$. The spectral type is on the scale of 0 is M0, 10 is L0, 20 is T0, and 30 is Y0. }
\figsetgrpend

\figsetgrpstart
\figsetgrpnum{8.293}
\figsetgrptitle{Light Curve for WISE2011m4812}
\figsetplot{WISE2011m4812_total_variable_SURF2022.pdf}
\figsetgrpnote{The light curve of WISE2011m4812 with \textit{Spitzer} ch2$_{aperture}$, ch2$_{PRF}$, and \textit{WISE} W2$_{PSF}$ photometry. The solid lines are the medians for each photometry type, while the dashed lines are $\pm$3$\sigma$. The spectral type is on the scale of 0 is M0, 10 is L0, 20 is T0, and 30 is Y0. }
\figsetgrpend

\figsetgrpstart
\figsetgrpnum{8.294}
\figsetgrptitle{Light Curve for WISE2012p7017}
\figsetplot{WISE2012p7017_total_variable_SURF2022.pdf}
\figsetgrpnote{The light curve of WISE2012p7017 with \textit{Spitzer} ch2$_{aperture}$, ch2$_{PRF}$, and \textit{WISE} W2$_{PSF}$ photometry. The solid lines are the medians for each photometry type, while the dashed lines are $\pm$3$\sigma$. The spectral type is on the scale of 0 is M0, 10 is L0, 20 is T0, and 30 is Y0. }
\figsetgrpend

\figsetgrpstart
\figsetgrpnum{8.295}
\figsetgrptitle{Light Curve for WISE2014p0424}
\figsetplot{WISE2014p0424_total_variable_SURF2022.pdf}
\figsetgrpnote{The light curve of WISE2014p0424 with \textit{Spitzer} ch2$_{aperture}$, ch2$_{PRF}$, and \textit{WISE} W2$_{PSF}$ photometry. The solid lines are the medians for each photometry type, while the dashed lines are $\pm$3$\sigma$. The spectral type is on the scale of 0 is M0, 10 is L0, 20 is T0, and 30 is Y0. }
\figsetgrpend

\figsetgrpstart
\figsetgrpnum{8.296}
\figsetgrptitle{Light Curve for WISE2015p6646}
\figsetplot{WISE2015p6646_total_variable_SURF2022.pdf}
\figsetgrpnote{The light curve of WISE2015p6646 with \textit{Spitzer} ch2$_{aperture}$, ch2$_{PRF}$, and \textit{WISE} W2$_{PSF}$ photometry. The solid lines are the medians for each photometry type, while the dashed lines are $\pm$3$\sigma$. The spectral type is on the scale of 0 is M0, 10 is L0, 20 is T0, and 30 is Y0. }
\figsetgrpend

\figsetgrpstart
\figsetgrpnum{8.297}
\figsetgrptitle{Light Curve for WISE2017m3421}
\figsetplot{WISE2017m3421_total_variable_SURF2022.pdf}
\figsetgrpnote{The light curve of WISE2017m3421 with \textit{Spitzer} ch2$_{aperture}$, ch2$_{PRF}$, and \textit{WISE} W2$_{PSF}$ photometry. The solid lines are the medians for each photometry type, while the dashed lines are $\pm$3$\sigma$. The spectral type is on the scale of 0 is M0, 10 is L0, 20 is T0, and 30 is Y0. }
\figsetgrpend

\figsetgrpstart
\figsetgrpnum{8.298}
\figsetgrptitle{Light Curve for WISE2018m1417}
\figsetplot{WISE2018m1417_total_variable_SURF2022.pdf}
\figsetgrpnote{The light curve of WISE2018m1417 with \textit{Spitzer} ch2$_{aperture}$, ch2$_{PRF}$, and \textit{WISE} W2$_{PSF}$ photometry. The solid lines are the medians for each photometry type, while the dashed lines are $\pm$3$\sigma$. The spectral type is on the scale of 0 is M0, 10 is L0, 20 is T0, and 30 is Y0. }
\figsetgrpend

\figsetgrpstart
\figsetgrpnum{8.299}
\figsetgrptitle{Light Curve for WISE2019m1148}
\figsetplot{WISE2019m1148_total_variable_SURF2022.pdf}
\figsetgrpnote{The light curve of WISE2019m1148 with \textit{Spitzer} ch2$_{aperture}$, ch2$_{PRF}$, and \textit{WISE} W2$_{PSF}$ photometry. The solid lines are the medians for each photometry type, while the dashed lines are $\pm$3$\sigma$. The spectral type is on the scale of 0 is M0, 10 is L0, 20 is T0, and 30 is Y0. }
\figsetgrpend

\figsetgrpstart
\figsetgrpnum{8.300}
\figsetgrptitle{Light Curve for WISE2030p0749}
\figsetplot{WISE2030p0749_total_variable_SURF2022.pdf}
\figsetgrpnote{The light curve of WISE2030p0749 with \textit{Spitzer} ch2$_{aperture}$, ch2$_{PRF}$, and \textit{WISE} W2$_{PSF}$ photometry. The solid lines are the medians for each photometry type, while the dashed lines are $\pm$3$\sigma$. The spectral type is on the scale of 0 is M0, 10 is L0, 20 is T0, and 30 is Y0. }
\figsetgrpend

\figsetgrpstart
\figsetgrpnum{8.301}
\figsetgrptitle{Light Curve for WISE2038m0649}
\figsetplot{WISE2038m0649_total_variable_SURF2022.pdf}
\figsetgrpnote{The light curve of WISE2038m0649 with \textit{Spitzer} ch2$_{aperture}$, ch2$_{PRF}$, and \textit{WISE} W2$_{PSF}$ photometry. The solid lines are the medians for each photometry type, while the dashed lines are $\pm$3$\sigma$. The spectral type is on the scale of 0 is M0, 10 is L0, 20 is T0, and 30 is Y0. }
\figsetgrpend

\figsetgrpstart
\figsetgrpnum{8.302}
\figsetgrptitle{Light Curve for WISE2043p6220}
\figsetplot{WISE2043p6220_total_variable_SURF2022.pdf}
\figsetgrpnote{The light curve of WISE2043p6220 with \textit{Spitzer} ch2$_{aperture}$, ch2$_{PRF}$, and \textit{WISE} W2$_{PSF}$ photometry. The solid lines are the medians for each photometry type, while the dashed lines are $\pm$3$\sigma$. The spectral type is on the scale of 0 is M0, 10 is L0, 20 is T0, and 30 is Y0. }
\figsetgrpend

\figsetgrpstart
\figsetgrpnum{8.303}
\figsetgrptitle{Light Curve for WISE2056p1459}
\figsetplot{WISE2056p1459_total_variable_SURF2022.pdf}
\figsetgrpnote{The light curve of WISE2056p1459 with \textit{Spitzer} ch2$_{aperture}$, ch2$_{PRF}$, and \textit{WISE} W2$_{PSF}$ photometry. The solid lines are the medians for each photometry type, while the dashed lines are $\pm$3$\sigma$. The spectral type is on the scale of 0 is M0, 10 is L0, 20 is T0, and 30 is Y0. }
\figsetgrpend

\figsetgrpstart
\figsetgrpnum{8.304}
\figsetgrptitle{Light Curve for WISE2057m1704}
\figsetplot{WISE2057m1704_total_variable_SURF2022.pdf}
\figsetgrpnote{The light curve of WISE2057m1704 with \textit{Spitzer} ch2$_{aperture}$, ch2$_{PRF}$, and \textit{WISE} W2$_{PSF}$ photometry. The solid lines are the medians for each photometry type, while the dashed lines are $\pm$3$\sigma$. The spectral type is on the scale of 0 is M0, 10 is L0, 20 is T0, and 30 is Y0. }
\figsetgrpend

\figsetgrpstart
\figsetgrpnum{8.305}
\figsetgrptitle{Light Curve for WISE2100m2931}
\figsetplot{WISE2100m2931_total_variable_SURF2022.pdf}
\figsetgrpnote{The light curve of WISE2100m2931 with \textit{Spitzer} ch2$_{aperture}$, ch2$_{PRF}$, and \textit{WISE} W2$_{PSF}$ photometry. The solid lines are the medians for each photometry type, while the dashed lines are $\pm$3$\sigma$. The spectral type is on the scale of 0 is M0, 10 is L0, 20 is T0, and 30 is Y0. }
\figsetgrpend

\figsetgrpstart
\figsetgrpnum{8.306}
\figsetgrptitle{Light Curve for WISE2114m1805}
\figsetplot{WISE2114m1805_total_variable_SURF2022.pdf}
\figsetgrpnote{The light curve of WISE2114m1805 with \textit{Spitzer} ch2$_{aperture}$, ch2$_{PRF}$, and \textit{WISE} W2$_{PSF}$ photometry. The solid lines are the medians for each photometry type, while the dashed lines are $\pm$3$\sigma$. The spectral type is on the scale of 0 is M0, 10 is L0, 20 is T0, and 30 is Y0. }
\figsetgrpend

\figsetgrpstart
\figsetgrpnum{8.307}
\figsetgrptitle{Light Curve for WISE2117m2940}
\figsetplot{WISE2117m2940_total_variable_SURF2022.pdf}
\figsetgrpnote{The light curve of WISE2117m2940 with \textit{Spitzer} ch2$_{aperture}$, ch2$_{PRF}$, and \textit{WISE} W2$_{PSF}$ photometry. The solid lines are the medians for each photometry type, while the dashed lines are $\pm$3$\sigma$. The spectral type is on the scale of 0 is M0, 10 is L0, 20 is T0, and 30 is Y0. }
\figsetgrpend

\figsetgrpstart
\figsetgrpnum{8.308}
\figsetgrptitle{Light Curve for WISE2121m6239}
\figsetplot{WISE2121m6239_total_variable_SURF2022.pdf}
\figsetgrpnote{The light curve of WISE2121m6239 with \textit{Spitzer} ch2$_{aperture}$, ch2$_{PRF}$, and \textit{WISE} W2$_{PSF}$ photometry. The solid lines are the medians for each photometry type, while the dashed lines are $\pm$3$\sigma$. The spectral type is on the scale of 0 is M0, 10 is L0, 20 is T0, and 30 is Y0. }
\figsetgrpend

\figsetgrpstart
\figsetgrpnum{8.309}
\figsetgrptitle{Light Curve for WISE2123m2614}
\figsetplot{WISE2123m2614_total_variable_SURF2022.pdf}
\figsetgrpnote{The light curve of WISE2123m2614 with \textit{Spitzer} ch2$_{aperture}$, ch2$_{PRF}$, and \textit{WISE} W2$_{PSF}$ photometry. The solid lines are the medians for each photometry type, while the dashed lines are $\pm$3$\sigma$. The spectral type is on the scale of 0 is M0, 10 is L0, 20 is T0, and 30 is Y0. }
\figsetgrpend

\figsetgrpstart
\figsetgrpnum{8.310}
\figsetgrptitle{Light Curve for WISE2127p7617}
\figsetplot{WISE2127p7617_total_variable_SURF2022.pdf}
\figsetgrpnote{The light curve of WISE2127p7617 with \textit{Spitzer} ch2$_{aperture}$, ch2$_{PRF}$, and \textit{WISE} W2$_{PSF}$ photometry. The solid lines are the medians for each photometry type, while the dashed lines are $\pm$3$\sigma$. The spectral type is on the scale of 0 is M0, 10 is L0, 20 is T0, and 30 is Y0. }
\figsetgrpend

\figsetgrpstart
\figsetgrpnum{8.311}
\figsetgrptitle{Light Curve for WISE2132p6901}
\figsetplot{WISE2132p6901_total_variable_SURF2022.pdf}
\figsetgrpnote{The light curve of WISE2132p6901 with \textit{Spitzer} ch2$_{aperture}$, ch2$_{PRF}$, and \textit{WISE} W2$_{PSF}$ photometry. The solid lines are the medians for each photometry type, while the dashed lines are $\pm$3$\sigma$. The spectral type is on the scale of 0 is M0, 10 is L0, 20 is T0, and 30 is Y0. }
\figsetgrpend

\figsetgrpstart
\figsetgrpnum{8.312}
\figsetgrptitle{Light Curve for WISE2137p0808}
\figsetplot{WISE2137p0808_total_variable_SURF2022.pdf}
\figsetgrpnote{The light curve of WISE2137p0808 with \textit{Spitzer} ch2$_{aperture}$, ch2$_{PRF}$, and \textit{WISE} W2$_{PSF}$ photometry. The solid lines are the medians for each photometry type, while the dashed lines are $\pm$3$\sigma$. The spectral type is on the scale of 0 is M0, 10 is L0, 20 is T0, and 30 is Y0. }
\figsetgrpend

\figsetgrpstart
\figsetgrpnum{8.313}
\figsetgrptitle{Light Curve for WISE2139p0427}
\figsetplot{WISE2139p0427_total_variable_SURF2022.pdf}
\figsetgrpnote{The light curve of WISE2139p0427 with \textit{Spitzer} ch2$_{aperture}$, ch2$_{PRF}$, and \textit{WISE} W2$_{PSF}$ photometry. The solid lines are the medians for each photometry type, while the dashed lines are $\pm$3$\sigma$. The spectral type is on the scale of 0 is M0, 10 is L0, 20 is T0, and 30 is Y0. }
\figsetgrpend

\figsetgrpstart
\figsetgrpnum{8.314}
\figsetgrptitle{Light Curve for WISE2141m5118}
\figsetplot{WISE2141m5118_total_variable_SURF2022.pdf}
\figsetgrpnote{The light curve of WISE2141m5118 with \textit{Spitzer} ch2$_{aperture}$, ch2$_{PRF}$, and \textit{WISE} W2$_{PSF}$ photometry. The solid lines are the medians for each photometry type, while the dashed lines are $\pm$3$\sigma$. The spectral type is on the scale of 0 is M0, 10 is L0, 20 is T0, and 30 is Y0. }
\figsetgrpend

\figsetgrpstart
\figsetgrpnum{8.315}
\figsetgrptitle{Light Curve for WISE2147m1029}
\figsetplot{WISE2147m1029_total_variable_SURF2022.pdf}
\figsetgrpnote{The light curve of WISE2147m1029 with \textit{Spitzer} ch2$_{aperture}$, ch2$_{PRF}$, and \textit{WISE} W2$_{PSF}$ photometry. The solid lines are the medians for each photometry type, while the dashed lines are $\pm$3$\sigma$. The spectral type is on the scale of 0 is M0, 10 is L0, 20 is T0, and 30 is Y0. }
\figsetgrpend

\figsetgrpstart
\figsetgrpnum{8.316}
\figsetgrptitle{Light Curve for WISE2151m2441}
\figsetplot{WISE2151m2441_total_variable_SURF2022.pdf}
\figsetgrpnote{The light curve of WISE2151m2441 with \textit{Spitzer} ch2$_{aperture}$, ch2$_{PRF}$, and \textit{WISE} W2$_{PSF}$ photometry. The solid lines are the medians for each photometry type, while the dashed lines are $\pm$3$\sigma$. The spectral type is on the scale of 0 is M0, 10 is L0, 20 is T0, and 30 is Y0. }
\figsetgrpend

\figsetgrpstart
\figsetgrpnum{8.317}
\figsetgrptitle{Light Curve for WISE2152p0937}
\figsetplot{WISE2152p0937_total_variable_SURF2022.pdf}
\figsetgrpnote{The light curve of WISE2152p0937 with \textit{Spitzer} ch2$_{aperture}$, ch2$_{PRF}$, and \textit{WISE} W2$_{PSF}$ photometry. The solid lines are the medians for each photometry type, while the dashed lines are $\pm$3$\sigma$. The spectral type is on the scale of 0 is M0, 10 is L0, 20 is T0, and 30 is Y0. }
\figsetgrpend

\figsetgrpstart
\figsetgrpnum{8.318}
\figsetgrptitle{Light Curve for WISE2154p5942}
\figsetplot{WISE2154p5942_total_variable_SURF2022.pdf}
\figsetgrpnote{The light curve of WISE2154p5942 with \textit{Spitzer} ch2$_{aperture}$, ch2$_{PRF}$, and \textit{WISE} W2$_{PSF}$ photometry. The solid lines are the medians for each photometry type, while the dashed lines are $\pm$3$\sigma$. The spectral type is on the scale of 0 is M0, 10 is L0, 20 is T0, and 30 is Y0. }
\figsetgrpend

\figsetgrpstart
\figsetgrpnum{8.319}
\figsetgrptitle{Light Curve for WISE2157p2659}
\figsetplot{WISE2157p2659_total_variable_SURF2022.pdf}
\figsetgrpnote{The light curve of WISE2157p2659 with \textit{Spitzer} ch2$_{aperture}$, ch2$_{PRF}$, and \textit{WISE} W2$_{PSF}$ photometry. The solid lines are the medians for each photometry type, while the dashed lines are $\pm$3$\sigma$. The spectral type is on the scale of 0 is M0, 10 is L0, 20 is T0, and 30 is Y0. }
\figsetgrpend

\figsetgrpstart
\figsetgrpnum{8.320}
\figsetgrptitle{Light Curve for WISE2159m4808}
\figsetplot{WISE2159m4808_total_variable_SURF2022.pdf}
\figsetgrpnote{The light curve of WISE2159m4808 with \textit{Spitzer} ch2$_{aperture}$, ch2$_{PRF}$, and \textit{WISE} W2$_{PSF}$ photometry. The solid lines are the medians for each photometry type, while the dashed lines are $\pm$3$\sigma$. The spectral type is on the scale of 0 is M0, 10 is L0, 20 is T0, and 30 is Y0. }
\figsetgrpend

\figsetgrpstart
\figsetgrpnum{8.321}
\figsetgrptitle{Light Curve for WISE2201p3222}
\figsetplot{WISE2201p3222_total_variable_SURF2022.pdf}
\figsetgrpnote{The light curve of WISE2201p3222 with \textit{Spitzer} ch2$_{aperture}$, ch2$_{PRF}$, and \textit{WISE} W2$_{PSF}$ photometry. The solid lines are the medians for each photometry type, while the dashed lines are $\pm$3$\sigma$. The spectral type is on the scale of 0 is M0, 10 is L0, 20 is T0, and 30 is Y0. }
\figsetgrpend

\figsetgrpstart
\figsetgrpnum{8.322}
\figsetgrptitle{Light Curve for WISE2203p4619}
\figsetplot{WISE2203p4619_total_variable_SURF2022.pdf}
\figsetgrpnote{The light curve of WISE2203p4619 with \textit{Spitzer} ch2$_{aperture}$, ch2$_{PRF}$, and \textit{WISE} W2$_{PSF}$ photometry. The solid lines are the medians for each photometry type, while the dashed lines are $\pm$3$\sigma$. The spectral type is on the scale of 0 is M0, 10 is L0, 20 is T0, and 30 is Y0. }
\figsetgrpend

\figsetgrpstart
\figsetgrpnum{8.323}
\figsetgrptitle{Light Curve for WISE2209p2711}
\figsetplot{WISE2209p2711_total_variable_SURF2022.pdf}
\figsetgrpnote{The light curve of WISE2209p2711 with \textit{Spitzer} ch2$_{aperture}$, ch2$_{PRF}$, and \textit{WISE} W2$_{PSF}$ photometry. The solid lines are the medians for each photometry type, while the dashed lines are $\pm$3$\sigma$. The spectral type is on the scale of 0 is M0, 10 is L0, 20 is T0, and 30 is Y0. }
\figsetgrpend

\figsetgrpstart
\figsetgrpnum{8.324}
\figsetgrptitle{Light Curve for WISE2209m2711}
\figsetplot{WISE2209m2711_total_variable_SURF2022.pdf}
\figsetgrpnote{The light curve of WISE2209m2711 with \textit{Spitzer} ch2$_{aperture}$, ch2$_{PRF}$, and \textit{WISE} W2$_{PSF}$ photometry. The solid lines are the medians for each photometry type, while the dashed lines are $\pm$3$\sigma$. The spectral type is on the scale of 0 is M0, 10 is L0, 20 is T0, and 30 is Y0. }
\figsetgrpend

\figsetgrpstart
\figsetgrpnum{8.325}
\figsetgrptitle{Light Curve for WISE2209m2734}
\figsetplot{WISE2209m2734_total_variable_SURF2022.pdf}
\figsetgrpnote{The light curve of WISE2209m2734 with \textit{Spitzer} ch2$_{aperture}$, ch2$_{PRF}$, and \textit{WISE} W2$_{PSF}$ photometry. The solid lines are the medians for each photometry type, while the dashed lines are $\pm$3$\sigma$. The spectral type is on the scale of 0 is M0, 10 is L0, 20 is T0, and 30 is Y0. }
\figsetgrpend

\figsetgrpstart
\figsetgrpnum{8.326}
\figsetgrptitle{Light Curve for WISE2211m4758}
\figsetplot{WISE2211m4758_total_variable_SURF2022.pdf}
\figsetgrpnote{The light curve of WISE2211m4758 with \textit{Spitzer} ch2$_{aperture}$, ch2$_{PRF}$, and \textit{WISE} W2$_{PSF}$ photometry. The solid lines are the medians for each photometry type, while the dashed lines are $\pm$3$\sigma$. The spectral type is on the scale of 0 is M0, 10 is L0, 20 is T0, and 30 is Y0. }
\figsetgrpend

\figsetgrpstart
\figsetgrpnum{8.327}
\figsetgrptitle{Light Curve for WISE2212m6931}
\figsetplot{WISE2212m6931_total_variable_SURF2022.pdf}
\figsetgrpnote{The light curve of WISE2212m6931 with \textit{Spitzer} ch2$_{aperture}$, ch2$_{PRF}$, and \textit{WISE} W2$_{PSF}$ photometry. The solid lines are the medians for each photometry type, while the dashed lines are $\pm$3$\sigma$. The spectral type is on the scale of 0 is M0, 10 is L0, 20 is T0, and 30 is Y0. }
\figsetgrpend

\figsetgrpstart
\figsetgrpnum{8.328}
\figsetgrptitle{Light Curve for WISE2215p2110}
\figsetplot{WISE2215p2110_total_variable_SURF2022.pdf}
\figsetgrpnote{The light curve of WISE2215p2110 with \textit{Spitzer} ch2$_{aperture}$, ch2$_{PRF}$, and \textit{WISE} W2$_{PSF}$ photometry. The solid lines are the medians for each photometry type, while the dashed lines are $\pm$3$\sigma$. The spectral type is on the scale of 0 is M0, 10 is L0, 20 is T0, and 30 is Y0. }
\figsetgrpend

\figsetgrpstart
\figsetgrpnum{8.329}
\figsetgrptitle{Light Curve for WISE2220m3628}
\figsetplot{WISE2220m3628_total_variable_SURF2022.pdf}
\figsetgrpnote{The light curve of WISE2220m3628 with \textit{Spitzer} ch2$_{aperture}$, ch2$_{PRF}$, and \textit{WISE} W2$_{PSF}$ photometry. The solid lines are the medians for each photometry type, while the dashed lines are $\pm$3$\sigma$. The spectral type is on the scale of 0 is M0, 10 is L0, 20 is T0, and 30 is Y0. }
\figsetgrpend

\figsetgrpstart
\figsetgrpnum{8.330}
\figsetgrptitle{Light Curve for WISE2230p2549}
\figsetplot{WISE2230p2549_total_variable_SURF2022.pdf}
\figsetgrpnote{The light curve of WISE2230p2549 with \textit{Spitzer} ch2$_{aperture}$, ch2$_{PRF}$, and \textit{WISE} W2$_{PSF}$ photometry. The solid lines are the medians for each photometry type, while the dashed lines are $\pm$3$\sigma$. The spectral type is on the scale of 0 is M0, 10 is L0, 20 is T0, and 30 is Y0. }
\figsetgrpend

\figsetgrpstart
\figsetgrpnum{8.331}
\figsetgrptitle{Light Curve for WISE2232m5730}
\figsetplot{WISE2232m5730_total_variable_SURF2022.pdf}
\figsetgrpnote{The light curve of WISE2232m5730 with \textit{Spitzer} ch2$_{aperture}$, ch2$_{PRF}$, and \textit{WISE} W2$_{PSF}$ photometry. The solid lines are the medians for each photometry type, while the dashed lines are $\pm$3$\sigma$. The spectral type is on the scale of 0 is M0, 10 is L0, 20 is T0, and 30 is Y0. }
\figsetgrpend

\figsetgrpstart
\figsetgrpnum{8.332}
\figsetgrptitle{Light Curve for WISE2236p5105}
\figsetplot{WISE2236p5105_total_variable_SURF2022.pdf}
\figsetgrpnote{The light curve of WISE2236p5105 with \textit{Spitzer} ch2$_{aperture}$, ch2$_{PRF}$, and \textit{WISE} W2$_{PSF}$ photometry. The solid lines are the medians for each photometry type, while the dashed lines are $\pm$3$\sigma$. The spectral type is on the scale of 0 is M0, 10 is L0, 20 is T0, and 30 is Y0. }
\figsetgrpend

\figsetgrpstart
\figsetgrpnum{8.333}
\figsetgrptitle{Light Curve for WISE2237p7228}
\figsetplot{WISE2237p7228_total_variable_SURF2022.pdf}
\figsetgrpnote{The light curve of WISE2237p7228 with \textit{Spitzer} ch2$_{aperture}$, ch2$_{PRF}$, and \textit{WISE} W2$_{PSF}$ photometry. The solid lines are the medians for each photometry type, while the dashed lines are $\pm$3$\sigma$. The spectral type is on the scale of 0 is M0, 10 is L0, 20 is T0, and 30 is Y0. }
\figsetgrpend

\figsetgrpstart
\figsetgrpnum{8.334}
\figsetgrptitle{Light Curve for WISE2239p1617}
\figsetplot{WISE2239p1617_total_variable_SURF2022.pdf}
\figsetgrpnote{The light curve of WISE2239p1617 with \textit{Spitzer} ch2$_{aperture}$, ch2$_{PRF}$, and \textit{WISE} W2$_{PSF}$ photometry. The solid lines are the medians for each photometry type, while the dashed lines are $\pm$3$\sigma$. The spectral type is on the scale of 0 is M0, 10 is L0, 20 is T0, and 30 is Y0. }
\figsetgrpend

\figsetgrpstart
\figsetgrpnum{8.335}
\figsetgrptitle{Light Curve for WISE2243m1458}
\figsetplot{WISE2243m1458_total_variable_SURF2022.pdf}
\figsetgrpnote{The light curve of WISE2243m1458 with \textit{Spitzer} ch2$_{aperture}$, ch2$_{PRF}$, and \textit{WISE} W2$_{PSF}$ photometry. The solid lines are the medians for each photometry type, while the dashed lines are $\pm$3$\sigma$. The spectral type is on the scale of 0 is M0, 10 is L0, 20 is T0, and 30 is Y0. }
\figsetgrpend

\figsetgrpstart
\figsetgrpnum{8.336}
\figsetgrptitle{Light Curve for WISE2249p3205}
\figsetplot{WISE2249p3205_total_variable_SURF2022.pdf}
\figsetgrpnote{The light curve of WISE2249p3205 with \textit{Spitzer} ch2$_{aperture}$, ch2$_{PRF}$, and \textit{WISE} W2$_{PSF}$ photometry. The solid lines are the medians for each photometry type, while the dashed lines are $\pm$3$\sigma$. The spectral type is on the scale of 0 is M0, 10 is L0, 20 is T0, and 30 is Y0. }
\figsetgrpend

\figsetgrpstart
\figsetgrpnum{8.337}
\figsetgrptitle{Light Curve for WISE2254m2652}
\figsetplot{WISE2254m2652_total_variable_SURF2022.pdf}
\figsetgrpnote{The light curve of WISE2254m2652 with \textit{Spitzer} ch2$_{aperture}$, ch2$_{PRF}$, and \textit{WISE} W2$_{PSF}$ photometry. The solid lines are the medians for each photometry type, while the dashed lines are $\pm$3$\sigma$. The spectral type is on the scale of 0 is M0, 10 is L0, 20 is T0, and 30 is Y0. }
\figsetgrpend

\figsetgrpstart
\figsetgrpnum{8.338}
\figsetgrptitle{Light Curve for WISE2255m5713}
\figsetplot{WISE2255m5713_total_variable_SURF2022.pdf}
\figsetgrpnote{The light curve of WISE2255m5713 with \textit{Spitzer} ch2$_{aperture}$, ch2$_{PRF}$, and \textit{WISE} W2$_{PSF}$ photometry. The solid lines are the medians for each photometry type, while the dashed lines are $\pm$3$\sigma$. The spectral type is on the scale of 0 is M0, 10 is L0, 20 is T0, and 30 is Y0. }
\figsetgrpend

\figsetgrpstart
\figsetgrpnum{8.339}
\figsetgrptitle{Light Curve for WISE2255m3118}
\figsetplot{WISE2255m3118_total_variable_SURF2022.pdf}
\figsetgrpnote{The light curve of WISE2255m3118 with \textit{Spitzer} ch2$_{aperture}$, ch2$_{PRF}$, and \textit{WISE} W2$_{PSF}$ photometry. The solid lines are the medians for each photometry type, while the dashed lines are $\pm$3$\sigma$. The spectral type is on the scale of 0 is M0, 10 is L0, 20 is T0, and 30 is Y0. }
\figsetgrpend

\figsetgrpstart
\figsetgrpnum{8.340}
\figsetgrptitle{Light Curve for WISE2256p4002}
\figsetplot{WISE2256p4002_total_variable_SURF2022.pdf}
\figsetgrpnote{The light curve of WISE2256p4002 with \textit{Spitzer} ch2$_{aperture}$, ch2$_{PRF}$, and \textit{WISE} W2$_{PSF}$ photometry. The solid lines are the medians for each photometry type, while the dashed lines are $\pm$3$\sigma$. The spectral type is on the scale of 0 is M0, 10 is L0, 20 is T0, and 30 is Y0. }
\figsetgrpend

\figsetgrpstart
\figsetgrpnum{8.341}
\figsetgrptitle{Light Curve for WISE2301p0216}
\figsetplot{WISE2301p0216_total_variable_SURF2022.pdf}
\figsetgrpnote{The light curve of WISE2301p0216 with \textit{Spitzer} ch2$_{aperture}$, ch2$_{PRF}$, and \textit{WISE} W2$_{PSF}$ photometry. The solid lines are the medians for each photometry type, while the dashed lines are $\pm$3$\sigma$. The spectral type is on the scale of 0 is M0, 10 is L0, 20 is T0, and 30 is Y0. }
\figsetgrpend

\figsetgrpstart
\figsetgrpnum{8.342}
\figsetgrptitle{Light Curve for WISE2302m7134}
\figsetplot{WISE2302m7134_total_variable_SURF2022.pdf}
\figsetgrpnote{The light curve of WISE2302m7134 with \textit{Spitzer} ch2$_{aperture}$, ch2$_{PRF}$, and \textit{WISE} W2$_{PSF}$ photometry. The solid lines are the medians for each photometry type, while the dashed lines are $\pm$3$\sigma$. The spectral type is on the scale of 0 is M0, 10 is L0, 20 is T0, and 30 is Y0. }
\figsetgrpend

\figsetgrpstart
\figsetgrpnum{8.343}
\figsetgrptitle{Light Curve for WISE2313m8037}
\figsetplot{WISE2313m8037_total_variable_SURF2022.pdf}
\figsetgrpnote{The light curve of WISE2313m8037 with \textit{Spitzer} ch2$_{aperture}$, ch2$_{PRF}$, and \textit{WISE} W2$_{PSF}$ photometry. The solid lines are the medians for each photometry type, while the dashed lines are $\pm$3$\sigma$. The spectral type is on the scale of 0 is M0, 10 is L0, 20 is T0, and 30 is Y0. }
\figsetgrpend

\figsetgrpstart
\figsetgrpnum{8.344}
\figsetgrptitle{Light Curve for WISE2317m4838}
\figsetplot{WISE2317m4838_total_variable_SURF2022.pdf}
\figsetgrpnote{The light curve of WISE2317m4838 with \textit{Spitzer} ch2$_{aperture}$, ch2$_{PRF}$, and \textit{WISE} W2$_{PSF}$ photometry. The solid lines are the medians for each photometry type, while the dashed lines are $\pm$3$\sigma$. The spectral type is on the scale of 0 is M0, 10 is L0, 20 is T0, and 30 is Y0. }
\figsetgrpend

\figsetgrpstart
\figsetgrpnum{8.345}
\figsetgrptitle{Light Curve for WISE2319m1844}
\figsetplot{WISE2319m1844_total_variable_SURF2022.pdf}
\figsetgrpnote{The light curve of WISE2319m1844 with \textit{Spitzer} ch2$_{aperture}$, ch2$_{PRF}$, and \textit{WISE} W2$_{PSF}$ photometry. The solid lines are the medians for each photometry type, while the dashed lines are $\pm$3$\sigma$. The spectral type is on the scale of 0 is M0, 10 is L0, 20 is T0, and 30 is Y0. }
\figsetgrpend

\figsetgrpstart
\figsetgrpnum{8.346}
\figsetgrptitle{Light Curve for WISE2320p1448}
\figsetplot{WISE2320p1448_total_variable_SURF2022.pdf}
\figsetgrpnote{The light curve of WISE2320p1448 with \textit{Spitzer} ch2$_{aperture}$, ch2$_{PRF}$, and \textit{WISE} W2$_{PSF}$ photometry. The solid lines are the medians for each photometry type, while the dashed lines are $\pm$3$\sigma$. The spectral type is on the scale of 0 is M0, 10 is L0, 20 is T0, and 30 is Y0. }
\figsetgrpend

\figsetgrpstart
\figsetgrpnum{8.347}
\figsetgrptitle{Light Curve for WISE2321p1354}
\figsetplot{WISE2321p1354_total_variable_SURF2022.pdf}
\figsetgrpnote{The light curve of WISE2321p1354 with \textit{Spitzer} ch2$_{aperture}$, ch2$_{PRF}$, and \textit{WISE} W2$_{PSF}$ photometry. The solid lines are the medians for each photometry type, while the dashed lines are $\pm$3$\sigma$. The spectral type is on the scale of 0 is M0, 10 is L0, 20 is T0, and 30 is Y0. }
\figsetgrpend

\figsetgrpstart
\figsetgrpnum{8.348}
\figsetgrptitle{Light Curve for WISE2325p4251}
\figsetplot{WISE2325p4251_total_variable_SURF2022.pdf}
\figsetgrpnote{The light curve of WISE2325p4251 with \textit{Spitzer} ch2$_{aperture}$, ch2$_{PRF}$, and \textit{WISE} W2$_{PSF}$ photometry. The solid lines are the medians for each photometry type, while the dashed lines are $\pm$3$\sigma$. The spectral type is on the scale of 0 is M0, 10 is L0, 20 is T0, and 30 is Y0. }
\figsetgrpend

\figsetgrpstart
\figsetgrpnum{8.349}
\figsetgrptitle{Light Curve for WISE2326p0201}
\figsetplot{WISE2326p0201_total_variable_SURF2022.pdf}
\figsetgrpnote{The light curve of WISE2326p0201 with \textit{Spitzer} ch2$_{aperture}$, ch2$_{PRF}$, and \textit{WISE} W2$_{PSF}$ photometry. The solid lines are the medians for each photometry type, while the dashed lines are $\pm$3$\sigma$. The spectral type is on the scale of 0 is M0, 10 is L0, 20 is T0, and 30 is Y0. }
\figsetgrpend

\figsetgrpstart
\figsetgrpnum{8.350}
\figsetgrptitle{Light Curve for WISE2327m2730}
\figsetplot{WISE2327m2730_total_variable_SURF2022.pdf}
\figsetgrpnote{The light curve of WISE2327m2730 with \textit{Spitzer} ch2$_{aperture}$, ch2$_{PRF}$, and \textit{WISE} W2$_{PSF}$ photometry. The solid lines are the medians for each photometry type, while the dashed lines are $\pm$3$\sigma$. The spectral type is on the scale of 0 is M0, 10 is L0, 20 is T0, and 30 is Y0. }
\figsetgrpend

\figsetgrpstart
\figsetgrpnum{8.351}
\figsetgrptitle{Light Curve for WISE2331m4718}
\figsetplot{WISE2331m4718_total_variable_SURF2022.pdf}
\figsetgrpnote{The light curve of WISE2331m4718 with \textit{Spitzer} ch2$_{aperture}$, ch2$_{PRF}$, and \textit{WISE} W2$_{PSF}$ photometry. The solid lines are the medians for each photometry type, while the dashed lines are $\pm$3$\sigma$. The spectral type is on the scale of 0 is M0, 10 is L0, 20 is T0, and 30 is Y0. }
\figsetgrpend

\figsetgrpstart
\figsetgrpnum{8.352}
\figsetgrptitle{Light Curve for WISE2332m4325}
\figsetplot{WISE2332m4325_total_variable_SURF2022.pdf}
\figsetgrpnote{The light curve of WISE2332m4325 with \textit{Spitzer} ch2$_{aperture}$, ch2$_{PRF}$, and \textit{WISE} W2$_{PSF}$ photometry. The solid lines are the medians for each photometry type, while the dashed lines are $\pm$3$\sigma$. The spectral type is on the scale of 0 is M0, 10 is L0, 20 is T0, and 30 is Y0. }
\figsetgrpend

\figsetgrpstart
\figsetgrpnum{8.353}
\figsetgrptitle{Light Curve for WISE2339p1352}
\figsetplot{WISE2339p1352_total_variable_SURF2022.pdf}
\figsetgrpnote{The light curve of WISE2339p1352 with \textit{Spitzer} ch2$_{aperture}$, ch2$_{PRF}$, and \textit{WISE} W2$_{PSF}$ photometry. The solid lines are the medians for each photometry type, while the dashed lines are $\pm$3$\sigma$. The spectral type is on the scale of 0 is M0, 10 is L0, 20 is T0, and 30 is Y0. }
\figsetgrpend

\figsetgrpstart
\figsetgrpnum{8.354}
\figsetgrptitle{Light Curve for WISE2343m7418}
\figsetplot{WISE2343m7418_total_variable_SURF2022.pdf}
\figsetgrpnote{The light curve of WISE2343m7418 with \textit{Spitzer} ch2$_{aperture}$, ch2$_{PRF}$, and \textit{WISE} W2$_{PSF}$ photometry. The solid lines are the medians for each photometry type, while the dashed lines are $\pm$3$\sigma$. The spectral type is on the scale of 0 is M0, 10 is L0, 20 is T0, and 30 is Y0. }
\figsetgrpend

\figsetgrpstart
\figsetgrpnum{8.355}
\figsetgrptitle{Light Curve for WISE2344m0733}
\figsetplot{WISE2344m0733_total_variable_SURF2022.pdf}
\figsetgrpnote{The light curve of WISE2344m0733 with \textit{Spitzer} ch2$_{aperture}$, ch2$_{PRF}$, and \textit{WISE} W2$_{PSF}$ photometry. The solid lines are the medians for each photometry type, while the dashed lines are $\pm$3$\sigma$. The spectral type is on the scale of 0 is M0, 10 is L0, 20 is T0, and 30 is Y0. }
\figsetgrpend

\figsetgrpstart
\figsetgrpnum{8.356}
\figsetgrptitle{Light Curve for WISE2344p1034}
\figsetplot{WISE2344p1034_total_variable_SURF2022.pdf}
\figsetgrpnote{The light curve of WISE2344p1034 with \textit{Spitzer} ch2$_{aperture}$, ch2$_{PRF}$, and \textit{WISE} W2$_{PSF}$ photometry. The solid lines are the medians for each photometry type, while the dashed lines are $\pm$3$\sigma$. The spectral type is on the scale of 0 is M0, 10 is L0, 20 is T0, and 30 is Y0. }
\figsetgrpend

\figsetgrpstart
\figsetgrpnum{8.357}
\figsetgrptitle{Light Curve for WISE2349p3458}
\figsetplot{WISE2349p3458_total_variable_SURF2022.pdf}
\figsetgrpnote{The light curve of WISE2349p3458 with \textit{Spitzer} ch2$_{aperture}$, ch2$_{PRF}$, and \textit{WISE} W2$_{PSF}$ photometry. The solid lines are the medians for each photometry type, while the dashed lines are $\pm$3$\sigma$. The spectral type is on the scale of 0 is M0, 10 is L0, 20 is T0, and 30 is Y0. }
\figsetgrpend

\figsetgrpstart
\figsetgrpnum{8.358}
\figsetgrptitle{Light Curve for WISE2354p0240}
\figsetplot{WISE2354p0240_total_variable_SURF2022.pdf}
\figsetgrpnote{The light curve of WISE2354p0240 with \textit{Spitzer} ch2$_{aperture}$, ch2$_{PRF}$, and \textit{WISE} W2$_{PSF}$ photometry. The solid lines are the medians for each photometry type, while the dashed lines are $\pm$3$\sigma$. The spectral type is on the scale of 0 is M0, 10 is L0, 20 is T0, and 30 is Y0. }
\figsetgrpend

\figsetgrpstart
\figsetgrpnum{8.359}
\figsetgrptitle{Light Curve for WISE2355p3804}
\figsetplot{WISE2355p3804_total_variable_SURF2022.pdf}
\figsetgrpnote{The light curve of WISE2355p3804 with \textit{Spitzer} ch2$_{aperture}$, ch2$_{PRF}$, and \textit{WISE} W2$_{PSF}$ photometry. The solid lines are the medians for each photometry type, while the dashed lines are $\pm$3$\sigma$. The spectral type is on the scale of 0 is M0, 10 is L0, 20 is T0, and 30 is Y0. }
\figsetgrpend

\figsetgrpstart
\figsetgrpnum{8.360}
\figsetgrptitle{Light Curve for WISE2356m4814}
\figsetplot{WISE2356m4814_total_variable_SURF2022.pdf}
\figsetgrpnote{The light curve of WISE2356m4814 with \textit{Spitzer} ch2$_{aperture}$, ch2$_{PRF}$, and \textit{WISE} W2$_{PSF}$ photometry. The solid lines are the medians for each photometry type, while the dashed lines are $\pm$3$\sigma$. The spectral type is on the scale of 0 is M0, 10 is L0, 20 is T0, and 30 is Y0. }
\figsetgrpend

\figsetgrpstart
\figsetgrpnum{8.361}
\figsetgrptitle{Light Curve for WISE2357p1227}
\figsetplot{WISE2357p1227_total_variable_SURF2022.pdf}
\figsetgrpnote{The light curve of WISE2357p1227 with \textit{Spitzer} ch2$_{aperture}$, ch2$_{PRF}$, and \textit{WISE} W2$_{PSF}$ photometry. The solid lines are the medians for each photometry type, while the dashed lines are $\pm$3$\sigma$. The spectral type is on the scale of 0 is M0, 10 is L0, 20 is T0, and 30 is Y0. }
\figsetgrpend

\figsetend

\section{New Technique for Identifying Binary Brown Dwarf Candidates}\label{binaries}
As discussed in \S\ref{trust} we use two different types of flux measurements for the \textit{Spitzer} ch2 photometry, both aperture and PRF-fit. A disadvantage of the aperture magnitude is that it can become cross contaminated from a secondary component, background source, or cosmic ray hit. The \textit{Spitzer} ch2 PRF-fit code, on the other hand, can attempt to fit the secondary, remove the background source, or ignore the cosmic ray to better estimate the true apparent magnitude for the object. However, the PRF-fit magnitude can be biased if the object is a partly resolved physical binary (within the MOPEX FWHM; later discussed in \S\ref{disbinary}) or a blend with a distant stellar object or extragalactic source. 

Our binary identification technique is a close variant of the widely used star-galaxy separation technique that interprets an excess of aperture flux relative to PSF flux as an indication of resolved rather than pointlike morphology. Such an approach has been used by major optical surveys, including Pan-STARRS (e.g., \citealt{Schlafly et al.(2012)}).

Within \textit{Spitzer's} ch2 PRF fitting code, there is a procedure called ``passive deblending", in which the code tries to separate what is believed to be two, or more, sources and then tries to fit multiple PRFs to the detection (\citealt{Makovoz Marleau(2005)}). Once multiple PRFs are fit to the detection, those are subtracted to get uncertainties, $\chi^2$, and the S/N to compare to a single PRF-fit. MOPEX then determines whether the single or multiple PRF-fit performed better. If more than two components are fit simultaneously, the flux measurements can be drastically off (\citealt{Makovoz Marleau(2005)}). The code, therefore, only tries to fit one other component. This is done in a 7$\times$7 pixel area, $7\farcs67\times7\farcs67$, compared to a single PRF fit in a 5$\times$5 pixel area, $5\farcs48\times5\farcs48$. Passive deblending was enabled for all of the 361 brown dwarfs (\citealt{Kirkpatrick et al.(2021)}), and we used whichever fit had better metrics. 

\subsection{Aperture-PRF Comparison Technique}\label{binary}
For some of the 361 objects, a trend is seen when comparing ch2$_{aperture}$ to ch2$_{PRF}$. Specifically, all the PRF points are dimmer than the aperture measurements. Of 361 objects, 126 (35$\%$) show this effect. A possibility is that a blended secondary component, which is being appropriately measured in the aperture's radius, causes a poor PRF fit measuring only the primary component. (The unTimely data were not used in this test because there is only a PSF fit with no corresponding aperture measurement.)

The 126 objects showing this trend have been divided into four different classes:

\begin{itemize}
\item{Class 0: Includes only those objects for which passive deblending in the \textit{Spitzer} PRF-fit photometry was triggered 100$\%$ of the time and N$_{spitzer}$ is \textgreater 5 (for statistical robustness). This class includes 16 objects.}

\item{Class 1: Includes objects for which $\tilde{ch2}_{aperture} + \sigma_{\tilde{ch2}_{aperture}}$ \textless \space $\tilde{ch2}_{PRF} - \sigma_{\tilde{ch2}_{PRF}}$ and the three rules below apply. This class includes 11 objects.}

\item{Class 2: Includes objects for which $\tilde{ch2}_{aperture}$ \textless \space $\tilde{ch2}_{PRF} - \sigma_{\tilde{ch2}_{PRF}}$ and the three rules below apply. This class includes 35 objects.}

\item{Class 3: Includes objects for which the Class 1 and 2 restriction on $\tilde{ch2}_{aperture}$ do not apply, but the three rules below are still met. This class includes 64 objects.}

\end{itemize}

\begin{figure}
        \centering
        \includegraphics[scale = 0.29]{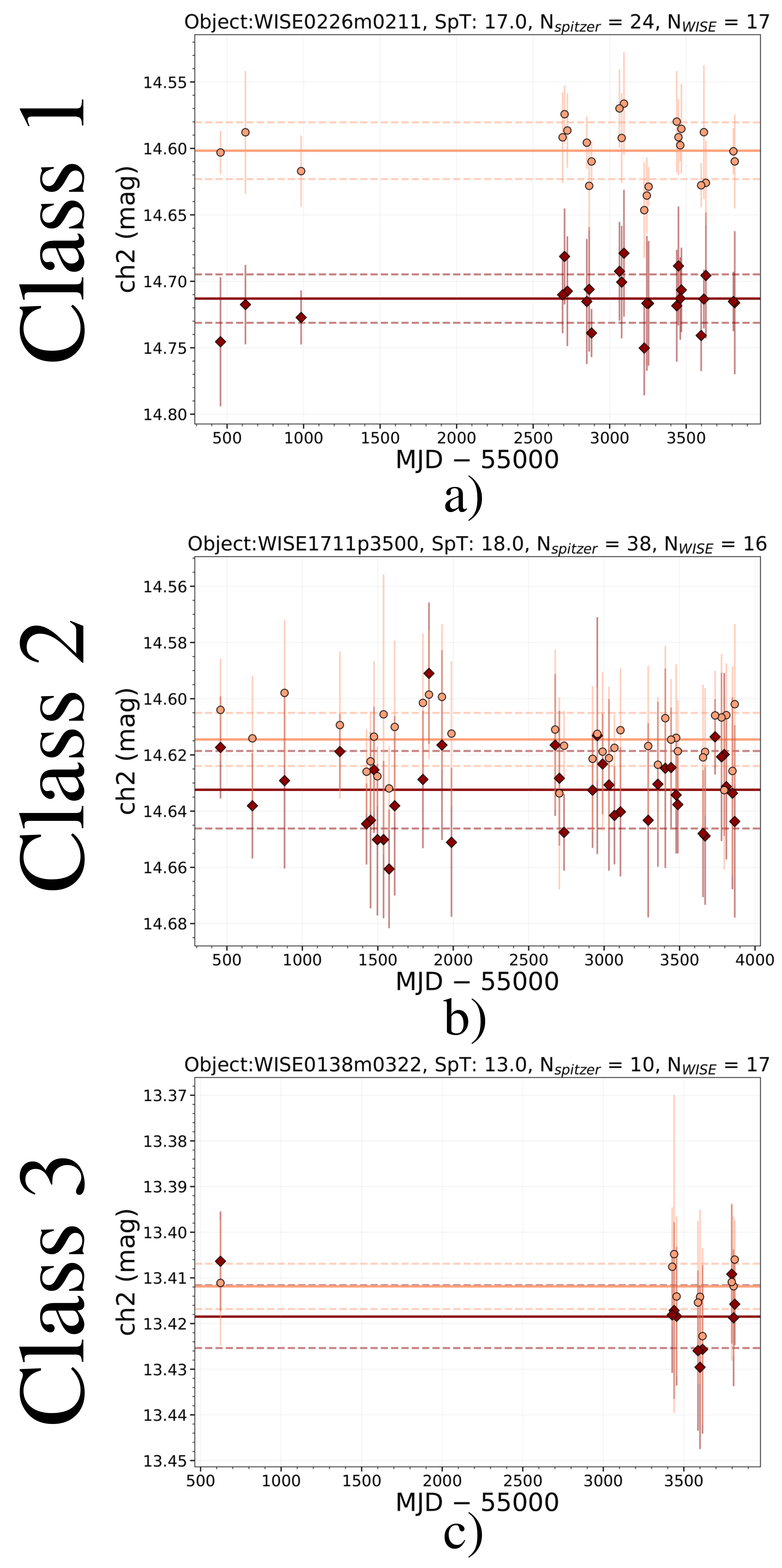}
        \caption{Light curves of three example objects. The x-axis is Modified Julian Date (MJD), with the y-axis being ch2$_{aperture}$ (orange circles) and ch2$_{PRF}$ (red diamonds). The objects in this plot are WISEPA J022623.98$-$021142.8 (subplot a), WISEPA J171104.60+350036.8 (subplot b), and WISEPC J013836.59$-$032221.2 (subplot c). The solid horizontal line is the median for each type of measurement, while the dashed lines show $\pm$1$\sigma$. All of these objects are known brown dwarf binaries. For Class 1 the lower orange dashed line lies brighter than the upper red dashed line. For Class 2 the orange solid line lies brighter than the upper red dashed line. In Class 3, the solid orange line falls below the upper red dashed line. For all three classes the other three rules, discussed in the text, also apply.}
        \label{Figure 8}
\end{figure}

For an object to fall in Class 1-3, it must have the following three characteristics:

\begin{itemize}
\item{\textit{Spitzer} aperture magnitude brighter than its PRF-fit magnitude at every epoch.}

\item{N$_{spitzer}$ \textgreater 5 (again for statistical robustness)}

\item{Passive deblending must have been triggered less than 100$\%$ of the time. }
\end{itemize}
The subplots in Figure \ref{Figure 8} show examples of Classes 1, 2, and 3. 

\begin{figure}
        \centering
        \includegraphics[scale = 0.18]{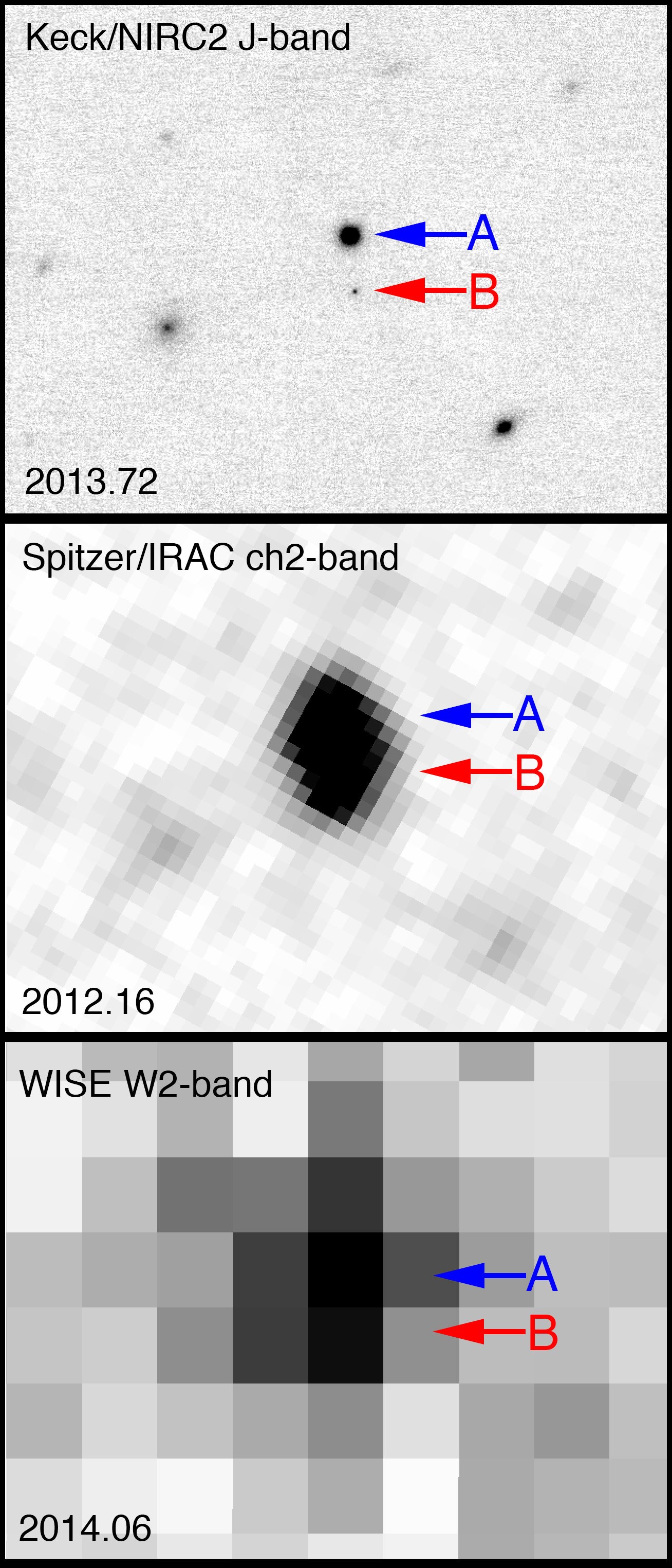}
        \caption{Images at different angular resolutions for the same object, WISE J022623.98$-$021142.8 (\citealt{Kirkpatrick et al.(2011)}). $Keck$/NIRC2 $J$ (top panel; \citealt{Gelino(2012)}), \textit{Spitzer}/IRAC ch2 (middle panel; \citealt{Kirkpatrick et al.(2021)}), and \textit{WISE}/unWISE W2 (bottom panel; \citealt{Lang(2014)}). The panels are 20$\times$30 arcsec with up being north and left being east. The primary is labeled with a blue ``A", while the secondary is a red ``B". See \S\ref{binary} for details.}
        \label{Figure 9}
\end{figure}

These four classes were considered for binary candidacy because a possible secondary could be affecting the flux measurements. This method has the potential to discover physical brown dwarf binaries with the components having the same distance, metallicity, and age. Figure \ref{Figure 9} shows various imaging of WISE J022623.98$-$021142.8 (\citealt{Kirkpatrick et al.(2011)}), which is known to have a secondary component. We see that the binary is clearly separated in $Keck$/NIRC2 $J$ (top panel; \citealt{Gelino(2012)}), while in \textit{Spitzer}/IRAC ch2 (middle panel; \citealt{Kirkpatrick et al.(2021)}) and \textit{WISE}/unTimely W2 (bottom panel; \citealt{Meisner et al.(2022)}) it is blended. Note that the secondary is significantly redder in $J-$ch2 than the primary. \cite{Kirkpatrick et al.(2021)} shows that for objects later than T5, the J$-$ch2 color quickly changes towards very large values. Figure 14d of the same paper shows that the decrease of M$_{ch2}$ with spectral type is less drastic. One can imagine a system like WISE J022623.98$-$021142.8 for which the secondary is even colder. In this case, the ground-based $J-$band image might not have the sensitivity to detect the companion, but at the 4.6$\mu$m wavelength of \textit{Spitzer}, the presence of the companion could be detected. Thus, this technique has the potential to identify new binaries in which the companion objects are too low in temperature to be found in ground-based, $J-$band follow-up. Additionally, it also provides a priority target list of nearby brown dwarfs that may not yet have seen any high-resolution imaging follow-up. For example, if a late-T or early-Y dwarf has a secondary candidate around it, there is a possibility it would be the first ever mid-late Y dwarf found.

For Figure \ref{Figure 6}, which compares ch2$_{PRF}$ and W2$_{PSF}$, we find that all the outliers were marked as Class 0. As illustrated in Figure \ref{Figure 9}, the poorer resolution of the \textit{WISE} data blends the secondary more thoroughly with the primary, meaning that the \textit{WISE} PSF fit is more successful at capturing the light of both components. In Figure \ref{Figure 7}, all but one outlier (mentioned in the caption) are identified as Class 1. This is because Class 1 objects are blended enough that a single PRF fit, though still poor, is not bad enough to trigger passive deblending, which would have made them Class 0 objects. Note that Classes 1-3 are numbered this way in order of the severity of the difference between the aperture and PRF-fit measurements. Class 0’s name results from its being a unique class compared to the rest.

\subsection{Possible Contamination by Subdwarfs}

The spectral energy distributions of subdwarfs show drastic differences compared to normal dwarfs within the wavelength ranges where they have been observed (\citealt{Lodieu et al.(2019)}). Subdwarfs are the metal-poor counterparts to normal metallicity brown dwarfs (\citealt{Lepine et al.(2003)}). This could affect how the point source looks on the detector. If there are large absorption bands in ch2/W2 that are drastically altering the effective wavelength through this bandpass, it would, because of the wavelength dependence of the MOPEX FWHM, change the full width half maximum value of the PRF. This difference between the expected PRF fit and the true PRF fit of a subdwarf would result in a high $\chi^2$ value. This is partly supported by the fact that one object in Class 3, 2MASS J06453153$-$6646120 (\citealt{Kirkpatrick et al.(2010)}), is an L subdwarf. We have compared a ``LOWZ"\footnote{https://doi.org/10.7910/DVN/SJRXUO} (\citealt{Meisner et al.(2021)}) spectral model of Z=$-$1.5 and T$_{eff}$=1300K to a solar-metallicity model at the same temperature. For the low-metallicity model, there is relatively more flux in the blue half of the W2 band compared to the solar-metallicity model. Based on this effect, we would expect a small change in the PRF width for the low-metallicity object, likely not large enough to support the hypothesis above. However, the models at these low-metallicities are completely untested in this wavelength regime. This effect can not be currently quantified because of the lack of observational spectra for subdwarfs at 4.6$\mu$m. Further data, particularly from \textit{JWST} (\citealt{Gardner et al.(2006)}), will be needed to test this hypothesis. Nonetheless, as subdwarfs are rare in a volume-limited census, it is unlikely that this effect is the cause for any other aperture/PRF discrepancies within our sample.

\subsection{Visual Inspection}\label{finder}


To better understand Classes 0, 1, 2, and 3, we made finder charts for each object that include images from the Digitized Sky Survey 1 and 2 $B$, $R$, and $IR$; the Pan-STARRS1 survey $g$, $r$, $i$, $z$, $y$; the UKIRT Infrared Deep Sky Survey $J$, $H$, and $K$ (\citealt{Lawrence et al.(2007)}); the UKIRT Hemisphere Survey $J$ (\citealt{Dye et al.(2018)}); the VISTA Hemisphere Survey $J$ and $K_{s}$ (\citealt{McMahon et al.(2021)}); the Two Micron All Sky Survey $J$, $H$, and $K_{s}$ (\citealt{Skrutskie et al.(2006)}); and \textit{WISE} W1, W2, W3, and W4. These finder charts span a long time baseline and show the background sky at different epochs to see if our target was contaminated by a distant source at the \textit{Spitzer} epochs. We also did a literature search for each object to check for binaries or subdwarfs. This led to the creation of 4 different subclasses: contaminated objects (C), potential binaries (P), known binaries (K), and subdwarfs (S).

\subsection{Checking for true and false positives}
Once the finder charts were made, it was then possible to search for an estimate of a true and false positive rate for the binary candidates. Although the finder charts provide a starting point for accessing false positives, high-resolution imaging is needed for more fully computing both the true and false positive rates. 

To estimate a true positive rate, we searched two high-resolution archives and the literature for known binaries in classes 0 through 3. We searched papers in SIMBAD \citep{Wenger et al.(2000)} to find published binary systems. The two high-resolution archives we selected are the Keck Observatory Archive (hereafter, KOA; {\url{https://koa.ipac.caltech.edu}}) and the \textit{Hubble Space Telescope}/NICMOS image archive studied in \cite{Factor Kraus(2022)}. The known binary systems found, 12 in total, are listed in Table \ref{Table K}. Comparing this to our K+P+S lists of 72 objects we find an estimated true positive rate of $\sim$17$\%$. The true positive rate is likely higher because there are other high-resolution archives that could be searched, and there are some objects that have yet to be observed at high-resolution. This exercise is left for a future paper.

For some of our objects, a real false positive rate is impossible without high-resolution images at the same wavelengths as our \textit{Spitzer} observations, 4.6$\mu$m. Consider a binary T8 (the most abundant spectral type in our sample) and Y1 at a distance of 10pc. Using Figures 16(a) and 16(d) from \cite{Kirkpatrick et al.(2021)}, we find apparent J magnitudes of $\sim$18 mag for the T8 and $\sim$24 mag for the Y1, and apparent ch2 magnitudes of $\sim$13.5 mag for the T8 and $\sim$16 mag for the Y1. These values of $\Delta$J$\approx$6 mag and J$\approx$24 mag would make the secondary difficult/impossible to image, even with adaptive-optics imaging from the ground (\citealt{Davies Kasper(2012)} and \citealt{Zhang et al.(2021)}). However, values of $\Delta$ch2$\approx$2.5 mag and ch2$\approx$16 mag are much easier with \textit{Spitzer} or \textit{JWST}. 

For earlier type primaries however, a false positive rate can be attempted because observations at $\sim$1$\mu$m can provide a $\Delta$ magnitude sufficient to rule out most secondaries detectable with our technique. Figure 4 of \cite{Gelino et al.(2011)} shows the contrast ratios achievable for typical high-resolution observations for brown dwarf binary pairs. These ratios depend on the separation between the two objects, the primary's brightness, and the dynamic range available for secondary detection. The contrast ratios start to plateau at binary separations of \textgreater$0\farcs2$. Since our ability to detect binaries is limited by the \textit{Spitzer}/IRAC ch2 FWHM of $1\farcs72$, we can assume this plateau value. It is typical for ground based and NASA/ESA \textit{Hubble Space Telescope} observations to have a depth of $\sim$22mag (\citealt{Lowrance et al.(2005)}, \citealt{Burgasser et al.(2006)}, and the Keck NIRC2 exposure time calculator\footnote{{\url{https://www2.keck.hawaii.edu/inst/nirc2/nirc2_snr_eff.html}}}). A typical contrast ratio at wide separations is $\Delta$H$\approx$4mag \citep{Gelino et al.(2011)}. At our average distance of 15.9pc, an apparent magnitude of H=22mag would correspond to an absolute H magnitude of 21mag, or a spectral type of Y0 (Figure 16(b) of \citealt{Kirkpatrick et al.(2021)}). A Y0 detection with $\Delta$H=4 mag is just achievable if the primary has a spectral type of T7.5. At primary spectral types earlier than this, a companion detectable via our \textit{Spitzer} technique would likely also be detectable via ground based or \textit{Hubble Space Telescope} observations.

In Table \ref{Table P} the final column, labeled ``High-Resolution Imaging", indicates whether our candidate binary has high-resolution imaging observations in KOA or in \cite{Factor Kraus(2022)}. There are 25 objects earlier than spectral type T7.5 with high-resolution imaging from these 2 archives. Using these data it is possible to estimate a false-positive rate. We will do this as follows. First, we compare our 25 false binaries to the total number of objects earlier than T7.5 in sub-classes K, P, and S (59) to provide a false positive rate of $\frac{25}{59}$ = 42$\%$. However, there are still 25 objects earlier than T7.5 not in Table \ref{Table K} and lacking high-resolution imaging in Table \ref{Table P}, so this value is actually a (loose) lower bound for the false-positive rate. Second, we instead compare the same 25 false binaries to the subset of objects earlier than T7.5 with high-resolution imaging observations in Tables \ref{Table P} and \ref{Table K}. This gives a new false-positive rate of $\frac{25}{34}$ = 74$\%$. However, this is likely an upper bound because the faintest companions to our 25 ``false" binaries may not be detectable by our ground-based observations. In reality, we must wait for \textit{JWST} 4.6$\mu$m imaging to calculate a real false-positive rate, as an object lacking a companion detection at J-band might yet be harboring a cold companion seen only at longer wavelengths. 

\section{Discussion}\label{dis}
\subsection{Variability}\label{defvari}
\begin{figure}
        \centering
        \includegraphics[scale = 0.28]{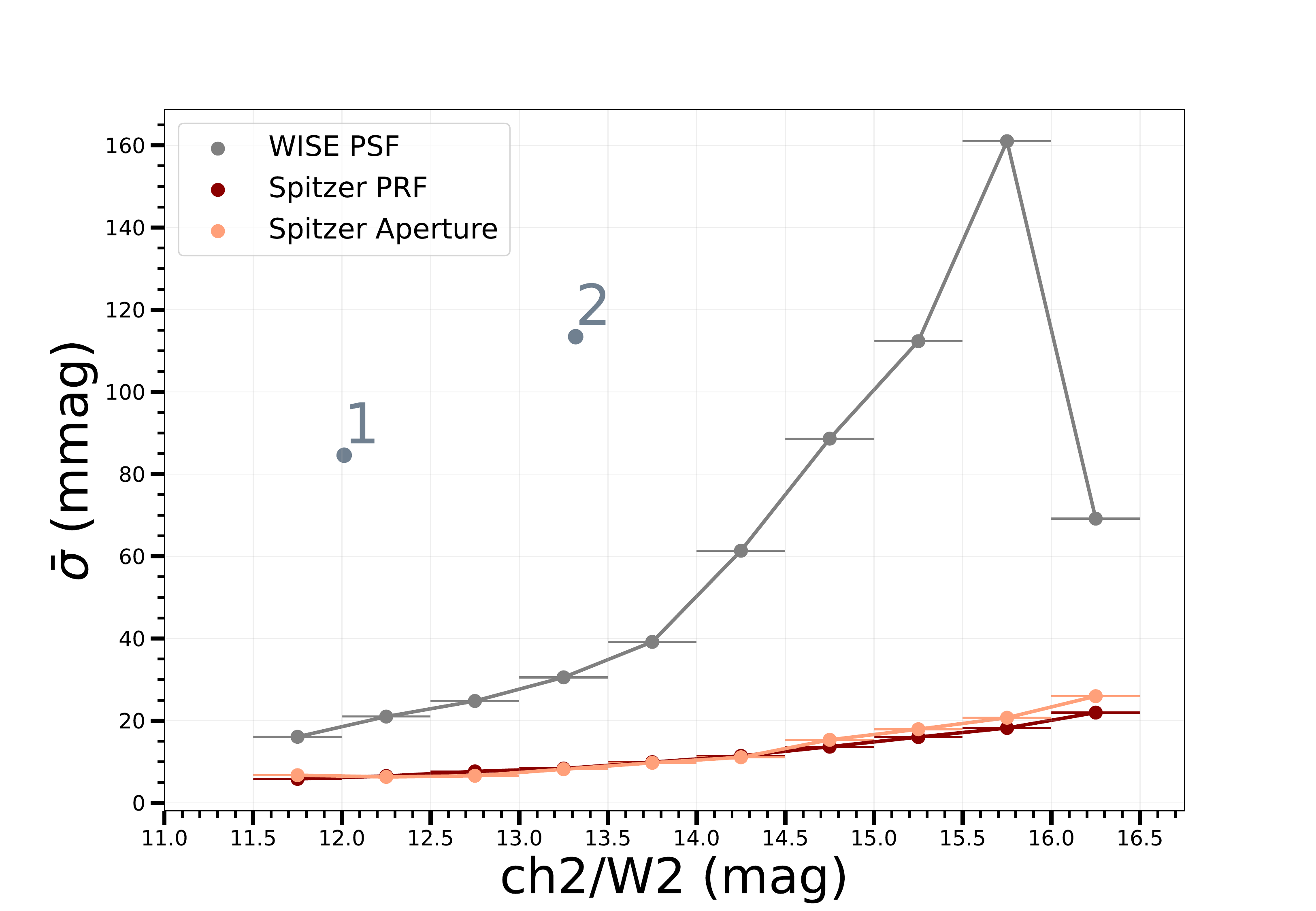}
        \caption{Photometric precision as a function of magnitude. The x-axis shows ch2$_{aperture}$, ch2$_{PRF}$, and W2$_{PSF}$ photometry in 0.5-mag bins, while the y-axis is the median uncertainties between all photometry points ($\sigma$) for each bin. Points 1 and 2 are known variables 2MASS J21392676+0220226 and 2MASS J22282889$-$4310262 (\citealt{Yang et al.(2016)}), respectively. The drop in \textit{WISE} sigma at 16-16.5 mag is a result of 4 out of 6 objects in this bin having lower than average standard deviations.}
        \label{Figure 10}
\end{figure}

Figure \ref{Figure 10} shows the median of the standard deviation as a function of ch2/W2 magnitude in 0.5-mag intervals. If an object has variability above the line, our analysis would have detected it. For example, if an object has a magnitude of 14.2, no variability can be detected below $\sim$12mmag (1$\sigma$) for the \textit{Spitzer} photometry or $\sim$60mmag (1$\sigma$) for \textit{WISE} photometry. Points 1 and 2 are known variable objects 2MASS J21392676+0220226 and 2MASS J22282889$-$4310262 (\citealt{Yang et al.(2016)}), respectively (not included in this study). Number 1 and 2 can be seen to be well above all three lines, which indicates that variability could be found in the unTimely photometry and easily detected in our \textit{Spitzer} photometry. Specifically, for 2MASS J21392676+0220226 we were able to find 14.5$\%$ variations in the unTimely W2 photometry, which is different from what is seen in \cite{Yang et al.(2016)} which found \textit{Spitzer} ch2 peak-to-peak variations of 26$\%$. Furthermore, this plot indicates that the unTimely data will only detect variability if it has relatively large amplitude, while the \textit{Spitzer} data should be able to detect much smaller variations. If we instead looked at the magnitude percentage variations, we are sensitive to variability of \textgreater 4$\%$ at 11.5 mag and \textgreater 13$\%$ at 16 mag, for the unTimely photometry. 

This study has shown no variability in brown dwarfs at $\sim$4.6$\mu$m over a time period of \textgreater10 years to the sensitivity limits shown in Figure \ref{Figure 10}. The large-scale variability to which our survey is sensitive has only been seen in a small number of published objects. This illustrates that large-amplitude variability is rare among brown dwarfs. At smaller amplitudes, however, \cite{Yang et al.(2016)}) found an increase in the percentage of objects with variability near the L/T transition compared to those in the T/Y transition. Figure \ref{Figure 3} shows that we have excellent statistics at the T/Y transition, yet we find no large-amplitude variability there. 

\subsection{Binarity}\label{disbinary}
Brown dwarf binaries are rare and hard to find. Two common ways to find brown dwarf binaries are obtaining a spectrum of a clear spectral composite object (\citealt{Bardalez Gagliuffi et al.(2014)}) or visually separating the components via imaging. Our Aperture-PRF technique provides a new way of finding these binaries, in much the same way that the PRF technique alone, through passive deblending, has previously been used to find secondary components (e.g., in circumstellar disks; \citealt{Martinez Kraus(2019)}). 

\begin{figure}
        \centering
        \includegraphics[scale = 0.3]{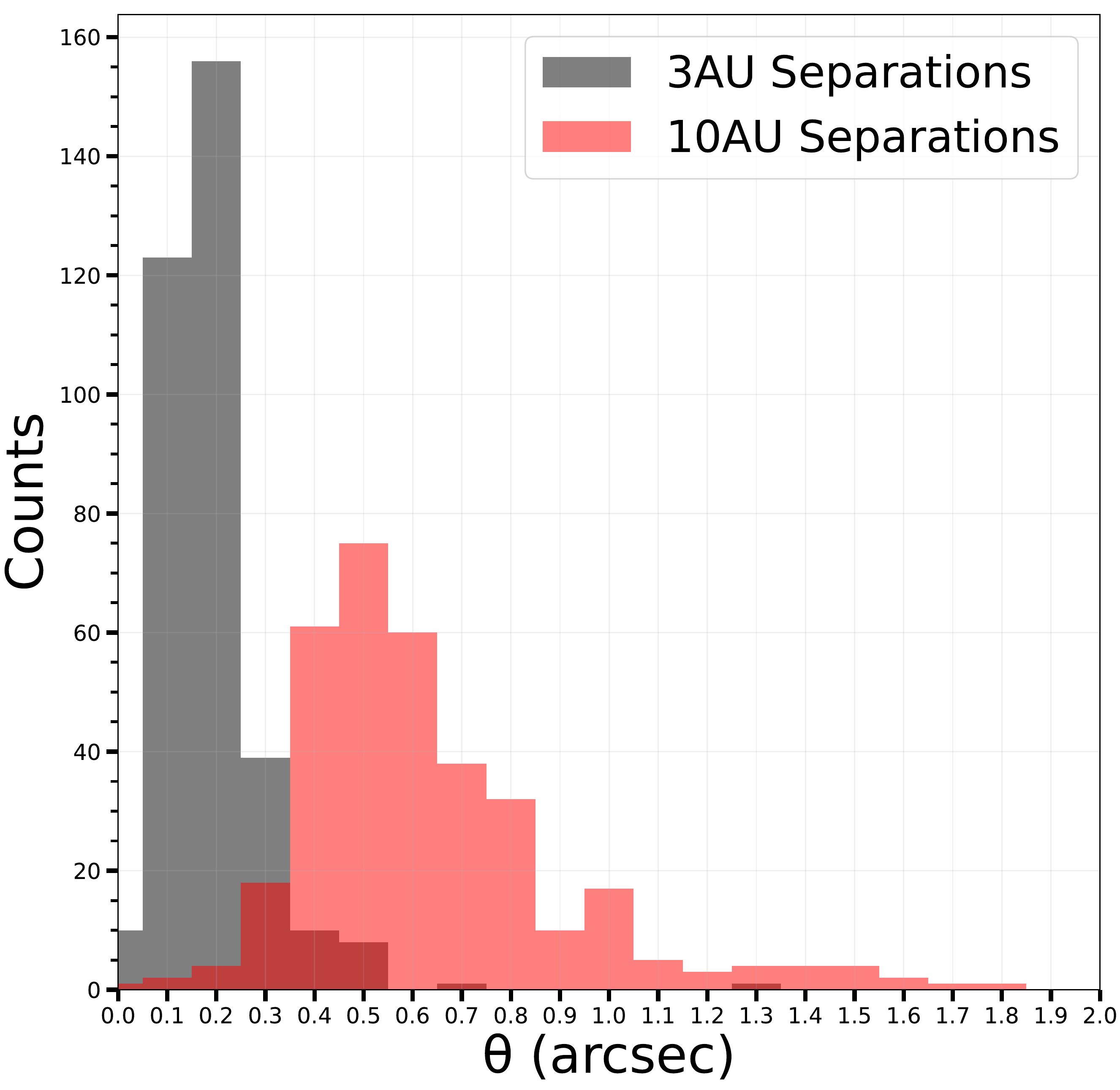}
        \caption{The distribution of angular separation for the 361 objects in our sample assuming a binary with 3 AU physical separation (black) or 10 AU physical separation (red). The 3 AU distribution peaks at $0\farcs$2 and the one at 10 AU peaks at $0\farcs$5.}
        \label{Figure 1888}
\end{figure}

We now compare the MOPEX FWHM of \textit{Spitzer} to the average separation of known nearby brown dwarf binaries, to test our binary hypothesis. The MOPEX FWHM of \textit{Spitzer} is $1\farcs72$ at ch2 (4.6$\mu$m). Figure 2 of \cite{Burgasser et al.(2007)} shows that the average brown dwarf binary has a physical separation between $\sim$3 $-$ 10 AU. When considering the average distance for our 361 objects is 15.9pc, this leads to an expected separation between $0\farcs19$ $-$ $0\farcs63$. The distribution of the 361 objects angular separation based on distance assuming 3 and 10AU binary separation can be seen in Figure \ref{Figure 1888}. We see that the objects clump around the predicted separations. However, there are a few objects that are outside of this separation, indicating that they could be angularly separated in the \textit{Spitzer} photometry. Objects of Class 0 must have separation \textgreater$1\farcs72$ in order for the code to passively deblend the object into two different PRF fits. This separation means that Class 0 is very unlikely to have physical binaries. Objects of Class 1, 2, and 3 are only marginally resolved or are within the MOPEX FWHM, meaning that Class 1-3 are the ones most likely to have hidden, faint secondaries.

\begin{table}

\caption{Class/Subclass Statistics}
\resizebox{8cm}{!}{
\begin{tabular}{l|llll|l}\label{table 2}
              & C  & K  & P  & S & Total \\
\hline
Class 0       & 10 & 1  & 5  & 0 & 16    \\
Class 1       & 3  & 3  & 5  & 0 & 11    \\
Class 2       & 18 & 3  & 14 & 0 & 35    \\
Class 3       & 23 & 5  & 35 & 1 & 64    \\
\hline
Total         & 54 & 12 & 59 & 1 & 126  
\end{tabular}
}
\begin{center}
\footnotesize{Reference Codes: (C) Contaminated objects, (K) Known binaries, (P) Potential binaries, and (S) Subdwarfs}
\end{center}
\end{table}

In Table \ref{table 2} we show how many objects fall in each class and subclass as defined in \S\ref{binary} and \S\ref{finder}. Details of all 126 objects can be found in Tables \ref{Table C}, \ref{Table P}, and \ref{Table K}. There was only one subdwarf found in this process, 2MASS J06453153$-$6646120, listed in Table \ref{Table K} with the known binaries.

We compare known binary/contaminated statistics below for each class: 
\begin{itemize}
    \item{Class 0: Only 1 known binary was found. Most known brown dwarf binaries have separation smaller than the separations at which passive deblending ($1\farcs33$ to $7\farcs67$) would take place. Contamination by background objects is the highest among all four classes ($\frac{10}{16}$ = 63$\%$).} 
    
    \item{Class 1: This is the highest percentage of known binaries of all four groups ($\frac{3}{11}$ = 27$\%$). The true binaries in this class are likely to have the smallest $\Delta$mag and/or largest separations, because they have the most discrepant aperture and PRF-fit measurements. Class 1 also has a significant percentage of background contamination ($\frac{3}{11}$ = 27$\%$).}
    
    \item{Class 2: This class has a much smaller percentage of known binaries ($\frac{3}{35}$ = 9$\%$). True binaries here will likely have slightly larger $\Delta$mag and smaller separations. Contamination by background objects is high ($\frac{18}{35}$ = 51$\%$). }
    
    \item{Class 3: This class has the smallest percentage of known binaries ($\frac{5}{64}$ = 8$\%$) and sizable percentage of background contamination ($\frac{23}{64}$ = 36$\%$). Given the smaller difference between aperture and PRF-fit measures, this group is likely to have more spurious binaries than the other classes, but could also have binaries with the tightest separation and/or largest $\Delta$mags. }
\end{itemize}

\section{Conclusion}\label{conclusion}
We examined two 4.6$\mu$m photometric datasets, \textit{Spitzer}/IRAC ch2 and unTimely W2, to analyze long-term variability in brown dwarfs. Using the \textit{Spitzer}/IRAC ch2 photometry, our study establishes that the typical ch2 variability on $\sim$10 year time scales is \textless8mmag for objects at ch2 = 11.5 mag and \textless 22mmag for objects at ch2 = 16 mag. Using the unTimely photometry, which has higher uncertainties than \textit{Spitzer}/IRAC ch2, we find that all variations must be \textless 4.5$\%$ at 11.5 mag and \textless 12.5$\%$ at 16 mag. Although a handful of objects in the literature have been reported with variations greater than this, we conclude that higher-amplitude variability must be the exception rather than the rule. 

We also demonstrate a new technique for the identification of brown dwarf binary candidates that uses a comparison of \textit{Spitzer's} aperture and PRF-fit photometry. We present a table of 59 previously unrecognized candidate binaries (Table \ref{Table P}) that constitute prime targets for high-resolution follow-up imaging. 

\section{Acknowledgement}
This Backyard Worlds research was supported by NASA grant 2017-ADAP17-0067. We thank the Student Astrophysics Society\footnote{\url{https://www.studentastrophysicssociety.com}} for providing the resources that enabled the pairing of high school and undergraduate students with practicing astronomers. This work makes use of data products from WISE/NEOWISE, which is a joint project of UCLA and JPL/Caltech, funding by NASA.

Work in this paper is based on observations made with the Spitzer Space Telescope, which is operated by JPL/Caltech, under a contract with NASA. Support for the original parallax work was provided to J.D.K. by NASA through a Cycle 14 award issued by JPL/Caltech. Some data presented here were obtained at the W. M. Keck Observatory, which is operated as a scientific partnership among Caltech, the University of California, and NASA. The authors wish to recognize and acknowledge the very significant cultural role and reverence that the summit of Maunakea has always had within the indigenous Hawaiian community. We are most fortunate to have the opportunity to conduct observations from this mountain. The Observatory was made possible by the generous financial support of the W. M. Keck Foundation. This research has made use of the Keck Observatory Archive (KOA), which is operated by the W. M. Keck Observatory and the NASA Exoplanet Science Institute (NExScI), under contract with the National Aeronautics and Space Administration. This research has made use of the NASA/IPAC Infrared Science Archive, which is funded by the National Aeronautics and Space Administration and operated by the California Institute of Technology. This research has made use of the SIMBAD database, operated at CDS, Strasbourg, France 

The Digitized Sky Survey was produced at the Space Telescope Science Institute under U.S. Government grant NAG W-2166. The images of these surveys are based on photographic data obtained using the Oschin Schmidt Telescope on Palomar Mountain and the UK Schmidt Telescope. The plates were processed into the present compressed digital form with the permission of these institutions. Our finder charts also used observations obtained as part of the VISTA Hemisphere Survey, ESO Progam, 179.A-2010 (PI: McMahon). The UHS is a partnership between the UK STFC, The University of Hawaii, The University of Arizona, Lockheed Martin and NASA.

The Pan-STARRS1 Surveys (PS1) and the PS1 public science archive have been made possible through contributions by the Institute for Astronomy, the University of Hawaii, the Pan-STARRS Project Office, the Max-Planck Society and its participating institutes, the Max Planck Institute for Astronomy, Heidelberg and the Max Planck Institute for Extraterrestrial Physics, Garching, The Johns Hopkins University, Durham University, the University of Edinburgh, the Queen's University Belfast, the Harvard-Smithsonian Center for Astrophysics, the Las Cumbres Observatory Global Telescope Network Incorporated, the National Central University of Taiwan, the Space Telescope Science Institute, the National Aeronautics and Space Administration under Grant No. NNX08AR22G issued through the Planetary Science Division of the NASA Science Mission Directorate, the National Science Foundation Grant No. AST-1238877, the University of Maryland, Eotvos Lorand University (ELTE), the Los Alamos National Laboratory, and the Gordon and Betty Moore Foundation.

\begin{center}

\begin{deluxetable}{lllll}
\tabletypesize{\scriptsize}
\caption{Contaminated Objects}\label{Table C}
\tablehead{
\colhead{Object Name}     & \colhead{RA (deg)}  & \colhead{Dec (deg)}    & \colhead{Class} & \colhead{Subclass} \\
}
\startdata
WISE J004945.61+215120.0     & 12.439581 & 21.855476    & 3       & C        \\
2MASSW J0051107$-$154417     & 12.79513  & $-$15.73814  & 2       & C        \\
WISEA J005811.69$-$565332.1  & 14.54965  & $-$56.892303 & 3       & C        \\
2MASSI J0103320+193536       & 15.88506  & 19.593263    & 3       & C        \\
CWISEP J010527.69$-$783419.3 & 16.370636 & $-$78.572554 & 0 & C        \\
CWISE J014837.51$-$104805.6  & 27.15694  & $-$10.801799 & 2       & C        \\
WISEPA J031325.96+780744.2   & 48.359129 & 78.129099    & 2       & C        \\
WISE J032301.86+562558.0     & 50.75963  & 56.431894    & 3       & C        \\
WISEPC J033349.34$-$585618.7 & 53.457085 & $-$58.939765 & 3       & C        \\
WISE J033515.01+431045.1     & 53.813599 & 43.177629    & 3       & C        \\
2MASS J03400942$-$6724051    & 55.034396 & $-$67.398518 & 1       & C        \\
2MASS J03582255$-$4116060    & 59.594508 & $-$41.267803 & 2       & C        \\
2MASS J04210718$-$6306022    & 65.281665 & $-$63.099437 & 2       & C        \\
2MASSI J0443058$-$320209     & 70.77379  & $-$32.034919 & 0 & C        \\
PSO J076.7092+52.6087        & 76.709485 & 52.608231    & 0 & C        \\
2MASSI J0512063$-$294954     & 78.026636 & $-$29.831269 & 2       & C        \\
WISE J052126.29+102528.4     & 80.360194 & 10.423512    & 0 & C        \\
CWISEP J063428.10+504925.9   & 98.618859 & 50.822681    & 2       & C        \\
WISE J064205.58+410155.5     & 100.52362 & 41.031459    & 3       & C        \\
2MASS J07414279$-$0506464    & 115.42829 & $-$5.11289   & 0 & C        \\
ULAS J074502.79+233240.3     & 116.26163 & 23.54453     & 0 & C        \\
WISE J105130.01$-$213859.7   & 162.87537 & $-$21.650142 & 2       & C        \\
SIMP J11322058$-$3809562     & 173.08717 & $-$38.166428 & 3       & C        \\
WISE J125715.90+400854.2     & 194.31615 & 40.14811     & 2       & C        \\
2MASS J14075361+1241099      & 211.97158 & 12.686486    & 3       & C        \\
CWISEP J145837.91+173450.1   & 224.65736 & 17.580799    & 0 & C        \\
2MASSI J1526140+204341       & 231.55718 & 20.726056    & 2       & C        \\
CWISEP J153859.39+482659.1   & 234.74805 & 48.44905     & 3       & C        \\
WISEA J162341.27$-$740230.4  & 245.92    & $-$74.04263  & 2       & C        \\
WISE J163940.86$-$684744.6   & 249.92274 & $-$68.798677 & 3       & C        \\
PSO J258.2413+06.7612        & 258.24107 & 6.761227     & 1       & C        \\
WISEA J173453.90$-$481357.9  & 263.72383 & $-$48.233351 & 2       & C        \\
WISEA J173551.56$-$820900.3  & 263.96544 & $-$82.150738 & 3       & C        \\
WISE J174102.78$-$464225.5   & 265.26168 & $-$46.707965 & 2       & C        \\
WISE J175510.28+180320.2     & 268.79212 & 18.056078    & 3       & C        \\
WISE J180952.53$-$044812.5   & 272.46884 & $-$4.804453  & 1       & C        \\
WISEA J181849.59$-$470146.9  & 274.70665 & $-$47.030892 & 0 & C        \\
CWISE J183207.94$-$540943.3  & 278.03261 & $-$54.162757 & 3       & C        \\
WISE J185101.83+593508.6     & 282.75777 & 59.586189    & 3       & C        \\
WISE J191915.54+304558.4     & 289.81597 & 30.767298    & 3       & C        \\
2MASS J19251275+0700362      & 291.30355 & 7.011064     & 0 & C        \\
WISENF J193656.08+040801.2   & 294.2323  & 4.131147     & 2       & C        \\
WISE J200050.19+362950.1     & 300.20898 & 36.497832    & 3       & C        \\
WISE J201404.13+042408.5     & 303.51622 & 4.40277      & 2       & C        \\
WISEA J201833.67$-$141720.3  & 304.64037 & $-$14.288602 & 3       & C        \\
2MASS J21265916+7617440      & 321.76234 & 76.29985     & 2       & C        \\
2MASS J21543318+5942187      & 328.637   & 59.70326     & 3       & C        \\
WISE J223617.59+510551.9     & 339.07618 & 51.098434    & 3       & C        \\
WISEP J223937.55+161716.2    & 339.90682 & 16.288033    & 3       & C        \\
2MASS J22490917+3205489      & 342.29281 & 32.095967    & 2       & C        \\
WISEA J230228.66$-$713441.7  & 345.61789 & $-$71.579123 & 3       & C        \\
2MASS J23254530+4251488      & 351.43842 & 42.861899    & 0 & C        \\
2MASS J23312378$-$4718274    & 352.84951 & $-$47.30787  & 3       & C        \\
2MASSI J2339101+135230       & 354.79477 & 13.869448    & 2       & C \\
\enddata
\end{deluxetable}

\begin{deluxetable}{lllccc}
\tabletypesize{\scriptsize}
\tablecaption{Potential Binaries}\label{Table P}    
\tablehead{
\colhead{Object Name} & \colhead{RA (deg)} & \colhead{Dec (deg)}    & \colhead{Class}   & \colhead{Subclass} & \colhead{High-Resolution Imaging\tablenotemark{b}} \\
}
\startdata
WISE J000517.48+373720.5\tablenotemark{c}     & 1.3243156 & 37.622133    & 3       & P  & Keck      \\
2MASS J00345157+0523050      & 8.7173945 & 5.385477     & 3       & P  & Keck      \\
WISE J011154.36$-$505343.2   & 17.975789 & $-$50.895856 & 3       & P        \\
2MASS J01550354+0950003      & 28.76651  & 9.832968     & 0 & P  & HST      \\
WISEA J020047.29$-$510521.4  & 30.197571 & $-$51.089537 & 3       & P        \\
2MASSW J0205034+125142       & 31.26636  & 12.86154     & 2       & P        \\
WISEA J030237.53$-$581740.3\tablenotemark{c}  & 45.656221 & $-$58.294742 & 3       & P        \\
WISEA J030919.70$-$501614.2  & 47.333383 & $-$50.270291 & 3       & P        \\
2MASS J03101401$-$2756452    & 47.55754  & $-$27.946302 & 2       & P     & Keck   \\
2MASS J03185403$-$3421292    & 49.727839 & $-$34.357738 & 1       & P     & HST   \\
SDSSp J033035.13$-$002534.5  & 52.64629  & $-$0.42628   & 3       & P        \\
PSO J052.7214$-$03.8409      & 52.72142  & $-$3.84092   & 3       & P   & Keck     \\
WISE J033651.90+282628.8     & 54.216403 & 28.44105     & 3       & P        \\
2MASSW J0337036$-$175807     & 54.26496  & $-$17.96886  & 2       & P  & Keck\\
WISE J045746.08$-$020719.2   & 74.442276 & $-$2.122342  & 3       & P  & Keck      \\
2MASS J06020638+4043588      & 90.528129 & 40.732372    & 3       & P  & Keck      \\
2MASS J07555430$-$3259589    & 118.97545 & $-$32.998815 & 3       & P        \\
SDSS J075840.33+324723.4     & 119.66821 & 32.79014     & 3     & P\\
WISEPC J075946.98$-$490454.0\tablenotemark{c} & 119.94553 & $-$49.081232 & 3       & P        \\
WISEA J082640.45$-$164031.8  & 126.66664 & $-$16.674235 & 3       & P  & Keck      \\
WISEPC J083641.12$-$185947.2\tablenotemark{c} & 129.17153 & $-$18.996518 & 2       & P    & Keck    \\
SDSS J085234.90+472035.0     & 133.14493 & 47.34106     & 2       & P    & Keck    \\
WISEPA J085716.25+560407.6\tablenotemark{c}   & 134.31581 & 56.068336    & 3       & P  & Keck      \\
SDSSp J085758.45+570851.4    & 134.4911  & 57.1464      & 3       & P        \\
2MASS J09054654+5623117      & 136.44392 & 56.38661     & 0 & P   & Keck     \\
WISE J092055.40+453856.3     & 140.23088 & 45.64886     & 3  & P   \\
2MASSI J1010148$-$040649     & 152.56001 & $-$4.113933  & 3       & P        \\
2MASSW J1036530$-$344138     & 159.22075 & $-$34.696486 & 2       & P        \\
WISE J103907.73$-$160002.9\tablenotemark{c}   & 159.78173 & $-$16.001139 & 3       & P  & Keck      \\
2MASS J10430758+2225236      & 160.7809  & 22.423186    & 2       & P  & HST      \\
2MASSI J1104012+195921       & 166.00555 & 19.989852    & 1       & P  & HST      \\
CFBDS J111807$-$064016       & 169.52805 & $-$6.669126  & 0 & P  & Keck      \\
WISEP J115013.88+630240.7\tablenotemark{c}    & 177.56027 & 63.044022    & 3       & P   & Keck     \\
SDSS J115553.86+055957.5     & 178.97306 & 5.999081     & 3       & P   & Keck     \\
SDSSp J120358.19+001550.3    & 180.98575 & 0.26256    & 1         & P     \\
2MASSI J1213033$-$043243     & 183.26275 & $-$4.545563  & 3       & P & HST       \\
SDSS J121951.45+312849.4     & 184.96319 & 31.480466    & 3       & P        \\
2MASS J123147.53+084733.1      & 187.94283 & 8.788455     & 3     & P        \\
PSO J201.0320+19.1072        & 201.03179 & 19.106953    & 2       & P        \\
2MASS J13243559+6358284\tablenotemark{a}      & 201.14508 & 63.974268    & 3       & P        \\
SDSSp J132629.82$-$003831.5  & 201.62282 & $-$0.642752  & 3       & P        \\
WISE J140035.40$-$385013.5   & 210.14758 & $-$38.83746  & 3       & P  & Keck      \\
SDSS J153453.33+121949.2     & 233.72292 & 12.330219    & 0       & P        \\
PSO J247.3273+03.5932        & 247.32757 & 3.592974     & 3       & P  & Keck      \\
WISE J165842.56+510335.0     & 254.67684 & 51.059217    & 3       & P & Keck       \\
WISE J172134.46+111739.4     & 260.39341 & 11.294473    & 2       & P   & Keck     \\
WISEA J172907.10$-$753017.0\tablenotemark{c}  & 262.27553 & $-$75.505234 & 3       & P        \\
2MASS J17461199+5034036      & 266.55258 & 50.567694    & 2       & P        \\
WISE J181329.40+283533.3\tablenotemark{c}     & 273.37128 & 28.591289    & 3       & P   & Keck      \\
2MASS J19010601+4718136      & 285.27586 & 47.305732    & 3       & P   & Keck     \\
WISE J203042.79+074934.7     & 307.68001 & 7.826027     & 2       & P   & Keck     \\
WISE J204356.42+622048.9     & 310.98754 & 62.347831    & 1       & P   & Keck     \\
PSO J319.3102$-$29.6682      & 319.31057 & $-$29.668531 & 3       & P & Keck       \\
CWISEP J213249.05+690113.7\tablenotemark{c}   & 323.20562 & 69.020736    & 2       & P        \\
2MASS J21373742+0808463      & 324.40963 & 8.146552     & 1       & P  & Keck      \\
WISE J214155.85$-$511853.1   & 325.48432 & $-$51.31501  & 2       & P        \\
2MASS J23174712$-$4838501    & 349.4483  & $-$48.646791 & 2       & P        \\
WISEPC J232728.75$-$273056.5 & 351.86997 & $-$27.515783 & 3       & P        \\
2MASS J23440624$-$0733282    & 356.02605 & $-$7.5584    & 0 & P   \\
\enddata
\tablenotetext{a}{Possible spectral binary \citep{Burgasser et al.(2010)}}
\tablenotetext{b}{Keck: High-resolution imaging from the Keck Observatory Archive ({\url{https://koa.ipac.caltech.edu}}), HST: High-resolution imaging from the HST/NICMOS image archive studied in \cite{Factor Kraus(2022)}}
\tablenotetext{c}{Object with spectral type T7.5 or later}
\end{deluxetable}

\begin{deluxetable}{llllll}
\tabletypesize{\scriptsize}
\tablecaption{Known Binaries and Subdwarfs}\label{Table K}  
\tablehead{
\colhead{Object Name}     & \colhead{RA (deg)}  & \colhead{Dec (deg)}    & \colhead{Class} & \colhead{Subclass} & \colhead{Ref} \\
}
\startdata
WISE J003110.04+574936.3     & 7.79433   & 57.826773    & 2       & K & 1   \\
WISEPC J013836.59$-$032221.2 & 24.652471 & $-$3.373118  & 3       & K & 2   \\
WISEPA J022623.98$-$021142.8 & 36.599284 & $-$2.195299  & 1       & K & 3   \\
WISEPA J045853.89+643452.9\tablenotemark{a}   & 74.725621 & 64.5818      & 3       & K & 4   \\
WISEA J064750.85$-$154616.4  & 101.96192 & $-$15.77122  & 0       & K & 1   \\
PSO J103.0927+41.4601        & 103.09287 & 41.460042    & 2       & K & 1   \\
WISEPC J121756.91+162640.2\tablenotemark{a}   & 184.48809 & 16.443144    & 1       & K & 5   \\
SIMP J1619275+031350         & 244.86502 & 3.229445     & 1       & K & 6   \\
WISEPA J171104.60+350036.8\tablenotemark{a}   & 257.76888 & 35.010238    & 2       & K & 5   \\
2MASS J21522609+0937575      & 328.10965 & 9.633177     & 3       & K & 7   \\
PSO J330.3214+32.3686        & 330.32183 & 32.368747    & 3       & K & 2   \\
2MASS J22551861$-$5713056    & 343.82535 & $-$57.2197   & 3       & K & 8   \\
2MASS J06453153$-$6646120    & 101.36693 & $-$66.76297  & 3       & S & 9   \\
\enddata
\tablenotetext{a}{Object with spectral type T7.5 or later}
\tablecomments{Reference Codes: (1) \cite{Best et al.(2021)}, (2) Unpublished data in the Keck Observatory Archive ({\url{https://koa.ipac.caltech.edu}}), (3) \cite{Kirkpatrick et al.(2019)}, (4) \cite{Gelino et al.(2011)}, (5) \cite{Liu et al.(2012)}, (6) \cite{Artigau et al.(2011)}, (7) \cite{Reid et al.(2006)}, (8) \cite{Reid et al.(2008)}, (9) \cite{Kirkpatrick et al.(2010)}}
\end{deluxetable}

\end{center}


\begin{thebibliography}{}
\bibitem[Allers et al.(2007)]{Allers et al.(2007)} Allers, K.~N., Jaffe, D.~T., Luhman, K.~L., et al.\ 2007, \apj, 657, 511. doi:10.1086/510845

\bibitem[Apai et al.(2017)]{Apai et al.(2017)} Apai, D., Karalidi, T., Marley, M.~S., et al.\ 2017, Science, 357, 683. doi:10.1126/science.aam9848

\bibitem[Artigau et al.(2009)]{Artigau et al.(2009)} Artigau, {\'E}., Bouchard, S., Doyon, R., et al.\ 2009, \apj, 701, 1534. doi:10.1088/0004-637X/701/2/1534

\bibitem[Artigau et al.(2011)]{Artigau et al.(2011)} Artigau, {\'E}., Lafreni{\`e}re, D., Doyon, R., et al.\ 2011, \apj, 739, 48. doi:10.1088/0004-637X/739/1/48

\bibitem[Artigau(2018)]{Artigau(2018)} Artigau, {\'E}.\ 2018, Handbook of Exoplanets, 94. doi:10.1007/978-3-319-55333-7\_94

\bibitem[Bailer-Jones \& Mundt(1999)]{Bailer 1999} Bailer-Jones, C.~A.~L. \& Mundt, R.\ 1999, \aap, 348, 800

\bibitem[Bardalez Gagliuffi et al.(2014)]{Bardalez Gagliuffi et al.(2014)} Bardalez Gagliuffi, D.~C., Burgasser, A.~J., Gelino, C.~R., et al.\ 2014, \apj, 794, 143.doi:10.1088/0004-637X/794/2/143

\bibitem[Bardalez Gagliuffi et al.(2015)]{Bardalez Gagliuffi et al.(2015)} Bardalez Gagliuffi, D.~C., Gelino, C.~R., \& Burgasser, A.~J.\ 2015, \aj, 150, 163. doi:10.1088/0004-6256/150/5/163

\bibitem[Bardalez Gagliuffi et al.(2019)]{Bardalez Gagliuffi et al.(2019)} Bardalez Gagliuffi, D., Ward-Duong, K., Faherty, J., et al.\ 2019, \baas, 51, 285

\bibitem[Beichman et al.(2013)]{Beichman et al.(2013)} Beichman, C., Gelino, C.~R., Kirkpatrick, J.~D., et al.\ 2013, \apj, 764, 101. doi:10.1088/0004-637X/764/1/101

\bibitem[Best et al.(2021)]{Best et al.(2021)} Best, W.~M.~J., Liu, M.~C., Magnier, E.~A., et al.\ 2021, \aj, 161, 42. doi:10.3847/1538-3881/abc893

\bibitem[Burgasser et al.(2006)]{Burgasser et al.(2006)} Burgasser, A.~J., Kirkpatrick, J.~D., Cruz, K.~L., et al.\ 2006, \apjs, 166, 585. doi:10.1086/506327

\bibitem[Burgasser et al.(2007)]{Burgasser et al.(2007)} Burgasser, A.~J., Reid, I.~N., Siegler, N., et al.\ 2007, Protostars and Planets V, 427

\bibitem[Burgasser et al.(2010)]{Burgasser et al.(2010)} Burgasser, A.~J., Cruz, K.~L., Cushing, M., et al.\ 2010, \apj, 710, 1142. doi:10.1088/0004-637X/710/2/1142

\bibitem[Cushing et al.(2016)]{Cushing et al.(2016)} Cushing, M.~C., Hardegree-Ullman, K.~K., Trucks, J.~L., et al.\ 2016, \apj, 823, 152. doi:10.3847/0004-637X/823/2/152

\bibitem[Davies \& Kasper(2012)]{Davies Kasper(2012)} Davies, R. \& Kasper, M.\ 2012, \araa, 50, 305. doi:10.1146/annurev-astro-081811-125447

\bibitem[Deacon et al.(2017)]{Deacon et al.(2017)} Deacon, N.~R., Magnier, E.~A., Best, W.~M.~J., et al.\ 2017, \mnras, 468, 3499. doi:10.1093/mnras/stx440

\bibitem[Deming et al.(2015)]{Deming et al.(2015)} Deming, D., Knutson, H., Kammer, J., et al.\ 2015, \apj, 805, 132. doi:10.1088/0004-637X/805/2/132

\bibitem[Ducrot et al.(2020)]{Ducrot et al.(2020)} Ducrot, E., Gillon, M., Delrez, L., et al.\ 2020, \aap, 640, A112. doi:10.1051/0004-6361/201937392

\bibitem[Dye et al.(2018)]{Dye et al.(2018)} Dye, S., Lawrence, A., Read, M.~A., et al.\ 2018, \mnras, 473, 5113. doi:10.1093/mnras/stx2622

\bibitem[Eriksson et al.(2019)]{Eriksson et al.(2019)} Eriksson, S.~C., Janson, M., \& Calissendorff, P.\ 2019, \aap, 629, A145. doi:10.1051/0004-6361/201935671

\bibitem[Esplin et al.(2016)]{Esplin et al.(2016)} Esplin, T.~L., Luhman, K.~L., Cushing, M.~C., et al.\ 2016, \apj, 832, 58. doi:10.3847/0004-637X/832/1/58

\bibitem[Factor \& Kraus(2022)]{Factor Kraus(2022)} Factor, S.~M. \& Kraus, A.~L.\ 2022, \aj, 164, 244. doi:10.3847/1538-3881/ac88d3

\bibitem[Fazio et al.(2004)]{Fazio et al.(2004)} Fazio, G.~G., Hora, J.~L., Allen, L.~E., et al.\ 2004, \apjs, 154, 10. doi:10.1086/422843

\bibitem[Gardner et al.(2006)]{Gardner et al.(2006)} Gardner, J.~P., Mather, J.~C., Clampin, M., et al.\ 2006, \ssr, 123, 485. doi:10.1007/s11214-006-8315-710.48550/arXiv.astro-ph/0606175

\bibitem[Ge et al.(2019)]{Ge et al.(2019)} Ge, H., Zhang, X., Fletcher, L.~N., et al.\ 2019, \aj, 157, 89. doi:10.3847/1538-3881/aafba7

\bibitem[Gelino \& Marley(2000)]{Gelino Marley(2000)} Gelino, C. \& Marley, M.\ 2000, From Giant Planets to Cool Stars, 212, 322

\bibitem[Gelino et al.(2011)]{Gelino et al.(2011)} Gelino, C.~R., Kirkpatrick, J.~D., Cushing, M.~C., et al.\ 2011, \aj, 142, 57. doi:10.1088/0004-6256/142/2/57

\bibitem[Gelino(2012)]{Gelino(2012)} Gelino, C.\ 2012, Keck Observatory Archive NIRC2, N099N2L

\bibitem[Gizis et al.(2017)]{Gizis et al.(2017)} Gizis, J.~E., Paudel, R.~R., Mullan, D., et al.\ 2017, \apj, 845, 33. doi:10.3847/1538-4357/aa7da0

\bibitem[Hirst \& Cardenes(2016)]{Hirst Cardenes(2016)} Hirst, P. \& Cardenes, R.\ 2016, \procspie, 9913, 99131E. doi:10.1117/12.2231833

\bibitem[Hora et al.(2012)]{Hora et al.(2012)} Hora, J.~L., Marengo, M., Park, R., et al.\ 2012, \procspie, 8442, 844239. doi:10.1117/12.926894

\bibitem[IRSA (2022)]{IRSA 2022} IRSA, 2022, Spitzer Heritage Archive, IPAC, doi:10.26131/IRSA543

\bibitem[Kellogg et al.(2017)]{Kellogg et al.(2017)} Kellogg, K., Metchev, S., Heinze, A., et al.\ 2017, \apj, 849, 72. doi:10.3847/1538-4357/aa8e4f

\bibitem[Kendall et al.(2007)]{Kendall et al.(2007)} Kendall, T.~R., Jones, H.~R.~A., Pinfield, D.~J., et al.\ 2007, \mnras, 374, 445. doi:10.1111/j.1365-2966.2006.11026.x

\bibitem[Kirkpatrick et al.(2010)]{Kirkpatrick et al.(2010)} Kirkpatrick, J.~D., Looper, D.~L., Burgasser, A.~J., et al.\ 2010, \apjs, 190, 100. doi:10.1088/0067-0049/190/1/100

\bibitem[Kirkpatrick et al.(2011)]{Kirkpatrick et al.(2011)} Kirkpatrick, J.~D., Cushing, M.~C., Gelino, C.~R., et al.\ 2011, \apjs, 197, 19. doi:10.1088/0067-0049/197/2/19

\bibitem[Kirkpatrick et al.(2016)]{Kirkpatrick et al.(2016)} Kirkpatrick, J.~D., Schneider, A., Fajardo-Acosta, S., et al.\ 2016, VizieR Online Data Catalog, J/ApJ/783/122

\bibitem[Kirkpatrick et al.(2019)]{Kirkpatrick et al.(2019)} Kirkpatrick, J.~D., Martin, E.~C., Smart, R.~L., et al.\ 2019, \apjs, 240, 19. doi:10.3847/1538-4365/aaf6af

\bibitem[Kirkpatrick et al.(2021)]{Kirkpatrick et al.(2021)} Kirkpatrick, J.~D., Gelino, C.~R., Faherty, J.~K., et al.\ 2021, \apjs, 253, 7. doi:10.3847/1538-4365/abd107

\bibitem[Kolb \& Baraffe(1999)]{Kolb 1999} Kolb, U. \& Baraffe, I.\ 1999, \mnras, 309, 1034. doi:10.1046/j.1365-8711.1999.02926.x

\bibitem[Lang(2014)]{Lang(2014)} Lang, D.\ 2014, \aj, 147, 108. doi:10.1088/0004-6256/147/5/108

\bibitem[Lawrence et al.(2007)]{Lawrence et al.(2007)} Lawrence, A., Warren, S.~J., Almaini, O., et al.\ 2007, \mnras, 379, 1599. doi:10.1111/j.1365-2966.2007.12040.x

\bibitem[Lee et al.(2018)]{Lee et al.(2018)} Lee, E.~K.~H., Blecic, J., \& Helling, C.\ 2018, \aap, 614, A126. doi:10.1051/0004-6361/201731977

\bibitem[Leggett et al.(2016)]{Leggett et al.(2016)} Leggett, S.~K., Cushing, M.~C., Hardegree-Ullman, K.~K., et al.\ 2016, \apj, 830, 141. doi:10.3847/0004-637X/830/2/141

\bibitem[L{\'e}pine et al.(2003)]{Lepine et al.(2003)} L{\'e}pine, S., Rich, R.~M., \& Shara, M.~M.\ 2003, \apjl, 591, L49. doi:10.1086/377069

\bibitem[Limbach et al.(2021)]{Limbach et al.(2021)} Limbach, M.~A., Vos, J.~M., Winn, J.~N., et al.\ 2021, \apjl, 918, L25. doi:10.3847/2041-8213/ac1e2d

\bibitem[Liu et al.(2012)]{Liu et al.(2012)} Liu, M.~C., Dupuy, T.~J., Bowler, B.~P., et al.\ 2012, \apj, 758, 57. doi:10.1088/0004-637X/758/1/57

\bibitem[Lodieu et al.(2019)]{Lodieu et al.(2019)} Lodieu, N., Allard, F., Rodrigo, C., et al.\ 2019, \aap, 628, A61. doi:10.1051/0004-6361/201935299

\bibitem[Lowrance et al.(2005)]{Lowrance et al.(2005)} Lowrance, P.~J., Becklin, E.~E., Schneider, G., et al.\ 2005, \aj, 130, 1845. doi:10.1086/432839

\bibitem[Luhman(2012)]{Luhman(2012)} Luhman, K.~L.\ 2012, \araa, 50, 65. doi:10.1146/annurev-astro-081811-125528

\bibitem[Mace(2014)]{Mace(2014)} Mace, G.~N.\ 2014, VizieR Online Data Catalog, V/144

\bibitem[Mainzer et al.(2011)]{Mainzer et al.(2011)} Mainzer, A., Bauer, J., Grav, T., et al.\ 2011, \apj, 731, 53. doi:10.1088/0004-637X/731/1/53

\bibitem[Mainzer et al.(2014)]{Mainzer et al.(2014)} Mainzer, A., Bauer, J., Cutri, R.~M., et al.\ 2014, \apj, 792, 30. doi:10.1088/0004-637X/792/1/30

\bibitem[Makovoz \& Marleau(2005)]{Makovoz Marleau(2005)} Makovoz, D. \& Marleau, F.~R.\ 2005, \pasp, 117, 1113. doi:10.1086/432977

\bibitem[Makovoz et al.(2006)]{Makovoz et al.(2006)} Makovoz, D., Roby, T., Khan, I., et al.\ 2006, \procspie, 6274, 62740C. doi:10.1117/12.672536

\bibitem[Marley et al.(2013)]{Marley et al.(2013)} Marley, M.~S., Ackerman, A.~S., Cuzzi, J.~N., et al.\ 2013, Comparative Climatology of Terrestrial Planets, 367. doi:10.2458/azu\_uapress\_9780816530595-ch015

\bibitem[Martinez \& Kraus(2019)]{Martinez Kraus(2019)} Martinez, R.~A. \& Kraus, A.~L.\ 2019, \aj, 158, 134. doi:10.3847/1538-3881/ab32e6

\bibitem[Mason et al.(2001)]{Mason et al.(2001)} Mason, B.~D., Wycoff, G.~L., Hartkopf, W.~I., et al.\ 2001, \aj, 122, 3466. doi:10.1086/323920

\bibitem[McMahon et al.(2021)]{McMahon et al.(2021)} McMahon, R.~G., Banerji, M., Gonzalez, E., et al.\ 2021, VizieR Online Data Catalog, II/367

\bibitem[Schlafly, Meisner, \& Green (2019)]{Meisner Schlafly(2019)} Meisner, A. \& Schlafly, E.\ 2019, Astrophysics Source Code Library. ascl:1901.004

\bibitem[Meisner et al.(2020)]{Meisner et al.(2020)} Meisner, A.~M., Faherty, J.~K., Kirkpatrick, J.~D., et al.\ 2020, \apj, 899, 123. doi:10.3847/1538-4357/aba633

\bibitem[Meisner et al.(2021)]{Meisner et al.(2021)} Meisner, A.~M., Schneider, A.~C., Burgasser, A.~J., et al.\ 2021, \apj, 915, 120. doi:10.3847/1538-4357/ac013c10.48550/arXiv.2106.01387

\bibitem[Meisner et al.(2022)]{Meisner et al.(2022)} Meisner, A.~M., Lang, D., Schlafly, E.~F., et al.\ 2022, arXiv:2204.05748

\bibitem[Metchev et al.(2015)]{Metchev et al.(2015)} Metchev, S.~A., Heinze, A., Apai, D., et al.\ 2015, \apj, 799, 154. doi:10.1088/0004-637X/799/2/154

\bibitem[Moore et al.(2019)]{Moore et al.(2019)} Moore, K., Scholz, A., \& Jayawardhana, R.\ 2019, \apj, 872, 159. doi:10.3847/1538-4357/aaff5c

\bibitem[M{\"u}ller et al.(2016)]{Muller et al.(2016)} M{\"u}ller, T.~G., Balog, Z., Nielbock, M., et al.\ 2016, \aap, 588, A109. doi:10.1051/0004-6361/201527371

\bibitem[Nelson et al.(1986)]{Nelson et al.(1986)} Nelson, L.~A., Rappaport, S.~A., \& Joss, P.~C.\ 1986, \apj, 311, 226. doi:10.1086/164767

\bibitem[Radigan et al.(2012)]{Radigan et al.(2012)} Radigan, J., Jayawardhana, R., Lafreni{\`e}re, D., et al.\ 2012, \apj, 750, 105. doi:10.1088/0004-637X/750/2/105

\bibitem[Radigan et al.(2014a)]{Radigan et al.(2014)a} Radigan, J., Lafreni{\`e}re, D., Jayawardhana, R., et al.\ 2014a, \apj, 793, 75. doi:10.1088/0004-637X/793/2/75

\bibitem[Radigan(2014b)]{Radigan(2014)b} Radigan, J.\ 2014b, \apj, 797, 120. doi:10.1088/0004-637X/797/2/120

\bibitem[Reach et al.(2005)]{Reach et al.(2005)} Reach, W.~T., Megeath, S.~T., Cohen, M., et al.\ 2005, \pasp, 117, 978. doi:10.1086/432670

\bibitem[Reid et al.(2006)]{Reid et al.(2006)} Reid, I.~N., Lewitus, E., Allen, P.~R., et al.\ 2006, \aj, 132, 891. doi:10.1086/505626

\bibitem[Reid et al.(2008)]{Reid et al.(2008)} Reid, I.~N., Cruz, K.~L., Burgasser, A.~J., et al.\ 2008, \aj, 135, 580. doi:10.1088/0004-6256/135/2/580

\bibitem[Rowe-Gurney et al.(2021)]{Rowe-Gurney et al.(2021)} Rowe-Gurney, N., Fletcher, L.~N., Orton, G.~S., et al.\ 2021, \icarus, 365, 114506. doi:10.1016/j.icarus.2021.114506

\bibitem[Schlafly et al.(2012)]{Schlafly et al.(2012)} Schlafly, E.~F., Finkbeiner, D.~P., Juri{\'c}, M., et al.\ 2012, \apj, 756, 158. doi:10.1088/0004-637X/756/2/158

\bibitem[Schlafly et al.(2019)]{Schlafly et al.(2019)} Schlafly, E.~F., Meisner, A.~M., \& Green, G.~M.\ 2019, \apjs, 240, 30. doi:10.3847/1538-4365/aafbea

\bibitem[Simon et al.(2021)]{Simon et al.(2021)} Simon, A., Orton, G.~S., \& Wong, M.~H.\ 2021, HST Proposal, 16790

\bibitem[Skrutskie et al.(2006)]{Skrutskie et al.(2006)} Skrutskie, M.~F., Cutri, R.~M., Stiening, R., et al.\ 2006, \aj, 131, 1163. doi:10.1086/498708

\bibitem[Softich et al.(2022)]{Softich et al.(2022)} Softich, E., Schneider, A.~C., Patience, J., et al.\ 2022, \apjl, 926, L12. doi:10.3847/2041-8213/ac51d8

\bibitem[Stauffer et al.(2016)]{Stauffer et al.(2016)} Stauffer, J., Marley, M.~S., Gizis, J.~E., et al.\ 2016, \aj, 152, 142. doi:10.3847/0004-6256/152/5/142

\bibitem[Tamburo et al.(2022)]{Tamburo et al.(2022)} Tamburo, P., Muirhead, P.~S., McCarthy, A.~M., et al.\ 2022, \aj, 163, 253. doi:10.3847/1538-3881/ac64aa

\bibitem[Tannock et al.(2021)]{Tannock et al.(2021)} Tannock, M.~E., Metchev, S., Heinze, A., et al.\ 2021, \aj, 161, 224. doi:10.3847/1538-3881/abeb67

\bibitem[Tran et al.(2004)]{Tran et al.(2004)} Tran, H.~D., Mader, J., Conrad, A., et al.\ 2004, \aas

\bibitem[Tsuji(2005)]{Tsuji(2005)} Tsuji, T.\ 2005, \apj, 621, 1033. doi:10.1086/427747

\bibitem[Vos et al.(2022)]{Vos et al.(2022)} Vos, J.~M., Faherty, J.~K., Gagn{\'e}, J., et al.\ 2022, \apj, 924, 68. doi:10.3847/1538-4357/ac4502

\bibitem[Wenger et al.(2000)]{Wenger et al.(2000)} Wenger, M., Ochsenbein, F., Egret, D., et al.\ 2000, \aaps, 143, 9. doi:10.1051/aas:2000332

\bibitem[Werner et al.(2004)]{Werner et al.(2004)} Werner, M.~W., Roellig, T.~L., Low, F.~J., et al.\ 2004, \apjs, 154, 1. doi:10.1086/422992

\bibitem[Wright et al.(2010)]{Wright et al. 2010} Wright, E.~L., Eisenhardt, P.~R.~M., Mainzer, A.~K., et al.\ 2010, \aj, 140, 1868. doi:10.1088/0004-6256/140/6/1868

\bibitem[Yang et al.(2016)]{Yang et al.(2016)} Yang, H., Apai, D., Marley, M.~S., et al.\ 2016, \apj, 826, 8. doi:10.3847/0004-637X/826/1/8

\bibitem[Zhang et al.(2017)]{Zhang et al.(2017)} Zhang, Z.~H., Pinfield, D.~J., G{\'a}lvez-Ortiz, M.~C., et al.\ 2017, \mnras, 464, 3040. doi:10.1093/mnras/stw2438

\bibitem[Zhang et al.(2021)]{Zhang et al.(2021)} Zhang, J., Cooke, J., Canalizo, G., et al.\ 2021, GRB Coordinates Network, Circular Service, No. 30858, 30858

\end{thebibliography}
\end{document}